\documentclass[
    a4paper,
    jou,
    american,
    floatsintext
]{apa6}
\pdfoutput=1

\usepackage{amsmath}
\usepackage{amssymb}
\usepackage{setspace}
\usepackage{booktabs}
\usepackage{placeins}
\usepackage{titletoc}
\usepackage{lineno}
\usepackage{wrapfig}
\usepackage{multirow}
\usepackage[american]{babel}
\usepackage{epstopdf}
\usepackage{csquotes}
\usepackage[hidelinks]{hyperref}
\usepackage{color}
\usepackage{apacite}
\usepackage{framed}
\usepackage{subcaption} 
\usepackage{caption,graphicx,newfloat}
\usepackage{xr}

\newcommand{\appendixA}{Validation of Reanalysis Method via Simulations}
\newcommand{\appendixB}{Optimality of Median Classifier}
\newcommand{\appendixC}{Estimating Sensitivities From Typically Reported Results}
\newcommand{\appendixD}{Estimating the Ratio $q^2$ of Between- vs. Within-Subject Variance}
\newcommand{\appendixE}{Details of Reanalyzed Studies}
\newcommand{\appendixF}{Cost of Dichotomization in Significance Testing and Bayesian Analyses}
\newcommand{\appendixG}{Glossary}

\newcommand*\linenomathpatchAMS[1]{%
  \expandafter\pretocmd\csname #1\endcsname {\linenomathAMS}{}{}%
  \expandafter\pretocmd\csname #1*\endcsname{\linenomathAMS}{}{}%
  \expandafter\apptocmd\csname end#1\endcsname {\endlinenomath}{}{}%
  \expandafter\apptocmd\csname end#1*\endcsname{\endlinenomath}{}{}%
}

\expandafter\ifx\linenomath\linenomathWithnumbers
  \let\linenomathAMS\linenomathWithnumbers
  \patchcmd\linenomathAMS{\advance\postdisplaypenalty\linenopenalty}{}{}{}
\else
  \let\linenomathAMS\linenomathNonumbers
\fi

\linenomathpatchAMS{gather}
\linenomathpatchAMS{multline}
\linenomathpatchAMS{align}
\linenomathpatchAMS{alignat}
\linenomathpatchAMS{flalign}

\newcommand{\ITA}{ITA}
\newcommand{\ITAs}{ITAs}
\newcommand{\totalCitations}{3277}
\newcommand{\totalReanalyzed}{15}

\DeclareMathOperator{\Var}{Var}

\newcommand\numberthis{\addtocounter{equation}{1}\tag{\theequation}}

\newcommand{\condon}{\, | \,}%
\newcommand{\R}{I\!\!R}

\newtheorem{propositionnn}{Proposition}

\def\ba#1\ea{\begin{align*}#1\end{align*}} %
\def\banum#1\eanum{\begin{align}#1\end{align}} %

\title{\textbf{Advancing Research on Unconscious Priming: When can Scientists Claim an Indirect Task Advantage?}}
\shorttitle{When can scientists claim an ITA?}

\fiveauthors{Sascha Meyen}{Iris A. Zerweck}{Catarina Amado}{Ulrike von Luxburg}{Volker H. Franz}

\fiveaffiliations{Experimental Cognitive Science, University of Tübingen}{Experimental Cognitive Science, University of Tübingen}{Experimental Cognitive Science, University of Tübingen}{Theory of Machine Learning, University of Tübingen \\ Max Planck Institute for Intelligent Systems, Tübingen}{Experimental Cognitive Science, University of Tübingen}

  \AtBeginDocument{%
  }

\rightheader{Advancing Research on Unconscious Priming}
\leftheader{Meyen, Zerweck, Amado, von Luxburg, \& Franz}

\abstract{
\singlespacing
Current literature holds that many cognitive functions can be
performed outside consciousness. Evidence for this view comes from
unconscious priming. In a typical experiment, visual stimuli are
masked such that participants are close to chance performance when
directly asked to which of two categories the stimuli belong. This
close-to-zero sensitivity is seen as evidence that participants cannot
consciously report the category of the masked stimuli. Nevertheless,
the category of the masked stimuli can indirectly affect responses to
other stimuli (e.g., reaction times or brain activity)---an effect
called priming. The priming effect is seen as evidence for a higher
sensitivity to the masked stimuli in the indirect responses as
compared to the direct responses. Such an apparent difference in
sensitivities is taken as evidence that processing occurred
unconsciously. But we show that this ``standard reasoning of
unconscious priming'' is flawed: Sensitivities are not properly
compared, creating the wrong impression of a difference in
sensitivities even if there is none. We describe the appropriate way
to determine sensitivities, replicate the behavioral part of a
landmark study, develop methods to estimate sensitivities from
reported summary statistics of published studies, and use these
methods to reanalyze 15 highly influential studies. Results show that
the interpretations of many studies need to be changed and that a
community effort is required to reassess the vast literature on
unconscious priming. This process will allow scientists to learn more
about the true boundary conditions of unconscious priming, thereby
advancing the scientific understanding of consciousness.

}


\begin{document}

\maketitle



\singlespacing
Research on consciousness and its cerebral substrates has far-reaching
implications and received substantial attention in recent years
\cite{Michel_etal_19}. A driving factor comes from reports that masked
stimuli that are not consciously perceived can nevertheless affect
behavioral responses and brain activity
\cite{Kouider_Dehaene_07,vanDenBussche_09}. The exciting claim here is
that unconscious processing might be more than a mere residue of
conscious processing and may be performed by different neuronal
processes than conscious processing. Such results impact current
theories about the functional role of consciousness
\cite{dehaene2017consciousness,Kouider_Dehaene_07,vanDenBussche_09,Sklar_etal_12,Hassin_13},
might suggest parallel neuronal routes for unconscious vs. conscious
processing \cite{Morris_etal_99}, and might support theories of
superior unconscious processing
\cite{Custers_Aarts_10,dijksterhuis2006making,tenBrinke_etal_16}.

Here, we scrutinize one of the most frequently used approaches in this
field. We show that the \textbf{standard reasoning} in the
dissociation paradigm
\cite{hannula2005imaging,Holender_86,Schmidt_Vorberg_06,simons2007behavioral}
is flawed for mathematical reasons. It fails to provide meaningful
interpretation of the data, and needs to be replaced by an
\emph{appropriate analysis}. Because many studies have used the
standard reasoning, a large body of literature needs reassessment.
This has the potential to drastically change our views on unconscious
processing and its neuronal underpinnings. The fallacy we expose
affects a wide range of research areas because the standard reasoning
has been employed on such diverse topics as, for example, unconscious
processing of semantic meaning \cite{Dehaene_etal_98}, motivation
\cite{Pessiglione_etal_07}, emotion \cite{Morris_etal_98}, cognitive
control \cite{van2010unconscious}, and detection of lies 
\cite{tenBrinke_14}.

To assess how seriously the literature is affected, we proceeded in
three strands:
(a) We replicated the behavioral part of a landmark
study~\cite{Dehaene_etal_98} and showed that the appropriate analysis
of the data does not support unconscious priming (in contrast to the
claims of the original study).
(b) We developed statistical methods to reanalyze published studies
based on the reported $t$ and $F$ statistics (because access to the
full trial-by-trial data is often lacking). We validated this approach
by showing that our reanalysis of the published data of
\citeA{Dehaene_etal_98} is consistent with the results of our
replication. 
(c) We used our methods to reanalyze \totalReanalyzed{} highly
influential studies (with a total of \totalCitations{} citations in
Web of Science). Even though all these studies used the standard
reasoning to infer unconscious processing, their data tell a different
story.


\section{The Standard Reasoning of Unconscious Priming}

As a typical example for a study using the standard reasoning,
consider the study by \citeA{tenBrinke_14} who reported that humans
can detect liars better unconsciously than consciously: ``[T]he
unconscious mind identifies and processes cues to deception ... more
efficiently and effectively than the conscious mind.'' (p.~1104).  In
the following, we will describe the specifics of this study as well as
the general aspects that are typical for studies using the standard
reasoning.

Participants of \citeA{tenBrinke_14} first watched videos of suspects
who were either lying or telling the truth.
Then participants performed two tasks: The direct and the indirect
task. These tasks were supposed to measure conscious and unconscious
lie detection, respectively.

\begin{figure*}[!ht]
\centering
\includegraphics[width=.7\textwidth]{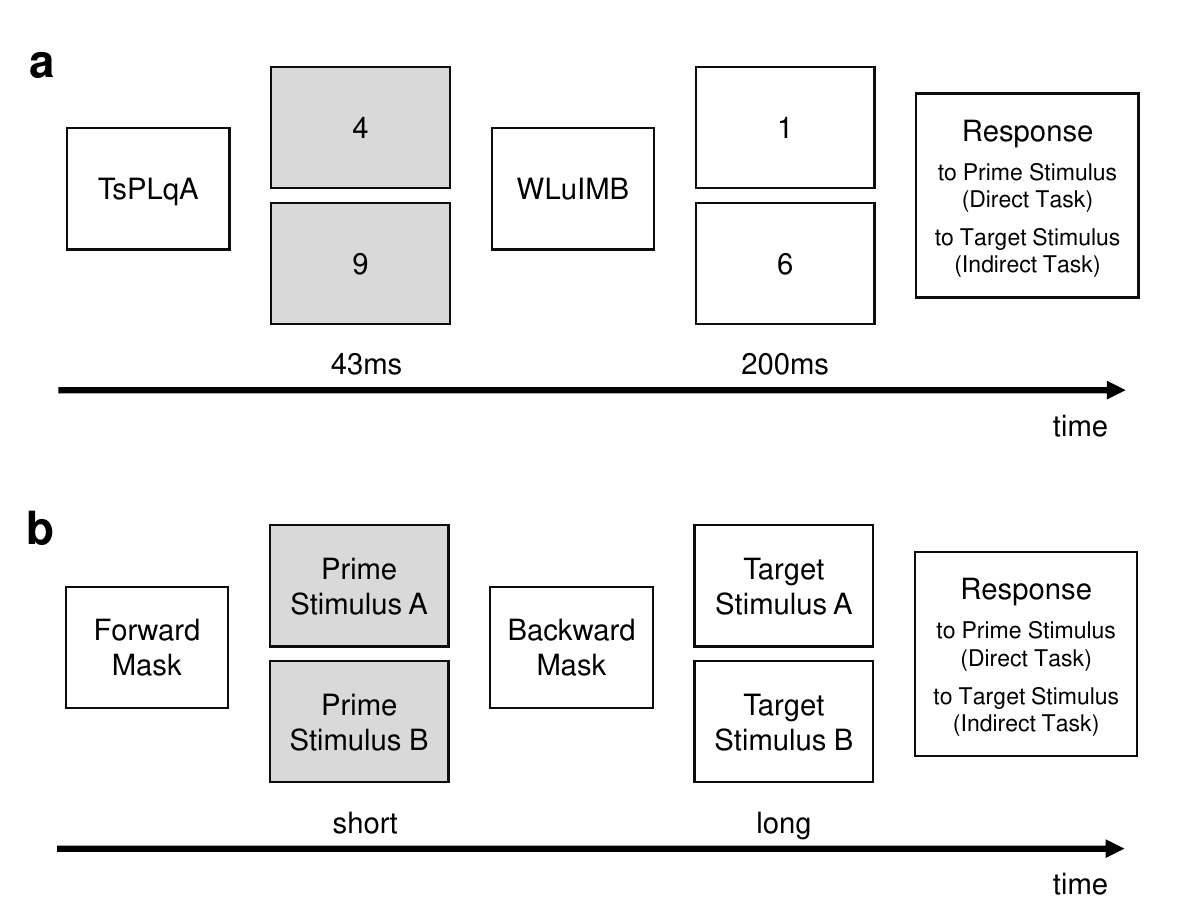}

\caption{\footnotesize 
\textbf{Typical study design to infer an indirect task advantage
(\ITA{}).} 
\textbf{(a)~Example study:} The study of
\protect\citeA{Dehaene_etal_98} is a prototypical example for
unconscious priming with number stimuli. In each trial, a masked
\emph{prime} stimulus is presented for a short duration followed by a
well visible target stimulus. In the direct task, participants
discriminated the primes and performance was close to chance level. In
the indirect task, participants discriminated the \emph{target}
stimuli by deciding whether they were smaller or larger than the
number~5. Reaction times (RTs) were faster and lateralization of brain
activity was larger when prime and target stimuli were congruent (both
smaller or both larger than~5) than when they were incongruent (one
larger one smaller). \protect\citeA{Dehaene_etal_98} followed the
standard reasoning to infer a higher sensitivity for the primes in the
indirect task than in the direct task (i.e., an \ITA{}) and conclude
that the primes were processed in the absence of conscious awareness.
\textbf{(b) General design:} In general, prime and target stimuli each
come from one of two categories, A or B. In the direct task,
participants discriminate the prime (e.g., guess whether it is from
category A or B) with a poor sensitivity. In the indirect task, the
same stimuli are presented and participants now discriminate the
target. In this task, the prime is shown to influence responses
resulting in faster RTs for congruent (A--A, or B--B) than incongruent
trials (incongruent: A--B, or B--A). From this pattern of results, the
standard reasoning infers an \ITA{} (cf.
Figure~\protect\ref{fig:ITAStandardReasoning}). 
}
\label{fig:ITAdesign}
\end{figure*}

In the \textbf{direct task}, participants judged which suspects had
been lying or telling the truth. Participants performed poorly with
an accuracy of only 49.62\%-correct (with chance level being 50\%),
which was taken by \citeA{tenBrinke_14} as evidence that participants
could not consciously detect liars with more than a poor sensitivity.
In the same way, studies using the standard reasoning typically let
participants directly discriminate stimuli belonging to one of two
categories (Figure~\ref{fig:ITAdesign}).  Participants'
performance---measured by the proportion of correct responses or by
the sensitivity index, $d'$, from Signal Detection Theory
\cite{Green_Swets_88}---is typically found to be close to chance
level. This result is then taken as evidence that conscious
discrimination of the presented stimuli is poor at best.

In the \textbf{indirect task} of \citeA{tenBrinke_14}, participants
categorized target--words, such as ``deceitful'' or ``honest'', into
two categories: lying or truth--telling. Before each target--word, a
masked picture of one of the suspects was briefly presented in order
to affect (or ``prime'') the responses to the target words (therefore
those masked stimuli are often called the ``primes''). ten Brinke et
al. found that participants' reaction times (RTs) to the target words
were faster when the primes were congruent with the targets (e.g., the
picture of a lying suspect before a lie--related word) than when the
primes were incongruent with the targets. That is,
\citeA{tenBrinke_14} found a congruency effect between primes and
targets in the indirect task. In the same way, studies using the
standard reasoning typically employ an indirect task attempting to
find such congruency effects (Figure~\ref{fig:ITAdesign}). These
congruency effects could be on RTs (as in the case
\citeNP{tenBrinke_14}), but also on other behavioral responses (e.g.,
skin conductance) or neurophysiological measures (e.g., in EEG or
fMRI).

\begin{figure*}[!ht]
\centering
\includegraphics[width=0.7\textwidth]{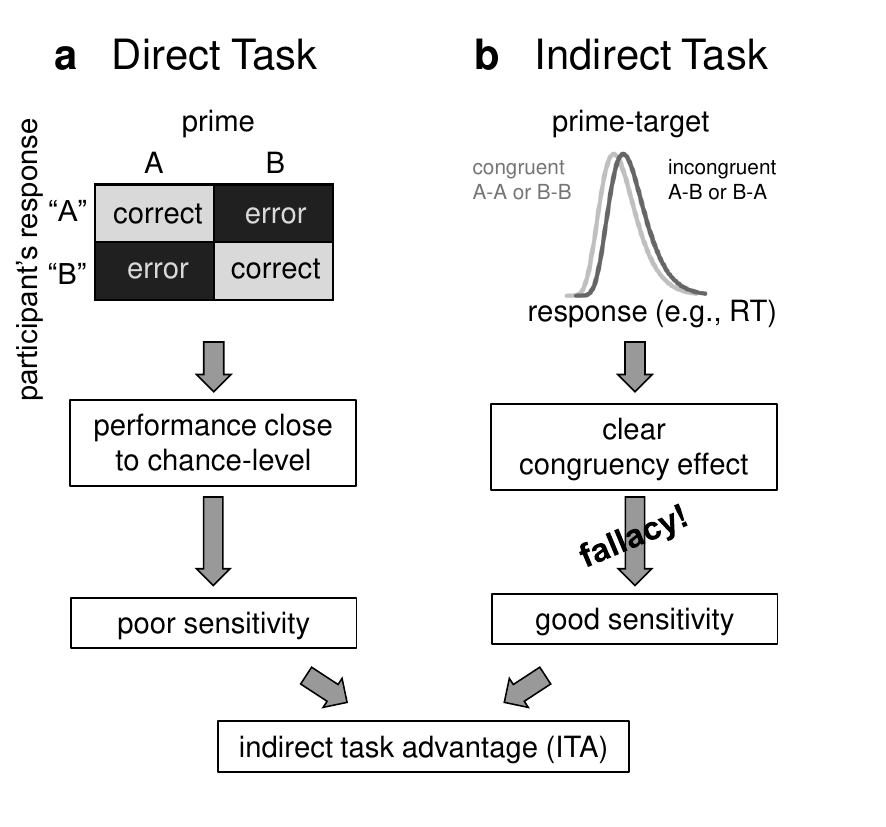}

\caption{\footnotesize 
\textbf{Standard reasoning to infer an indirect task advantage
(\ITA{}).}
\textbf{(a)} In the \emph{direct task}, the standard reasoning infers
from close-to-chance performance that there was poor sensitivity for
the primes, if any at all. \textbf{(b)} In the \emph{indirect task},
the standard reasoning infers from a clear congruency effect that
sensitivity was relatively good. Based on this pattern of results the
standard reasoning makes two inference steps: In Step~1, it
incorrectly infers that participants' responses in the indirect task
were more sensitive to the primes than responses in the direct task
(\ITA{}). In Step~2, it attributes this difference to unconscious
processing. However, already Step~1 of this reasoning is flawed
because a clear congruency effect does not necessarily indicate good
sensitivity. It could be caused by a sensitivity as poor as (or even
worse than) the sensitivity in the direct task! Because Step~1 is
independent of any (sometimes contentious) assumptions about conscious
vs. unconscious processing, our critique is also independent of any
such assumptions.
}
\label{fig:ITAStandardReasoning} 
\end{figure*}

Taken together, \citeA{tenBrinke_14} found the typical pattern of
results for the unconscious priming paradigm: (a) a poor accuracy, or
sensitivity in the direct task and (b) a clear congruency effect in
the indirect task. Based on this pattern, they concluded that
participants' indirect task revealed more accurate lie detection than
the direct task: ``[I]ndirect measures of deception detection are
significantly more accurate than direct measures'' (p.~1098,
Abstract). In the same way, studies using the standard reasoning infer
from such a pattern of results better sensitivity for the primes in
the indirect task than in the direct task
(Figure~\ref{fig:ITAStandardReasoning}). We dubbed this situation the
\textbf{indirect task advantage}, or short \textbf{\ITA{}}.  It is
important to note that the claim of an \ITA{} is, in this phase of the
reasoning, independent of any considerations about conscious or
unconscious processing. We call this descriptive phase of the standard
reasoning \textbf{Step~1}.

In \textbf{Step~2} of the standard reasoning, \citeA{tenBrinke_14}
used the presumed \ITA{} to conclude superior unconscious processing:
``[A]lthough humans cannot consciously discriminate liars from truth
tellers, they do have a sense, on some less-conscious level, of when
someone is lying'' (p.~1103). The authors thereby followed the typical
assumption that direct and indirect tasks measure conscious and
unconscious processing, respectively. Based on the supposed \ITA{}
from Step~1, these assumptions lead to the typical conclusion that
participants processed the category of the masked stimuli better
unconsciously than they can consciously report.

The standard reasoning is summarized for example by
\citeA{dell1999unconscious}: ``The present work follows the tradition
of providing evidence for a dissociation between direct and indirect
effects of unconsciously presented stimuli (Greenwald, Klinger \&
Schuh, 1995; Draine \& Greenwald, 1998). More specifically, null
effects are sought in direct measures (i.e. where subjects respond
directly to the unconsciously presented stimuli) accompanied by
non-null indirect effects (i.e. priming effects)'' (p.~B2).
For further description of the standard reasoning see
also \citeA{merikle1992perception} and \citeA{simons2007behavioral}.
Even though some studies may not state an \ITA{} as explicitly as
shown here, it is nevertheless necessarily implied when claims about
unconscious processing are made because Step~1 is a necessary
condition for Step~2.

But note that the standard reasoning infers better sensitivity in the
indirect task than in the direct task (i.e., an \ITA{}) without ever
calculating sensitivity (or accuracy) in the indirect task to compare
against that in the direct task. For example, \citeA{tenBrinke_14}
only demonstrated a congruency effect on RTs. \textbf{However, if this
congruency effect indicated accurate unconscious lie detection, we
should be able to use the RT data to determine which of the suspects
were lying with a higher accuracy than in the direct task.} Otherwise
the congruency effect does not truly provide evidence for better 
accuracy in the indirect than in the direct task (i.e., for an
\ITA{}).

Because \citeA{tenBrinke_14} laudably followed an open-data policy,
\citeA{Franz_vonLuxburg_15_PS} were able to reanalyze how much
evidence the RT data truly provided for better accuracy in the
indirect than in the direct task. To assess this, they determined
statistically optimal classifiers, used the RT of each trial in the
indirect task to classify (``predict'' in the nomenclature of
statistical learning) which of the suspects were lying, and found the
accuracy in the indirect task to be only at 50.6\%-correct ($SEM =
0.3\%$; see below for more details on the methods used). This value is
very similar to---and not significantly different from---the accuracy
in the direct task (which was 49.62\%-correct; $SEM = 1.4\%$).
Therefore, ten Brinke et al.'s inference in Step~1 was flawed: Their
data did not provide evidence for better accuracy in the indirect than
in the direct task. In our words, there was no evidence for an \ITA{}.
Because the existence of an \ITA{} in Step~1 is a necessary condition
for Step~2 of the standard reasoning, inferences about unconscious
processing were not warranted.

In the following section, we show in detail why claiming an \ITA{}
based on the standard reasoning is flawed. Note, that our critique
focuses on how an \ITA{} is established in Step~1 and is therefore
independent of any assumptions about conscious vs.  unconscious
processing, which are relevant only in Step~2 and for which different,
sometimes contentious approaches exist (e.g.,
\citeNP{Eriksen_60,Erdelyi_86,Holender_86,Reingold_Merikle_88,reingold1990inter,Schmidt_Vorberg_06}).
We avoid these discussions by focusing on an empirical investigation
of Step~1 which makes our critique very general.


\section{The Standard Reasoning is Flawed and Fails to Provide
Evidence for an \ITA{}} \label{sec:examples}
The standard reasoning is intuitively very appealing, which seems to
be one reason for its popularity. The colloquial version of the
arguments to infer an \ITA{} in Step~1 goes like this: ``Participants
have a very hard time to discriminate the masked stimuli in the direct
task. They are very close to zero sensitivity and usually not
significantly above chance. Nevertheless we find clear and highly
significant congruency effects in the indirect task. Therefore, it
seems obvious, that the indirect task responses are more sensitive to
the masked stimuli than the direct task responses.''

However, this intuition is misguided. To see this, consider what
happens if we increased the number of observations (number of
participants or trials). The poor sensitivity in the direct task
(Figure~\ref{fig:ITAStandardReasoning}a) will only be measured more
precisely but will still be poor. In contrast, the congruency effect
in the indirect task (Figure~\ref{fig:ITAStandardReasoning}b) becomes
clearer because it is based on the difference between congruent and
incongruent condition means: With more observations, the variability
of the two means becomes smaller, such that the difference between
them becomes clearer. Therefore, a clear congruency effect can be
generated by a good underlying sensitivity (corresponding to, say, $d'
= 5$ or 99\%-correct) but it can also be generated by a very poor
sensitivity (say, $d' = 0.05$ or 51\%-correct).
In cases where the sensitivity of the indirect task is as poor as in
the direct task, there is no ITA and further interpretations about
unconscious processing are unwarranted. Not recognizing this is the
\textbf{main fallacy of the standard reasoning}. We demonstrate this
problem by using a toy example.


\subsection{Toy Example With Baby Weights} 
\label{sec:toyExampleBabies}

To illustrate the problem of the standard reasoning, consider an
example in which responses in the direct and indirect tasks
are based on the \emph{exact same} underlying sensitivity.
Nevertheless, the standard reasoning would erroneously infer that
responses in the indirect task were \textit{more} sensitive than
responses in the direct task (i.e., an \ITA{}).

\begin{figure*}[!ht]
\centering
\includegraphics[width=0.8\textwidth]{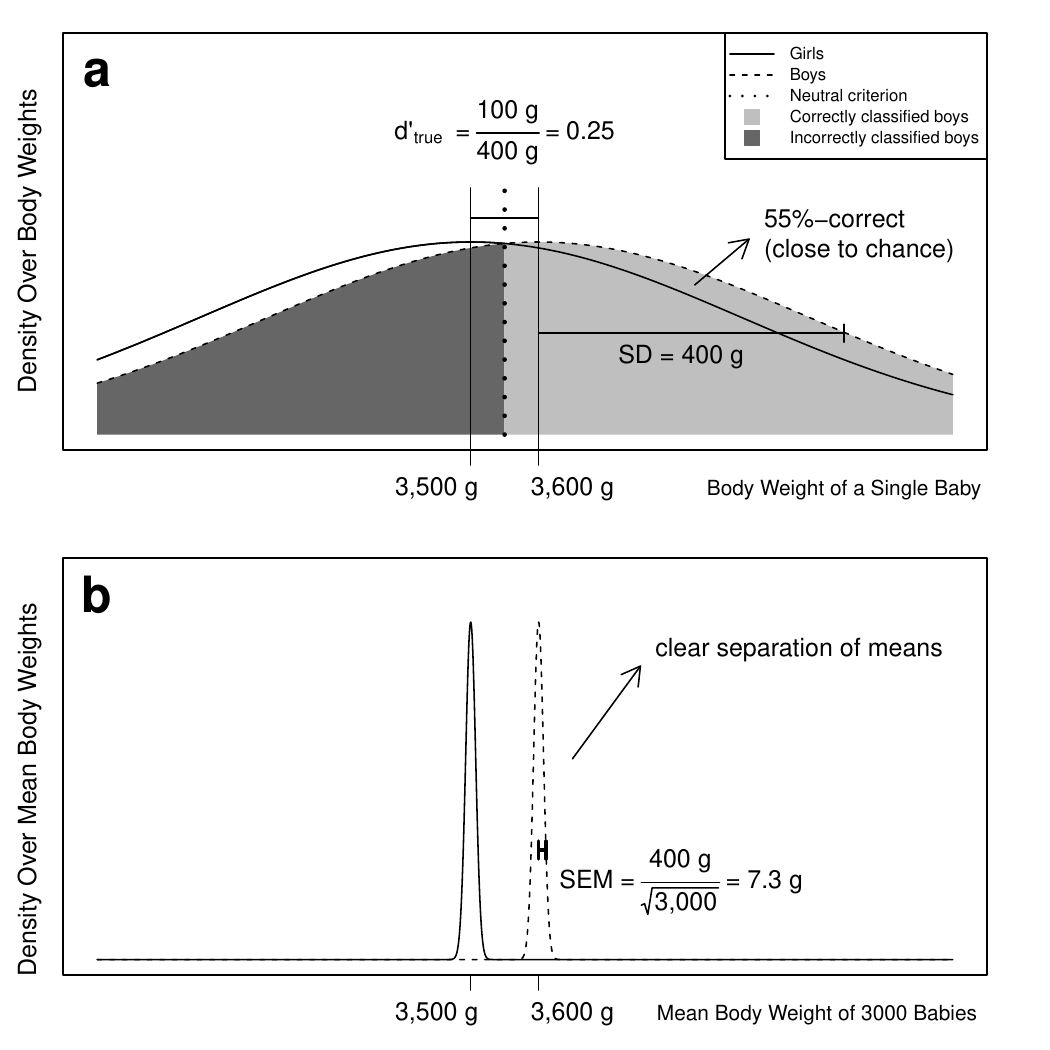}

\caption{\footnotesize 
\textbf{Toy-example demonstrating fallacy of standard reasoning.} 
We show that even when responses in the direct and indirect tasks are
based on the \emph{exact same} information the standard reasoning
would nevertheless infer an indirect task advantage (\ITA{}): a
\emph{higher} sensitivity in the indirect as compared to the direct
task. Consider participants of a hypothetical experiment measured the
birth weight of babies but did not know the babies' sex. \textbf{(a)}
In the \emph{direct task}, participants used the weight of an
individual baby to guess whether it is a girl or a boy. The weight
distributions overlap heavily such that sensitivity would be poor
($d'_\text{true} = 0.25$; corresponding to~$55\%$-correct).
\textbf{(b)} In the \emph{indirect task}, participants responded by
simply stating the measured weights. The experimenter would average
those responses across many trials (e.g., across 3000 girls and 3000
boys). The resulting group means are much less variable than the
individual weights such that the experimenter would obtain a clear
difference between the two group means (this corresponds to a clear
congruency effect in the priming paradigm). Based on this result, the
standard reasoning would erroneously infer that participants had
relatively good sensitivity about whether a baby was a girl or a boy
in the indirect task---better than in the direct task. That is, the
standard reasoning would infer an \ITA{} even though the \emph{exact
same} information created the responses in both tasks. Weight--data
based on \protect\citeA{janssen2007standards}.
}
\label{fig:toyExample}
\end{figure*}

Consider participants measured the birth weights of newborn girls
(category A) and boys (category B), such that they only knew the
weight of the babies but not the biological sex. This would be all the
information participants had in both, direct and indirect, tasks.

In the direct task, participants would use this weight information to
guess whether a baby is a girl or a boy (newborn girls weigh a little
less than boys). Due to the large overlap between the weight
distributions (Figure \ref{fig:toyExample}a), participants would be
correct in only approximately $55\%$ of the cases even when using an
ideal decision criterion. This corresponds to a poor performance that
is close-to-chance level (50\%). Following the standard reasoning, an
experimenter would correctly infer a poor sensitivity in this direct
task (Figure~\ref{fig:ITAStandardReasoning}a).

In the indirect task, participants would simply report the numerically
measured weight of the babies. The experimenter would \emph{average}
those responses across groups of baby girls and boys and would
calculate the difference of the \emph{mean} responses to those two
groups. With increasing group sizes, the experimenter would eventually
find a clear difference (corresponding to the clear congruency effect
in the priming paradigm). Figure \ref{fig:toyExample}b shows this for
$3000$ observations in each group, which is a typical number of
observations in the indirect task (e.g., when 10 participants perform
300 trials in each condition, the number of observations per condition
is $10~\times~300 = 3000$). Following the standard reasoning, the
experimenter would incorrectly infer a good sensitivity in this
indirect task (Figure~\ref{fig:ITAStandardReasoning}b).

\textbf{Here is the catch:} The standard reasoning would incorrectly
interpret this pattern of results as evidence for better sensitivity
in the indirect task than in the direct task (i.e., for an \ITA{}).
However, this inference is wrong because participants gave responses
in both tasks based on exactly the same information: In both tasks
they knew only the weight of the babies. The illusion of an \ITA{} is
generated by the different data-analysis strategies of the
experimenter in the two tasks and by the fact that the experimenter
never attempted to estimate the sensitivity in the indirect task.

\subsection{Further Details on the Standard Reasoning}

We have shown that the standard reasoning is flawed because it infers
an \ITA{} in Step~1 even when there is none. The problem is that the
standard reasoning calculates two very different things in the direct
and indirect tasks: In the direct task, it calculates how well each
stimulus can be classified on a trial-by-trial level. In the indirect
task, it assesses whether there is a difference in mean
responses. These are two very different things and it is a priori to
be expected that the sensitivity in single trials can be poor while
mean responses can nevertheless be clearly separated between the two
categories given enough trials. A more appropriate analysis to
determine whether there is an \ITA{} would need to estimate
sensitivities in both tasks and compare them. Before we present such
an analysis, we want to first discuss some details of the standard
reasoning.

\subsubsection{True Zero-Sensitivity in the Direct Task} 

Consider that the true sensitivity in the direct task were known to be
exactly zero and that there were at the same time a clear congruency
effect in the indirect task. This ideal situation is typically
sought---but typically not fully achieved---in the dissociation
paradigm
\cite{Schmidt_Vorberg_06,hannula2005imaging,simons2007behavioral}. In
this case (and only in this case), the standard reasoning would be
justified in claiming that responses in the indirect task were somehow
more sensitive than responses in the direct task. This is so, because
a positive (larger than zero) sensitivity---even if it is minute---is
required to produce a congruency effect and therefore the indirect
task sensitivity must be larger than zero.  However, there are a
number of problems with this scenario:
(a) It is unrealistic. Typically, studies either find some small,
residual sensitivity in the direct task or they do not find a
congruency effect \cite{zerweck2020testing}.
(b) One cannot be certain of a true zero sensitivity. Instead,
sensitivity in the direct task always needs to be measured (and is
therefore affected by measurement error). Thus, we would still need to
establish that the sensitivity in the indirect task is indeed larger
than that in the direct task (e.g., by a significance test on the
difference).
(c) The sensitivity in the indirect task could still be so low, that
it would be close--enough to the zero sensitivity of the direct task
to not allow for strong conclusions (e.g., consider a sensitivity that
corresponded to 50\%-correct in the direct task and to 51\%-correct in
the indirect task).

\subsubsection{Significance Testing vs. Bayesian methods}
Until now, we purposefully did not talk about statistical significance
testing because we wanted to focus on the main fallacy of the standard
reasoning. Because significance testing and its applications have been
heavily---and often rightfully---criticized since the very inception
of the concept
\cite{Boring_1919,Morrison_Henkel_70,Dienes_11,Cumming_14}, it might
be tempting to attribute the main fallacy of the standard reasoning
also to significance testing. However, the problem of the standard
reasoning is not so much that the statistical tools were wrong, but
that the wrong statistical question is asked for the indirect task:
The standard reasoning asks whether there is a true difference in
means between congruent and incongruent conditions. However, the
correct question to ask would be what the sensitivity in the indirect
task is and whether this sensitivity is higher than in the direct task
(such that an \ITA{} can be concluded). Therefore, it would not help
to simply replace the frequentist significance testing by Bayesian
methods. Because researchers are interested in establishing an ITA
(i.e., a difference in sensitivities) it does not suffice to evaluate
both tasks in isolation. We must test directly for a difference in
sensitivities between the two tasks. Failure to do so can lead to
serious errors no matter whether we used significance testing (cf.
Appendix~B of \citeNP{Franz_Gegenfurtner_08}, and
\citeNP{Nieuwenhuis_etal_11}) or Bayesian methods (cf.
Supplement~\ref{supp:bayes}, and \citeNP{palfi2020bayesian}).

\subsubsection{Direct Task is Typically Underpowered}
An additional problem in the application of the standard reasoning
arises from the widespread use of seriously underpowered direct tasks
\cite{buchner2000reliability,vadillo2016underpowered,vadillo2020unconscious}.
When the direct task is sampled with fewer participants and trials
than the indirect task (as is often the case), a non-significant
direct task result may not indicate that the true sensitivity is close
to or exactly zero but rather that statistical power is low. Moreover,
participants are required to give binary responses in the direct task
in contrast to the continuous measures in the indirect task (e.g.,
RTs). Since participants have some continuous sense (confidence) about
their responses \cite{rausch2018confidence,zehetleitner2013being}, the
binary response format forces them to discard this information
\cite{Cohen_83}, which further decreases the statistical power in the
direct task. Therefore, even if the same sensitivity underlies
responses in both tasks, it is a priori to be expected that the direct
task produces less often significant results than the indirect task.  



\section{Appropriate Analysis: Calculate Sensitivities and Test for a Difference}
\label{sec:correctAnalysis}

We have shown that the standard reasoning is flawed and that
researchers must compare sensitivities of both tasks if they want to
infer an \ITA{}. In this section, we describe more appropriate
analyses. First, we assume that trial--by--trial data are available
(this analysis was used by \citeNP{Franz_vonLuxburg_15_PS}). Then we
describe our newly developed method to
reanalyze studies when only summary statistics are available. For
detailed mathematical derivations see the online supplementary
materials.

In deriving our methods, we unavoidably were confronted with degrees
of freedom when choosing the details of our analysis strategy. In
these cases, we chose strategies that favored finding an \ITA{}. That
is, we followed a \textbf{benefit-of-the-doubt} approach, thereby
increasing the chances of confirming an \ITA{}. We adopted this
approach because we are criticizing a large body of literature.
Therefore, it seemed necessary and reasonable to adopt such a liberal
bias in confirming \ITAs{} (and thereby being conservative in our
critique) at this stage of the scientific discussion. It is
understandable that researchers who have spent years using the
standard reasoning might be reluctant to accept our arguments if our
methods were too restrictive. This approach makes our arguments even
stronger when we nevertheless do not find evidence for \ITAs{}.

\subsection{Sensitivity Comparison When Trial-By-Trial Data are
Available} 

The appropriate method directly compares sensitivities in the direct
and indirect tasks. Different than the standard reasoning, the
appropriate analysis equates analysis steps for both tasks such that
the calculated statistics are comparable. Then, a test of the
difference between the two tasks is applied. Similar approaches have
been used in previous---albeit very few---studies
\cite{Dulaney_Eriksen_59,Klotz_Neumann_99,kunst1980affective,Schmidt_02,Franz_vonLuxburg_15_PS}
in accordance with the long standing (but often ignored) request for
both tasks to be measured using the ``same metric''
\cite{Reingold_Merikle_88}.

In both tasks, we compute $d'$ using Signal Detection Theory
\cite{Green_Swets_88} and then test for a difference between them. In
the \textbf{direct task}, participants typically classify the primes
in each trial and a $d'$ value is often already reported by the
studies using the standard reasoning. In the \textbf{indirect task},
however, the standard reasoning computes a congruency effect on
continuous measures (e.g., RTs or brain activity as measured by EEG or
fMRI). For a proper comparison, we have to transform these continuous
measures into classifications (predictions) for each trial. There are
different ways to achieve this. We suggest to use the optimal
classifier for the given setup. This gives the indirect task the best
possible performance and increases the chances of finding an \ITA{}
following the \textbf{benefit-of-the-doubt} approach.

Which classifier is best? We have shown that under typical conditions
with equal number of congruent and incongruent trials, the
median-split classifier is optimal (\citeNP{Franz_vonLuxburg_15_PS};
see our Supplement~\ref{supp:medianClassifier} for details and proof).
The classification proceeds as follows: For each participant, we
determine the median RT and classify (``predict'' in the nomenclature
of statistical learning) all trials with smaller RTs as congruent, and
trials with larger RTs as incongruent. Then, we compare these
classifications to the true labels (congruent/incongruent) evaluating
for each trial whether the classification was correct or not, and we
then compute a $d'$ value as in the direct task.  Finally, we compare
the $d'$ values of the direct and indirect task and test for a
difference.

\textbf{Some details:}
(a)~Instead of computing $d'$ values, the analysis could also be based
on \%-correct values. Assuming a neutral observer predicting both
categories equally often in the direct task, both approaches
produce the same results and we later report both measures to foster
intuition. 
(b)~Dichotomization of the continuous, indirect measures will result
in a loss of information \cite{Cohen_83}. However, the direct task
also requires participants to give binary responses. Converting
indirect task responses into a binary response format using our median
split approach only equates this dichotomization to make responses in
both tasks comparable.
(c)~We classify the trials of the indirect task according to the
labels congruent/incongruent and not according to the prime category
A/B, as is typically asked in the direct task. This is so because
studies typically find a congruency effect between prime and target
(and not a mere effect of the prime being in category A or B). For a
comparison to the direct task, we would ideally transform the
congruency classification into a classification of the prime category
(A vs. B). For simplicity, we assume an optimal transformation here
(without errors). This is plausible, because the target stimuli are
typically fully visible to the participants, such that errors are
rare. Again, our approach increases the chances of finding an \ITA{}
following the benefit-of-the-doubt approach.

\subsection{Sensitivity Comparison When Only Summary Statistics are 
Available}
\label{sec:method}

Because the standard reasoning to infer an \ITA{} is flawed, many
already published studies on unconscious priming need reassessment.
However, the appropriate analysis as described in the previous section
would require full trial-by-trial data. Unfortunately, trial-by-trial
data can be difficult or impossible to obtain for published studies
\cite{wicherts2006poor}. For the older---but nevertheless
influential---studies, those data might not even exist anymore.
Therefore, we developed an approach that allows to estimate the
results of the appropriate analysis without access to trial-by-trial
data and solely based on the typically reported statistics from the
standard reasoning. Here, we sketch the central approach of this
analysis; details are given in
Supplement~\ref{supp:reanalysisFormulas}.

The overall aim of this reanalysis is, again, to estimate
sensitivities for the direct and indirect tasks (i.e., to either
calculate $d'$ from Signal Detection Theory or \%-correct assuming a
neutral observer). The direct task typically already provides $d'$ or
\%-correct values. In the indirect task, studies typically report $t$
or $F$ values from a repeated measures design for the congruency
effect. In this design, we show how $F$ values can be translated to
$t$ values. We then derive an estimator for the underlying sensitivity
that takes the form of a constant $c_{N,K,q^2}$ multiplied onto the
reported $t$ value. This constant will include the number of
participants $N$ and trials $K$ from the indirect task because $t$
values become larger the more observations are made. Additionally,
because this reanalysis can only use the reported statistics, one free
parameter needs to be estimated: the ratio of between- vs.
within-subject variances, which we denote by $q^2$. We estimated this
parameter based on (a) our own replication experiment (b) a literature
review, and (c) extensive simulations (see
Supplement~\ref{supp:roleOfq}). By assuming the largest plausible
value for $q^2$, we again maximize the estimated sensitivity, $d'$, in
the indirect task and therefore increase the likelihood of confirming
an \ITA{}. Here, we again follow the benefit-of-the-doubt approach.



\section{Replication of a Landmark Study Finds no \ITA{}}

We are now equipped with the appropriate tools that allow us to
analyze typical settings and tasks that have been investigated in the
context of unconscious priming. In this section, we will focus on one
highly influential study on unconscious semantic priming of numbers
\cite{Dehaene_etal_98}. We will first describe the study and how its
conclusions depend crucially on the flawed standard reasoning. Then,
we will describe a replication experiment of the behavioral part of
this study and analyze the trial-by-trial data. In the next section,
we will then reanalyze the published results of this and other studies
(\totalReanalyzed{} in total). Overall, we will conclude that the
results of our replication are similar to those of the original
study. Both, our replication and our reanalysis of the original study,
give reason to seriously doubt the existence of an \ITA{}, questioning
the authors' interpretation in the original study.

\citeA{Dehaene_etal_98} were interested in the question of whether the
semantic meaning of numbers can be processed outside conscious
awareness. They employed a prototypical priming experiment with
stimuli shown in Figure~\ref{fig:ITAdesign}a and applied the standard
reasoning: In the direct task, participants discriminated features of
masked numbers and performed poorly ($d'=0.2$; corresponding to
$54\%$-correct). In the indirect task, participants were again
presented with the masked numbers (now serving as primes), but decided
whether subsequent target numbers were smaller or larger than five.
Participants responded approximately $24$~ms faster when prime and
target were congruent (both larger or smaller than five) than when
they were incongruent (one smaller and one larger than five). Similar
congruency effects were found for brain activity in EEG and fMRI
(i.e., larger lateralization of brain activity in congruent than
incongruent trials).

\citeA{Dehaene_etal_98} interpreted these results according to the
standard reasoning:
In \textbf{Step~1}, they inferred an \ITA{}. That is, higher
sensitivity in the indirect task than in the direct task:
``[participants] could neither reliably report [the prime's] presence
or absence nor discriminate it from a nonsense string [...]
Nevertheless, we show here that the prime is processed to a high
cognitive level.'' (p. 597). 
In \textbf{Step~2}, they argued that ``the prime was unconsciously
processed'' (p.~597) because participants were at chance performance
in the direct task. Overall, they concluded: ``By showing that a large
amount of cerebral processing, including perception, semantic
categorization and task execution, can be performed in the absence of
consciousness, our results narrow down the search for its cerebral
substrates'' (p.~599). In short, \citeA{Dehaene_etal_98} employed a
prototypical version of the standard reasoning to infer an \ITA{} and
unconscious processing exactly as described above. To assess the
validity of these claims, we first replicate the behavioral part of
that study, later we will reanalyze the published data.

\subsection{Disclosures}
\subsubsection{Data, Materials, and Online Resources}

The experimental material, data and the scripts for the analyses
reported in this article have been made available on the Open Science
Framework (OSF), at \url{https://osf.io/kp59h} (Meyen, Zerweck,
Amado, von Luxburg, \& Franz, 2020)\nocite{ita_osf_link}. We also
provide an online tool to apply our reanalysis to other data at
\url{http://www.ecogsci.cs.uni-tuebingen.de/ITAcalculator/}.

\subsubsection{Reporting}

We report how we determined our sample size, all data exclusions, and
all measures in the study.

\subsection{Methods}

Twenty-four volunteers participated in our study (13 female, 5
left-handed, age range: 19--27 years; $M = 21.5$, $SD = 1.9$).  All
had normal or corrected-to-normal vision, signed written informed
consent and were naive to the purpose of the experiment. In the
original study by \citeA{Dehaene_etal_98}, six and seven participants
took part in the first and second direct task, respectively, and 12
participants took part in the indirect task.

We took great care to make stimuli and timings as similar as possible
to those of the original study. Each trial consisted of: fixation
cross ($417$~ms), forward mask ($67$~ms), prime ($42$~ms), backward
mask ($67$~ms), and target ($200$~ms). In the original study, those
values were: forward mask ($71$~ms), prime ($43$~ms), backward mask
($71$~ms), and target ($200$~ms; cf. Figure~\ref{fig:ITAdesign}a).
Slight differences in timing are due to slightly different refresh
rates of the monitors used. The prime duration of $43$~ms was chosen
by the original authors because it was the longest duration that
produced non-significant results in the direct tasks.
Primes and targets were numbers (1, 4, 6 or 9) that were either
presented as digit (e.g., ``1'') or word (e.g., ``EINS''; German for
``ONE''). The original study used the same numbers in English, a
follow--up used French \cite{Kouider_Dehaene_09}. As in the original
study, primes and targets could be congruent (both smaller or both
larger than 5) or incongruent (one smaller, one larger). Masks were
composed of seven randomly drawn characters from $\{\text{a-z, A-Z}\}$
mirroring the original study's masks.
Participants were seated in front of a monitor (VIEWPixx~/3D, VPixx
Technologies Inc., Montreal, Canada), effective refresh rate $120$~Hz
at a viewing distance of approximately $60$~cm in a sound- and
light-protected cabin.
In the original study, the monitor was a cathode-ray tube (CRT) with a
refresh rate of $70$~Hz. Stimuli were presented centrally as white
text ($69$~cd/m$^2$; character height: $1^\circ$; width: $0.5^\circ$
visual angle; font: Helvetica) on a black background ($0.1$~cd/m$^2$).
These luminance values were not specified in the original study so 
that we chose the most plausible settings for our experiment.

In the direct task, participants classified whether the prime was
smaller or larger than five. We used this particular task because the
original authors argued in a subsequent study that it is ``better
matched with the [indirect] task''
\cite[p.~227]{Naccache_Dehaene_01_novelStimuli}. In the original study
by \citeA{Dehaene_etal_98}, two direct tasks were used, that produced
similar results: In their first direct task, the prime stimulus was
omitted in some trials and participants had to discriminate their
presence vs. absence. In the second direct task, the prime stimuli
were replaced by random letter strings and participants had to
discriminate between numbers vs. random strings.

In the indirect task, participants decided as quickly as possible
whether the target was smaller or larger than five; as was the case in
the original study. Each participant performed $256$ trials per task,
preceded by $16$ practice trials in each task. In contrast, in the
original study, participants performed only $96$ and $112$ trials in
the first and second direct tasks, respectively, and $512$ trials in
the indirect task.

We repeated indirect task trials with RTs that were too slow ($>1$~s)
or too fast ($<100$~ms). The original study also rejected too slow
trials ($>1$~s) but was more restrictive in terms of fast trials: They
rejected with RT $<250$~ms. However, we only found 8 out of
6144~trials in our data to be above 100 ms but below 250 ms so that we
obtained very similar results when applying their criterion. The
indirect task was performed before the direct task (as is common
practice in this paradigm) to prevent participants attending to the
prime in the indirect task. In the original study, the direct and
indirect tasks were performed by different groups.

The number of participants and trials were chosen to produce a
statistical power of above $95$\% to find a difference between
sensitivities and confirm an \ITA{} if it is there (see
Supplement~\ref{supp:validation}). For this power estimation, we
assumed a true sensitivity of $d'_\text{true, direct} = 0$ in the
direct task vs. $d'_\text{true, indirect} = 0.25$ in the indirect task
(values based on our reanalysis of \citeNP{Dehaene_etal_98}, see
below). To our knowledge, the original study did not perform a power
analysis. A post hoc power analysis revealed that the original study
had a statistical power of only $46$\% to find an \ITA{} using the
appropriate analysis (again, assuming $d'_\text{true, direct} = 0$ and
$d'_\text{true, indirect} = 0.25$). This low power is due to a small
number of direct task participants and trials.

\subsection{Results and Discussion}
\label{sec:replicationDehaene}

Our analysis proceeded in two strands: First, we perform the
traditional analysis which forms the basis for the standard reasoning.
Second, we perform the appropriate analysis.

\subsubsection{Standard Reasoning} 
\begin{figure*}[!ht]
\centering
\includegraphics[width=1\textwidth]{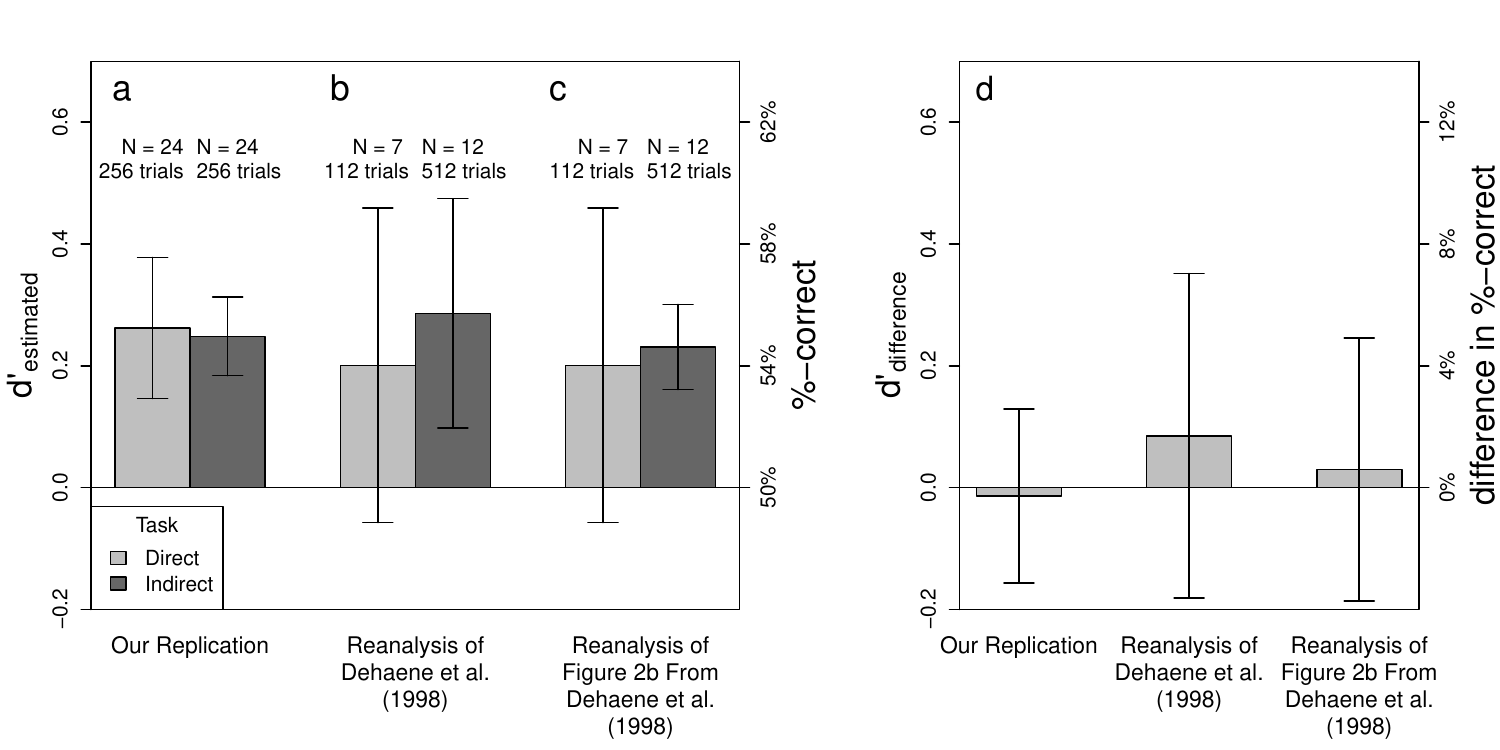}

\caption{\footnotesize 
\textbf{Sensitivities in the \protect\citeA{Dehaene_etal_98} setting}.
\textbf{(a)} Results of our replication study, based on our full
trial-by-trial data. \textbf{(b)} Results of our reanalysis approach
based on the published statistics from
\protect\citeA{Dehaene_etal_98}. \textbf{(c)} Reanalysis results from
digitizing Figure~2b from \protect\citeA{Dehaene_etal_98} showing
histograms of indirect task's RT data. For the comparison, we used the
same direct task results herein (c) as we used in (b). Comparing
(a)--(c) we see that our replication closely matches the results of
the original study. \textbf{(d)} Difference in sensitivities between
direct and indirect tasks: There is no significant difference in
sensitivities in our replication study or in our reanalyses of
\protect\citeA{Dehaene_etal_98}. That is, there is no evidence for an
\ITA{}. The reanalysis result from (b) is also shown in the large
summary in Figure~\protect\ref{fig:reanalysis}. Error bars indicate
95\% confidence intervals.
}
\label{fig:replicationComparison}
\end{figure*}

The direct task sensitivity was
$d' = 0.26$~($SD = 0.27$), $t(23) = 4.68$, $p < .001$, corresponding
to an accuracy in prime identification of $M=54.87$\%-correct ($SD =
4.9$, $t(23) = 4.82$, $p < .001$). This is exactly in the range of
sensitivities reported in the original study's direct tasks ($d' =
0.3$ in the first and $d'=0.2$ in the second direct task). For a
graphical depiction of these results, compare the bars corresponding
to the direct tasks in Figures~\ref{fig:replicationComparison}a
and~\ref{fig:replicationComparison}b.

Note that, in contrast to the original study, our direct task
sensitivity is significantly above zero. This is so, because we
sampled much more participants and trials than in the original study.
Therefore, we had much higher statistical power. To simulate the lower
power of the original study, we discarded data from participants and
trials to match the same number of observations as in the original
study: We kept only the first $N=7$ participants and the first 112
trials of each participant. This leads to a non-significant result,
$d' = 0.31$ ($SD=0.39$), $t(6) = 2.06$, $p = 0.085$, as was the case
in the original study. Therefore, it is plausible that it was the low
statistical power in the original study (and not the sensitivity being
exactly at zero) that was the reason for the nonsignificant result in
the direct task of the original study.

In the indirect task, the congruent condition yielded faster RTs
($M = 445$~ms, $SD = 42$) than the incongruent condition
($M = 457$~ms, $SD = 37$), resulting in a clear and highly significant
congruency effect of $M=12$~ms ($SD = 11.8$), $t(23) = 4.95$,
$p < .001$. That is, we found a highly significant congruency effect
on RTs, as did the original study.

There is one potential caveat here: The congruency effect in the
original study was larger than that in our replication ($24~ms$ vs.
$12~ms$, respectively). However, we will show that sensitivities in
our replication and our reanalysis of the original study are very
consistent, see Figures~\ref{fig:replicationComparison}a
and~\ref{fig:replicationComparison}b. This can be explained by larger 
trial-by-trial variability in the original study counteracting the
larger RT effect: The original study, despite using 512 trials per
participant, observed $SD = 13.5$ while we observed $SD = 11.8$ in our
replication with only 256 trials per participant. Generally, more
trials per participant should make individual RT effects more
precisely measured. Thus, the standard deviation in the original study
should be smaller than in our replication. But the opposite is the
case! This can be explained by a larger trial-by-trial variance in the
original study. Larger effect and larger variability in the original
study cancel out such that sensitivities are in fact quite comparable
to our replication, see below. Further research employing systematic
variation of stimulus parameters can further clarify this situation.
For example, we are currently determining the role of an \ITA{} in the
particular setting of \citeA{Dehaene_etal_98} in a more extensive
study, see \citeA{zerweck2020testing}.

In summary, we found a similar pattern of results as in the original
study: A very poor direct task performance and a clear congruency
effect in the indirect task. Based on this pattern of results many
researchers would have applied the standard reasoning and inferred an
\ITA{}.

\subsubsection{Sensitivity Comparison} The
appropriate analysis compares sensitivities in direct and indirect
tasks. We have already described in the last section that the direct
task in our experiment yielded a sensitivity of $d' = 0.26$~($SD =
0.27$), corresponding to an accuracy of $M=54.87$\%-correct. For the
indirect task, we obtained a sensitivity of $d'=0.25$ ($SD = 0.15$),
corresponding to an accuracy of $M=54.93$\%-correct ($SD = 3.03$).

Inspection of Figure~\ref{fig:replicationComparison}a shows that these
sensitivities in direct and indirect tasks are very similar, see their
difference plot in Figure~\ref{fig:replicationComparison}d. We found
virtually no difference between these sensitivities, $M = -0.01$
($SD=0.34$), $t(23) = -0.2$, $p = 0.844$. That is, there is no
indication for an \ITA{}.

In conclusion, our results are similar to the typical pattern of
results found by \citeA{Dehaene_etal_98} and many researchers would
have inferred an \ITA{}. However, the appropriate analysis yields no
evidence for an \ITA{}: The sensitivities in both tasks are
essentially identical.


\section{Reanalysis of \protect\totalReanalyzed{} Influential Studies Finds Hardly any \ITA{}}

After having demonstrated that the problems of the widely used
standard reasoning are indeed serious, we now apply our approach to a
sample of \totalReanalyzed{} highly relevant studies in the field of
unconscious priming.

\subsection{Methods}

\subsubsection{Selection Criteria for Reanalyzed Studies}
\label{selectionCriteriaForStudies}

We focused on studies that applied the standard reasoning and claimed
an \ITA{}. First, we selected eight studies by hand that are
particularly relevant. These studies and their number of citations in
Web of Science (Clarivate Analytics, Philadelphia, U.S.A.) are:
\citeA[56 citations]{Finkbeiner_Palermo_09},
\citeA[13 citations]{finkbeiner2011subliminal},
\citeA[76 citations]{mattler2003priming},
\citeA[352 citations]{Pessiglione_etal_07},
\citeA[34 citations]{sumner2008mask},
\citeA[154 citations]{van2010unconscious},
\citeA[0 citations]{wang2017role},
\citeA[1 citations]{wojcik2019unconscious},

Second, we searched for English articles in Web of Science with the
topic ``unconscious priming''. We selected all studies with more than
150 citations that applied the standard reasoning and claimed an ITA.
This resulted in seven additional studies:
\citeA[178 citations]{Damian_01},
\citeA[662 citations]{Dehaene_etal_98},
\citeA[770 citations]{Dehaene_etal_01},
\citeA[237 citations]{Kiefer_02},
\citeA[217 citations]{kunde2003conscious},
\citeA[214 citations]{Naccache_Dehaene_01_novelStimuli},
\citeA[313 citations]{Naccache_etal_02}. 
Overall, these 15 studies received a total of \totalCitations{}
citations. See Supplement~\ref{supp:reportedResults} for details on
these studies. 

\subsubsection{Details of Analysis When Only Summary Statistics are Available}

Our reanalysis method estimates and compares sensitivities for direct
and indirect tasks. Here, we sketch some technical details of the
analysis. A detailed account with mathematical derivations is given in
Supplement~\ref{supp:reanalysisFormulas}.

We denote the estimated sensitivities in the direct and indirect tasks
by $d'_\text{estimated, direct}$ and $d'_\text{estimated, indirect}$,
respectively. For the direct task, the typically reported statistics
are average $d'$ or \%-correct values. Therefore, our estimate is
simply the measured sensitivity,
\[ d'_\text{estimated, direct} = d' ,\]
or a well-known conversion of \%-correct values to $d'$ values
assuming neutral observers \cite{Green_Swets_88},
\[d'_\text{estimated, direct} = 2\Phi^{-1}(\text{\%-correct})\text{,}\] 
where $\Phi^{-1}$ is the inverse of the normal cumulative density
function.

In the indirect task, statistics for the congruency effect are
typically reported by $t$ values from a paired $t$ test or $F$ values
from a repeated--measures ANOVA. In this setting, $F$ values can be
translated into $t$ values by $|t| = \sqrt{F}$. From a $t$ value, we
estimate the sensitivity by
\[
d'_\text{estimated, indirect} = t\cdot c_{N,K,q^2}\text{, with}
\]
\[
  c_{N,K,q^2}
        =
        \sqrt{\frac{q^2 + \frac{4}{K}}{N}} \sqrt{\frac{2}{N-1}}~\frac{\Gamma\left(\frac{N-1}{2}\right)}{
        \Gamma\left( \frac{N-2}{2} \right)},
\]
where $\Gamma$ is the gamma distribution. The constant $c_{N,K,q^2}$
corrects for the fact that $t$ values increase with increasing number
of participants ($N$), increasing number of trials ($K$), and that
they depend on the ratio of between- and within-subject variance,
which we denote by $q^2$.

The parameter $q^2$ is the only free parameter we need to estimate for
our approach. It is reasonable to assume that this ratio is at most
$q^2 = 0.0225$ given our replication study, a literature review (see
Supplement~\ref{supp:roleOfq}) and extensive simulations (see
Supplement~\ref{supp:validation}). Assuming the largest plausible
value for $q^2$, increases the likelihood of finding an \ITA{} thereby
following the benefit-of-the-doubt approach.

From the estimated sensitivities, we compute the difference
\[
d'_\text{difference} = d'_\text{estimated, indirect} - d'_\text{estimated, direct}
\]
and construct a 95\% confidence interval using the  corresponding
standard errors (derived in Supplement~\ref{supp:reanalysisFormulas}).
This allows to test for an \ITA{}: If the confidence interval lies
above $0$ (that is, it has the form $[a,b]$ with $a>0$), the reported
result is significant and an \ITA{} is confirmed, otherwise there is
not sufficient evidence to claim an \ITA{}. 

We demonstrate in the Appendix that
confidence intervals based on our reanalysis method are quite
comparable to those based on the trial-by-trial
analysis. For the study of \citeA{tenBrinke_14}, the
trial-by-trial analysis versus our reanalysis method using summary
statistics produced 95\%~CI~[-0.07; 0.23] and [-0.11; 0.25],
respectively (Figure~\ref{fig:validationTenBrinke}). In our
replication, the two methods produced 95\%~CI~[-0.15; 0.12] and
[-0.20; 0.06], respectively (Figure~\ref{fig:validationDehaene}).
Thus, our reanalysis method produces consistent results with the
analysis based on trial-by-trial data. 

\begin{figure*}[!htbp]
\centering
\includegraphics[width=\textwidth, height=17.8cm]{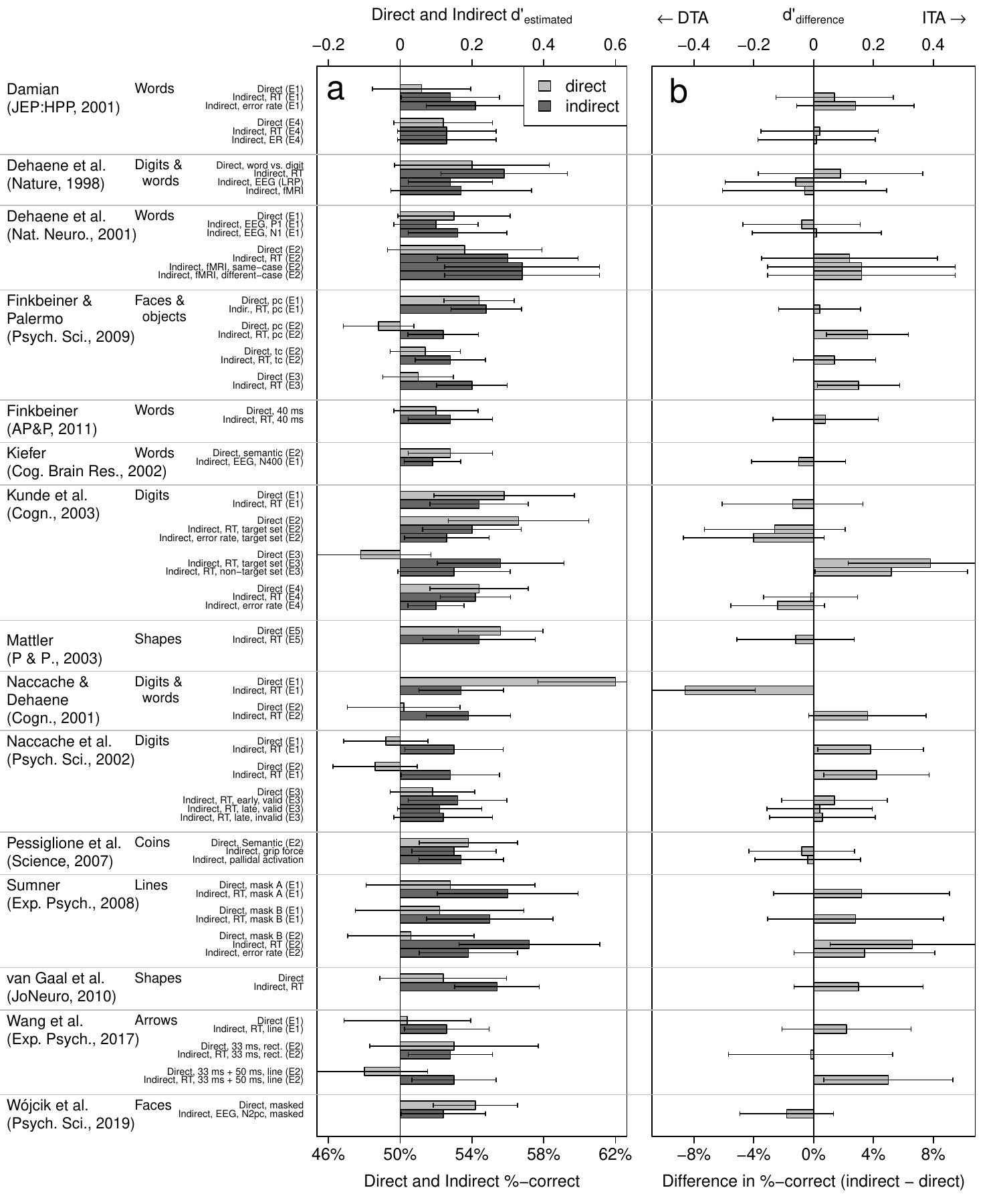}

\caption{\footnotesize 
\textbf{Reanalysis of influential studies reporting indirect task
advantages (\ITAs{}).}
The 15 studies used the
standard reasoning to infer an \ITA{} in 44 conditions.
\textbf{(a)} We reanalyzed the sensitivities and, to foster intuition,
we also show \%-correct values assuming a neutral observer.
\textbf{(b)} We reanalyzed the difference in sensitivities: In each
group of bars from (a), the indirect task is compared to the
corresponding direct task yielding the differences shown in (b). Only
if a confidence interval (error bars) around the difference lies to
the right and does not contain $0$, there is evidence for an \ITA{}.
Only in very few cases (8 out of 44), there is evidence for an \ITA{}, 
while in most cases (35 out of 44) there is no evidence. There is even
one case with a significant opposite result, an advantage of the direct
task (DTA). Not a single study provides consistent evidence for \ITAs{}
across its experiments and conditions in which it claimed \ITAs{}. 
Moreover, these results are obtained under most favorable conditions
for finding an \ITA{}: Our reanalysis overestimates the indirect task
sensitivities and therefore the evidence for an \ITA{} due to our
conservative choice of analysis strategies. Additionally, some of the
reanalyzed studies apply problematic methodology that further biases
the results towards finding an \ITA{} even if there is none, see
Discussion.
This pattern of results casts serious doubts on the existence of
\ITAs{} in most, if not all, of the studies.
Error bars represent 95\%-confidence intervals. 
}
\label{fig:reanalysis}
\end{figure*}

\subsection{Results and Discussion}

We first describe our reanalysis in detail for the study of
\citeA{Dehaene_etal_98} and then use the same methods for all the
other studies.

\subsubsection{Reanalysis of Dehaene et al. (1998)}

As discussed in our replication, the study reported two direct tasks
with sensitivities of $d' = 0.2$ and $d' = 0.3$, respectively. We used
the results of the first task, because it had the smaller sensitivity,
thereby, increasing the chances of our reanalysis to confirm an \ITA{}
and following the benefit-of-the-doubt approach.

In this direct task, $N=7$ participants were sampled in $K = 112$
trials and a sensitivity of $d' = 0.2$ was
reported, see light gray bar
in Figure~\ref{fig:replicationComparison}b. From these values, our
reanalysis method estimates the standard error to be $SE = 0.11$.

In the indirect task, the study reported on average a congruency
effect of $24$~ms with a standard deviation of $13.5$~ms in a sample
of $N=12$ participants sampled in $K=512$ trials each. This equals a
$t$ value of $t = 24~\text{ms}/(13.5~\text{ms}/\sqrt{12}) = 6.12$ from
which our reanalysis method estimates the sensitivity to be
$d'_\text{estimated, indirect} = t \cdot c_{N,K,q^2} = 0.29$ ($SE =
0.09$), see dark gray bar in Figure~\ref{fig:replicationComparison}b.

Taken together, the sensitivities in both tasks are very similar with
no clear difference between them, $d'_\text{difference} = 0.09$, $SE =
0.14$, see Figure~\ref{fig:replicationComparison}d. The confidence
interval for the difference includes zero, $95\%~$CI$~=[-0.18, 0.35]$,
thereby indicating that the sensitivity difference did not deviate
significantly from zero. That is, there is no evidence for an \ITA{}.

We were able to reanalyze the results from \citeA{Dehaene_etal_98} in
an additional way. They depicted summary histograms of RTs in their
Figure~2b visualizing that congruent and incongruent RT distributions
are similar in shape but only shifted because incongruent RTs were
slower than congruent RTs. Despite the shift, RT distributions largely
overlap. We digitized the histogram and split RTs along the  median as
described in the appropriate analysis section. From this, we estimated
the indirect task sensitivity to be $d'=0.23$ ($SE = 0.03$). Again, we
find no difference to their first direct task's sensitivity ($d' =
0.2$, $SE = 0.11$) since zero is included in the confidence interval
of the difference, $95\%$~CI~$[-0.19; 0.25]$, see
Figure~\ref{fig:replicationComparison}c and
\ref{fig:replicationComparison}d. Note that this approach deviated
from our appropriate analysis in that it does not compute the median
for each individual participant but uses a grand median across
participants because the published histogram pools all participants'
RT data. This approach ignores between-subject variance leading to a
slight underestimation of the indirect task's sensitivity.
Nevertheless, this additional reanalysis provides converging evidence
complementing our previous results.

The results from our reanalysis of the original study
(Figure~\ref{fig:replicationComparison}b and
\ref{fig:replicationComparison}c) and the results from our replication
experiment (Figure~\ref{fig:replicationComparison}a) are very
consistent. Estimates for the sensitivities are very stable. This
corroborates the validity of our reanalysis approach as well as of our
replication experiment (see Supplement~\ref{supp:validation} for
further validation of our reanalysis approach).

To summarize, both, our reanalysis of \citeA{Dehaene_etal_98} as well
as our replication of the behavioral responses, suggest that there is
no \ITA{} in the behavioral part of that study. This demonstrates the
fundamental flaw of the standard reasoning and suggests that similar
problems might exist in other studies.

\subsubsection{Reanalysis of all \protect\totalReanalyzed{} studies}

We now apply our reanalysis in a similar way to all other studies. For
this, we present the data in a more compact fashion in
Figure~\ref{fig:reanalysis}. For example, what we showed in
Figures~\ref{fig:replicationComparison}b and
\ref{fig:replicationComparison}d for the study of
\citeA{Dehaene_etal_98} now corresponds to the lines 7 and 8 in
Figure~\ref{fig:reanalysis}, showing the sensitivities for each task
in Figure~\ref{fig:reanalysis}a and the difference of sensitivities in
Figure~\ref{fig:reanalysis}b.

When evaluating this figure, it is important to be aware that we used
our benefit-of-the-doubt approach. For example,
\citeA{Dehaene_etal_98} had two direct tasks, resulting in $d' = 0.2$
and $d' = 0.3$, respectively. As described above, we used the smaller
of those values, thereby increasing the chances of finding an \ITA{},
which makes our arguments stronger if we nevertheless do not find an
\ITA{} (cf. General Discussion).

Inspecting the figure shows that in most studies the sensitivities of
direct and indirect tasks have comparable sensitivities, such that the
differences are small and not significantly different from zero. This
is the case for 35 of the 44 differences between direct and indirect
tasks (Figure~\ref{fig:reanalysis}b). This is in stark contrast to the
fact that all studies claimed \ITA{}s in all these cases.

Only in 8 of the 44 differences there is a significant difference in
the direction of an \ITA{}, such that the indirect task has higher
sensitivity than the direct task. These results are, however,
intermixed with inverted differences in the same studies. For example,
although \citeA{kunde2003conscious} have two significant differences
in the direction of an \ITA{}, there are five differences pointing in
the opposite direction within the same study (albeit those are not
significantly different from zero).

Finally, the largest of all differences is even inverted: In
Experiment~1 of \citeA{Naccache_Dehaene_01_novelStimuli} there is a
significantly higher sensitivity in the direct task than in the
indirect task, just the opposite of an \ITA{}.

To summarize, our reanalysis found significant \ITAs{} in only 8 out
of 44 instances, which are spread across five different studies
\cite{Finkbeiner_Palermo_09,kunde2003conscious,Naccache_etal_02,sumner2008mask,wang2017role}.
Note that for multiple hypothesis testing, one would expect at least
some false positive results. These results are intermixed with 35
inconclusive results and even an opposite result where the direct task
showed significantly higher sensitivity than the indirect task
\cite{Naccache_Dehaene_01_novelStimuli}. Inspecting
Figure~\ref{fig:reanalysis} shows that there is no consistent evidence
for an \ITA{} in any of the reanalyzed studies. Not a single study
showed significant \ITAs{} in all conditions, albeit all studies
claimed ITAs{} for all reanalyzed conditions.

Let us stress that our goal was not to investigate whether there
exists a ``general'' \ITA{} across all studies with their vastly
different stimuli, experimental setups, tasks and scientific
questions. Therefore, we did not perform a meta-analysis or correct
for multiple testing. This had several reasons. First, our reanalysis
favored finding an \ITA{} by using our benefit-of-the-doubt approach.
Second, there are additional methodological issues in the reanalyzed
studies that introduce further biases, and for which we cannot correct
in our reanalysis (see General Discussion). Considering these two
biases towards finding an \ITA{}, a meta-analysis could misleadingly
produce the impression that there is a slight \ITA{} present across
all reanalyzed studies. An \ITA{} might exist but perhaps only for
some particular stimuli and setups. Given that the evidence for an
\ITA{} in each individual study is now in question, the research goal
should be to differentiate under which conditions a reliable \ITA{}
can be obtained and under which conditions this is not possible. A
meta-analysis would not serve this differentiating purpose.

In summary, reanalyzing the results from studies on unconscious
priming shows that there is little to no evidence for \ITAs{} in those
studies despite them claiming \ITAs{} for all conditions.
Sensitivities in the indirect tasks are not consistently larger than
sensitivities in the direct task as one would expect, given that
unconscious processing was inferred using the standard reasoning that
necessarily implies \ITA{}s. This demonstrates how seriously the
literature on unconscious priming is affected by the flaws of the
standard reasoning.


\section{General Discussion}

Many studies on consciousness that investigate a wide range of
cognitive functions are based on the flawed standard reasoning. The
main fallacy occurs when the standard reasoning infers an \ITA{}. That
is, a higher sensitivity for masked stimuli in the indirect task as
compared to the direct task. In an earlier reanalysis of
\citeA{tenBrinke_14} by \citeA{Franz_vonLuxburg_15_PS}, in our
replication of the behavioral part of \citeA{Dehaene_etal_98}, and in
our reanalysis of 15 highly influential studies, we found that none of
these studies can overall truly claim evidence for an \ITA{}. To the
contrary, responses in the indirect task often show a similar
sensitivity as compared to the direct task. This casts serious doubt
on the evidence for unconscious processing that exceeds conscious
reportability in these studies.

The fallacy of the standard reasoning has serious consequences for the
trustworthiness of the scientific literature on consciousness. It also
takes away from the appeal of many claims in the field: For example,
it would be an interesting result if lie detection and semantic
meaning of numbers could be processed outside of awareness. But such
strong claims require substantive empirical evidence, which we did not
find because the reanalyzed studies employed the flawed standard
reasoning. The appropriate analysis yields results that may be
considered as less exciting because---under scrutiny---participants'
responses did not seem to be affected by processing beyond what they
can consciously report.

Besides theoretical issues, there are also additional methodological
problems that can systematically bias the results and lead to claims
of an \ITA{} even if the true underlying sensitivities in the direct
and the indirect task are perfectly equal.

First, a common practice is to exclude participants with a good direct
task sensitivity.  The researchers' motivation here is to avoid
including the subset of participants who are consciously aware of the
masked stimuli. However, this practice bears the problem of regression
to the mean
\cite{barnett2004regression,schmidt2015invisible,shanks2017regressive}.
Thus, this practice is biased towards finding a smaller sensitivity in
the direct task and thus biased towards finding an \ITA{} even if
there is none. Several studies in our reanalysis have this problem
\cite{finkbeiner2011subliminal,mattler2003priming,Pessiglione_etal_07,sumner2008mask,van2010unconscious}.
This can explain why these studies produced some of the largest
differences in our reanalysis in Figure~\ref{fig:reanalysis}.

Second, in some experimental procedures participants have to respond
to the target stimulus (indirect task) first and only then respond to
the masked stimulus (direct task) all within the same trial (see
\citeNP{Finkbeiner_Palermo_09,peremen2014conscious}). Because the
cognitive impact of a masked stimulus decays quickly after $300$~ms
\cite{mattler2005inhibition,wolfe1999inattentional}, this procedure
makes the direct task more difficult. Participants have to memorize
the masked stimulus while performing the indirect task until they can
give a direct task response. This may decrease the direct task
sensitivity due to the additional difficulty, which can produce
misleading \ITA{} results. It is somewhat impressive that, even under
these favorable circumstances, none of these reanalyzed studies
provide consistent evidence for an \ITA{}.

Nevertheless, our results do not necessarily rule out the possibility
that \ITAs{} exist in some cases. But the existence of an \ITA{} may
depend on the particular task and stimuli used. It might not be as
ubiquitous as previously thought. Albeit the long standing request to
use the same metric for both tasks \cite{Reingold_Merikle_88} has
often been ignored, there are some studies that provide evidence for
an \ITA{} using the appropriate analysis. For example, the setting of
\citeA{Schmidt_02}---color stimuli served as primes and
targets---found a distinct \ITA{} result. Another example is the study
by \citeA{kunst1980affective} using geometric shapes \cite<but see
also>{de2013exposure,seamon1983affective}.

Therefore, we do not claim that there are no instances in which an
\ITA{} exists. Such a claim would be far beyond the scope of a single
scientific study. But we do claim that one of the most prevalent
methods in the wide research area of unconscious priming is
fundamentally flawed. This flaw affects and potentially invalidates
interpretations of many studies. As a consequence, the field has to
reassess the situation of \ITAs{} by applying the appropriate analysis
to substantiate or refute previously made 
claims.

In deriving our appropriate methods, we have chosen strategies that
favored finding an \ITA{}. That is, we have followed the
benefit-of-the-doubt approach to increase the chances of confirming an
\ITA{}. From such an approach, one would have expected clear evidence
for an \ITA{} in each of the reanalyzed studies. But since we
nevertheless did not find consistent evidence for \ITAs{}, having
followed the benefit-of-the-doubt approach makes our arguments even
stronger.

However, in future research, we hope that the benefit-of-the-doubt
approach will no longer be necessary because it has a drawback: It
would be inappropriate to simply revert the reasoning and use our
liberal method to establish evidence \emph{for} an ITA. To provide
convincing evidence for an \ITA{}, we would need a more balanced
approach, one that might have not convinced researchers in the current
situation (because they might have rejected it for being too
conservative in terms of finding an \ITA{}). For example, we used a
clearly fail-safe estimate for $q^2$ in our reanalysis, that was
chosen to be \emph{larger} than all reported values on which this
estimate is based. A more balanced approach would use a smaller
estimate, which would reduce the chances to find an \ITA{}, see our
additional reanalyses in Supplement~\ref{supp:roleOfq} for a figure
like Figure~\ref{fig:reanalysis} but with a more balanced estimate of
$q^2$. Of course, trial-by-trial data should be used whenever
possible.

To summarize, what we suggest is a research program: Given the
tremendous interest in unconscious priming and the far-reaching
inferences based on studies using the standard reasoning, researchers
should reinvestigate the most relevant cases of claimed \ITAs{} and
clarify to which degree the claims in those studies are truly
warranted. In those cases where an \ITA{} is properly established,
researchers can then start to draw further reaching conclusions about
conscious vs. unconscious processing
\cite{Eriksen_60,Erdelyi_86,Holender_86,Reingold_Merikle_88,Schmidt_Vorberg_06}.
An \ITA{} is only a prerequisite but not a sufficient condition for
the inferences that are typically drawn about unconscious processing.

In short, the literature needs a serious and concerted reassessment
that would go well beyond the scope of a single study and will also
require---in critical cases---the collection of new data. In many
cases where superior unconscious processing already seemed an
established fact \cite<e.g.,>{Hassin_13}, we expect that this view
needs to be revised. In other cases, researchers might still be able
to establish such a relationship---which will then be even more
interesting and foster the theoretical understanding of when exactly
conscious processing is vital for a cognitive function and when it is
not.

\section{Context}

Unconscious processing has been investigated for a long time. A common
notion is that more information is processed unconsciously than
consciously accessible. A main body of evidence for this comes from
unconscious priming, where a standard reasoning is used to provide
evidence for unconscious processing that exceeds consciously
reportable processing. We show that the standard reasoning is flawed
for statistical reasons. We introduce an analysis that is more
appropriate. Using this analysis, we find that interpretations about
unconscious processing break down. That is, even though the standard
reasoning produced the notion that participants process more
information unconsciously than they can consciously report, we show
that there is inconsistent evidence for this in many studies. This
lack of supposed evidence for superior unconscious processing has
far-reaching consequences: It questions the idea that conscious
processing is not necessary for many cognitive processes. We call for
a community effort to apply the appropriate analysis and differentiate
between situations in which processing can occur without being
consciously reportable, that is, unconsciously.

\section*{Availability of data and material}
The data set and analysis scripts supporting the conclusions of this article is available in the Open Science Framework repository (doi: \url{https://doi.org/10.17605/OSF.IO/KP59H}), \url{https://osf.io/kp59h}.

\section*{Funding}
This project is supported by the Deutsche Forschungsgemeinschaft (DFG,
German Research Foundation) through the CRC 1233 ``Robust Vision'',
project number 276693517; 
the Institutional Strategy of the University
of T{\"u}bingen (DFG, ZUK 63); and the Cluster of Excellence “Machine
Learning: New Perspectives for Science”, EXC 2064/1, project number
390727645. 

\section{Author Contributions}

S. Meyen, V. H. Franz, and U. von Luxburg developed the methods. I. A.
Zerweck conducted the experiment. All authors wrote the manuscript.

\section{Conflicts of Interest}

The authors declared no conflicts of interest with respect to the
authorship or the publication of this article.

\newpage

\bibliographystyle{apacite}


\newpage
\onecolumn
\section{Appendix}

\setcounter{secnumdepth}{1}
\def\thesection{\Alph{section}}

\label{sec:appendix}

\renewcommand{\thefigure}{A\arabic{figure}}
\setcounter{figure}{0}

\begin{figure}[!ht]
\centering
  \begin{minipage}[b]{.45\textwidth}
    \vspace{1cm}
    \includegraphics[width=\textwidth]{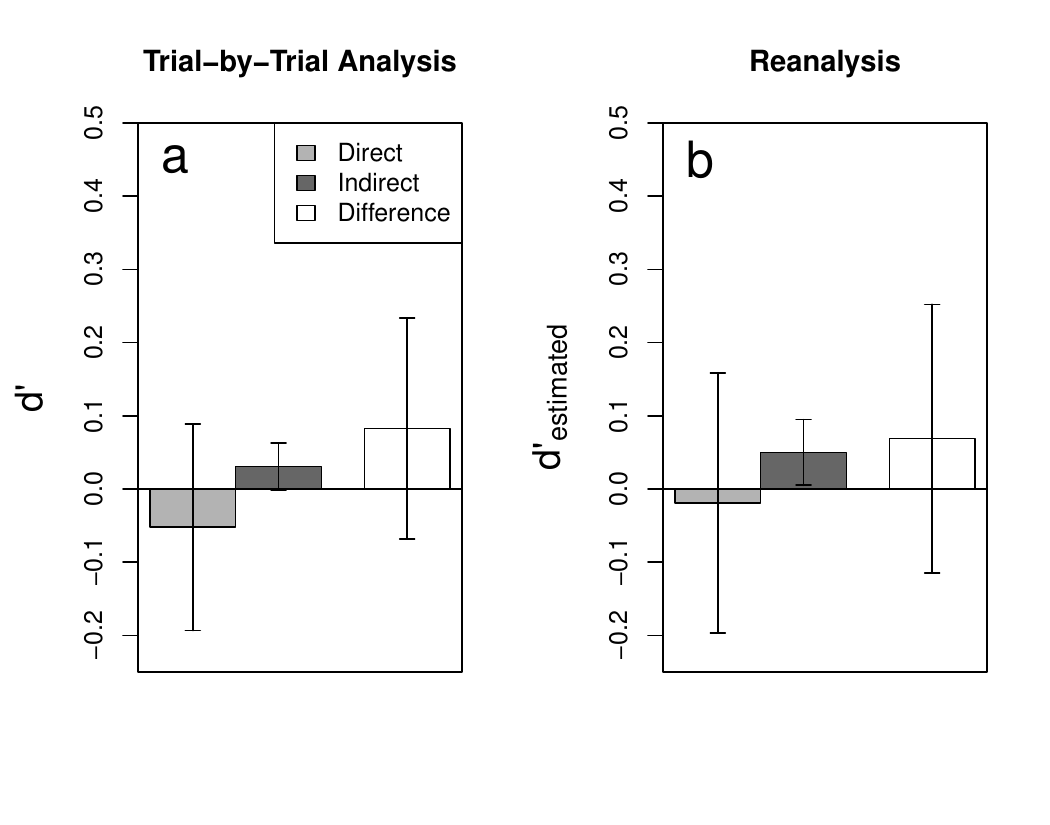}
    \vspace{-1cm}
  \end{minipage}
  \begin{minipage}[t]{.05\textwidth}
    ~
  \end{minipage}
  \begin{minipage}[b]{.45\textwidth}
    \vspace{1cm}
    \includegraphics[width=\textwidth]{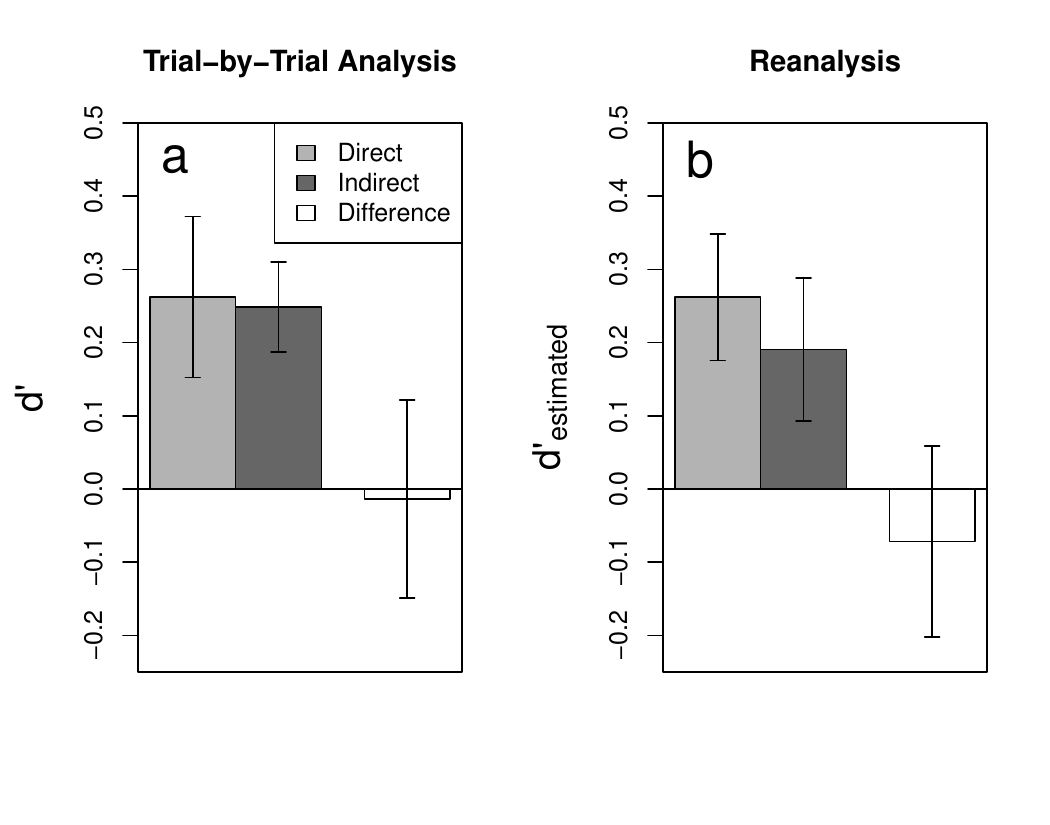}
    \vspace{-1cm}
  \end{minipage}
  \vspace{-0.5cm}
  \begin{minipage}[t]{.45\textwidth}
    \captionof{figure}{\footnotesize \textbf{Appropriate analysis
applied to \protect\citeA{tenBrinke_14}} using the full,
trial-by-trial data in (a) and using our reanalysis method in (b). Our
reanalysis using only the typically reported statistics produced
approximately the same results as the trial-by-trial analysis. In both
cases, there is no evidence for an indirect task advantage. Error bars
indicate 95\% confidence intervals.}
  \label{fig:validationTenBrinke}
  \end{minipage}
  \begin{minipage}[t]{.05\textwidth}
    ~
  \end{minipage}
  \begin{minipage}[t]{.45\textwidth}
  \captionof{figure}{\footnotesize \textbf{Appropriate analysis
applied to our replication of \protect\citeA{Dehaene_etal_98}} using
the full, trial-by-trial data in (a) and using our reanalysis method
in (b). Our reanalysis using only the typically reported statistics
produced approximately the same results as the trial-by-trial
analysis. In both cases, there is no evidence for an indirect task
advantage. Note that the indirect task sensitivity in our reanalysis
is smaller than in the trial-by-trial analysis. This is not a
contradiction to our claim that \emph{in expectation} the indirect
task sensitivity is overestimated by our reanalysis. Estimates of
individual studies can vary as indicated by the error bars indicating
95\% confidence intervals.}
    \label{fig:validationDehaene}
  \end{minipage}

\end{figure}

We demonstrate in two studies that the appropriate
analysis based on the full trial-by-trial data is well approximated by
our reanalysis based only on the typically reported statistics (e.g.,
a $t$ value for the congruency effect in the indirect task). We
compare the appropriate analysis using the full, trial-by-trial data
on one hand and our reanalysis method based on only the reported
summary statistics on the other hand. We apply both approaches to the
original data from \citeA[see
Figure~\ref{fig:validationTenBrinke}]{tenBrinke_14} and to our
replication of \citeA[see
Figure~\ref{fig:validationDehaene}]{Dehaene_etal_98}. Results from the
two analyses are very comparable confirming the validity of our
reanalysis.

\newpage


\setcounter{secnumdepth}{0}
\section[Supplemental Materials]{\textbf{Supplemental Materials\\ Advancing Research on Unconscious Priming: When can Scientists Claim an Indirect Task Advantage\\ by S. Meyen, I. A. Zerweck, C. Amado, U. v. Luxburg, \& V. H. Franz}}

\setcounter{secnumdepth}{1}
\setcounter{section}{0}
\setcounter{figure}{0}
\setcounter{table}{0}
\renewcommand{\thefigure}{S\arabic{figure}}
\renewcommand{\thetable}{S\arabic{table}}

\vspace{3cm}

\section{\textbf{Overview}}
\label{supp:overview}

\noindent \ref{supp:overview}. Overview \dotfill \pageref{supp:overview} \\
\noindent \ref{supp:validation}. \appendixA \dotfill \pageref{supp:validation} \\
\noindent \ref{supp:medianClassifier}. \appendixB \dotfill \pageref{supp:medianClassifier} \\
\noindent \ref{supp:reanalysisFormulas}. \appendixC \dotfill \pageref{supp:reanalysisFormulas} \\
\noindent \ref{supp:roleOfq}. \appendixD \dotfill \pageref{supp:roleOfq} \\
\noindent \ref{supp:reportedResults}. \appendixE \dotfill \pageref{supp:reportedResults} \\
\noindent \ref{supp:bayes}. \appendixF \dotfill \pageref{supp:bayes}\\
\noindent \ref{supp:symbols}. \appendixG \dotfill \pageref{supp:symbols} \\
\noindent \ref{supp:references}. References \dotfill \pageref{supp:references}

\newpage

\section{\textbf{\appendixA}}
\label{supp:validation}

We conducted multiple simulations to validate that our reanalysis
method appropriately controls for statistical errors (type~I and
type~II). Each simulation was repeated $10,000$ times. In each run, we
generated a trial-by-trial data set with a direct and an indirect task
according to the standard repeated measures model outlined in Appendix
\ref{supp:preliminaries}. We simulated $N$ participants with
sensitivities, $d'_{\text{true},i}$, independently and randomly drawn
from normal distributions with expected value $d'_{\text{true}}$ and
variance $q^2$ (see Appendix \ref{supp:roleOfq} for why $q^2$ is the
variance of individual true sensitivities). Note that we sampled
$d'_{\text{true},i}$ for each participant independently in the direct
and indirect task to avoid making additional assumptions on their
correlation between tasks. Applying Signal Detection Theory, each of
these individual sensitivities implies two normal distributions
separated by $d'_{\text{true},i}$ standard deviations. From these
normal distributions, we sampled a total of $K$ trials for each
participant, $K/2$ in each condition. We did this twice, once for each
task. In the direct task, we compared each response to the true
median: If the response lied on the same side as the normal
distribution it was sampled from, the simulated binary decision by the
participant in this trial was correct, otherwise it was wrong. In the
indirect task, we simply treated the drawn responses as the indirect
measures (e.g., RTs). We then applied the traditional analysis used in
the standard reasoning and the appropriate analyses, first based on
the full, trial-by-trial analysis and second our reanalysis based on
typically reported summary statistics. We obtained similar results
with log-normal distributions and only report normal distribution
results for brevity.

In each simulation, we varied $N$, $K$, $d'_{\text{true}}$ and $q^2$.
If not declared otherwise, the same $q^2$ was used for data simulation
and reanalysis. Only in simulations 5 and 6, we varied the true $q^2$
with which the data was simulated and used a different $q^2$ for our
reanalysis in order to see how getting this parameter wrong would
affect our results.

Simulations 1-3 demonstrate that the standard reasoning applied to the
traditional analysis miserably fails when applied to the study of
\citeA{Dehaene_etal_98}. Simulation 4 shows that our replication has
sufficient statistical power to find an \ITA{} if it was there.
Simulations 5 and 6 show how our reanalysis would be affected, if the
true $q^2$ was different than what we assumed. We then summarize
additional 108 simulations showing that our estimators, even though
they use simplifying approximations, are approximately unbiased.

\subsection{Simulations}

\subsubsection{Simulation 1: Controlling type I errors}
\label{sec:underpower}

We used the same number of participants in the direct ($N = 7$) vs.
indirect ($N=12$) task as well as the same number of trials per
condition (direct $K=112$ vs. indirect $K=512$) as the original study
of \citeA{Dehaene_etal_98}. Assuming no \ITA{}, we set sensitivities
in both tasks to be equal (direct $d'_\text{true} = 0.25$ vs. indirect
$d'_\text{true} = 0.25$). We assumed $q^2 = 0.0225$ for this simulation.

Even though the same sensitivity underlies both tasks, the direct task
fails to reach significance half of the time ($51.2\%$) while the
indirect task is almost always significant ($99.5\%$). This is not
surprising and shows how seriously underpowered the direct task was
due to fewer samples, $N$ and $K$. When applying the standard
reasoning, a scientist would erroneously conclude an \ITA{} from a
non-significant direct task result and a significant indirect task
effect in $48.6$\% of the experiments. In other words: The widely used
standard reasoning would infer an \ITA{} half of the time even though
there is no \ITA{} present!

Since there is no \ITA{} present, our reanalysis should find an \ITA{}
only as often as prespecified by the significance level $\alpha=5\%$.
Indeed, we find a difference between the two tasks only in 4.7\% of
the runs. This demonstrates that our reanalysis approach controls
appropriately for type I errors.

\subsubsection{Simulation 2: Controlling for type II errors with an
underpowered direct task} 

\label{supp:sim2}

We use the same settings as in Simulation~1 except that we now assume
there exists an \ITA{} (direct $d'_\text{true} = 0$ vs. indirect
$d'_\text{true} = 0.25$). Since there is an \ITA{} present, a high
statistical power is desired to detect it and avoid type~II errors.
Typically, a power above $1-\beta = 80\%$ is desired. However, our
reanalysis found the \ITA{} in only $46.2\%$ of the runs. Using the
full trial-by-trial data to test for a difference (instead of only
using the reported $t$ value from the indirect task) also produced a
test power of only $45.9\%$. There is simply not enough data in the
direct task to provide sufficient evidence for an \ITA{}. The problem
with lacking statistical power is not located in our reanalysis
because the analysis based on the trial-by-trial data also has a low
statistical power. Instead, the problem is the low sample size in the
direct task.

\subsubsection{Simulation 3: Controlling for type II errors with
sufficient samples in the direct task}

\label{supp:sim3}

We repeated Simulation 2 but increased the number of participants and
trials in the direct task to match the ones of the indirect task
($N=12$ and $K=512$). This is most sensible when testing for a
difference because a balanced design maximizes statistical power.
Here, our reanalysis method detects the \ITA{} in 78.3\% of the runs,
which is close to the desired 80\%. Using the full trial-by-trial data
provides a power of 84.2\%. This demonstrates that our reanalysis
method provides sufficient power given sufficient samples.

\subsubsection{Simulation 4: Statistical power in our replication}

\label{supp:sim4}

We repeated Simulation 3 but used the same number of participants and
samples as in our replication study, $N=24$ and $K=256$ in both tasks.
There, we have the same amount of observations as
\citeA{Dehaene_etal_98} (double the participants, half the trials).
Here, our reanalysis detects the \ITA{} in 96.5\% of the runs. The
analysis using trial-by-trial data instead of only a $t$ value
achieves 97.0\%. The increase in statistical power compared to
Simulation 3 comes from sampling more participants which is more
efficient than sampling more trials given a fixed total number of
observations \cite{rouder2018power}.

\subsubsection{Simulation 5: Overestimating parameter 
\texorpdfstring{$q^2$}{}}

\label{supp:sim5}

We repeated Simulation 3, the balanced \citeA{Dehaene_etal_98} setting
with an \ITA{}, but generated the data with $q^2=0.01$. We still use
$q^2=0.0225$ for the reanalysis, thus, we overestimate the true $q^2$.
Our reanalysis now successfully detects the \ITA{} in 99.6\% of the
runs and so does the appropriate analysis with 99.2\%. We detect more
\ITAs{} here than in Simulation 3 because we make our reanalysis more
liberal by choosing a larger $q^2$.

\subsubsection{Simulation 6: Underestimating parameter
\texorpdfstring{$q^2$}{}}
\label{supp:sim6}

Repeating Simulation 5, we now simulated the data with $q^2=0.09$ and
kept the parameter of our reanalysis at $q^2=0.0225$, that is, we now
underestimate the true $q^2$. Individual sensitivities vary a lot now.
Even though the mean true direct task sensitivity is $d'_\text{true} =
0$ (50\%-correct), due to a large standard deviation of $q = 0.3$,
95\% of participants' true sensitivities range between -0.6
(38\%-correct) and 0.6 (62\%-correct). The assumption $q=0.3$ poses a
problem from a theoretical perspective because some participants can
discriminate the masked stimuli relatively well (above 60\%-correct).
In this case, our reanalysis is more conservative and detects an
\ITA{} in only 62.2\% of the runs. However, the analysis based on the
trial-by-trial data also only achieves a power of 69.2\% due to the
large variability: Even in this case, our reanalysis would not be too
conservative.

\subsubsection{Additional Simulations} 

We conducted additional simulations, one for each combination of the
following parameters: $N\in\{5, 10, 20\}$, $K\in\{100, 200, 400\}$,
$d'_\text{true} = \{0, 0.1, 0.2, 0.5\}$, and $q^2\in\{0.01, 0.0225,
0.09\}$. In all these simulations, the average, absolute deviation
between true and estimated sensitivities was small, $|d'_\text{true} -
d'_\text{estimated}| \leq 0.01$. A deviation of $0.01$ in terms of
sensitivity translates into a deviation as small as 0.2\%-correct,
which can be considered negligible in this setting---and deviations in
simulations with $N\geq 10$ are substantially smaller.

We computed the standard deviation of $d'_\text{estimated}$ (denoted
by $SD[d'_\text{estimated}]$) across the 10,000 simulations of each
parameter combination. We compared this with the estimated standard
error, $SE$. For this purpose, we squared $SE$ of each run, averaged
the values and took the square root of the average, which is the
standard procedure to average standard errors. For the direct task,
the difference between actual variability and our estimates was again
$|SD[d'_\text{estimated}] - SE|\leq 0.01$. For the indirect task, the
same was true when $N\geq 10$. However, for very small sample sizes
($N=5$) our reanalysis deviated to some degree but the absolute
difference between actual standard deviation and our estimates still
was $|SD[d'_\text{estimated}] - SE|\leq 0.05$. Since all reanalyzed
studies use sample sizes of $N\geq 10$ in the indirect task, our
reanalysis produced approximately unbiased estimates. Overall, our
reanalysis approximates the appropriate analysis sufficiently well in
the context we applied it to.


\section{\textbf{\appendixB}}
\label{supp:medianClassifier}

In the appropriate analysis to infer an \ITA{}, one needs to transform
continuous measurements of the indirect task (e.g., RTs) into a binary
classification response. In this step it is important to use the best
possible classifier, in order to achieve the highest $d'$ or
\%-correct values and thereby increase the chance to establish an
\ITA{}. Depending on the type of measurement that is taken in the
indirect task (e.g., RT, brain activity, grip force, etc.), this best
classifier can have different forms. In many cases, the median
classifier is a suitable choice. For example with RT data (as in our
replication based on \citeNP{Dehaene_etal_98}), the classifier
computes for each participant the median RT across all trials and
classifies a trial as congruent if the RT is faster than the median
and as incongruent if the RT is slower. Below, we prove that the
median classifier is optimal in this setting. The proof requires two
assumptions: 

\begin{enumerate} 
\item[(1)] The indirect measure follows a normal or log-normal
distribution with an additive shift between congruent and incongruent
conditions. In our case, this assumption is justified because it is
well known that RT distributions are well approximated by log-normal
distributions \cite{Ulrich_Miller_93,palmer2011shapes}.
\item[(2)] An equal number of observations need to be drawn in both
conditions, which is satisfied by the typical experimental design.
\end{enumerate}

Note, that \citeA{Franz_vonLuxburg_15_PS} also applied nonparametric
machine learning classifiers with similar results.

\paragraph*{General form of the optimal classifier.}  Consider a
classification task where the input is a real-valued number $x$ (e.g.,
a reaction time, RT), and the classifier is supposed to predict one of
two labels $y$ (e.g., 'congruent' or 'incongruent'; for simplicity we
use labels 1 and 2 in the following). Following the standard setup in
statistical decision theory \cite[section 1.5]{bishop2006pattern} we
assume that the input data $X$ and the output data $Y$ are drawn
according to some fixed (but unknown) probability distribution $P$. 
This distribution can be described uniquely by the class-conditional
distributions $P( X \condon Y = 1)$ and $P(X \condon Y = 2)$ and the
class priors $\pi_1 = P(Y = 1)$ and $\pi_2 = P(Y=2)$. A classifier is
a function $f:\R \to \{1,2\}$ that assigns a label $y$ to each input
$x$. The classifier that has the smallest probability of error is
called the Bayes classifier. In case the classes have equal weight,
that is $\pi_1 = \pi_2$, the Bayes classifier has a particularly
simple form: It classifies an input point $x$ by the class that has
the higher class-conditional density at this point. Formally, this
classifier is given by

\banum
\label{eq-fopt}
f_{opt}(x) := \begin{cases} 1 & \text{ if } P(X = x \condon Y =1 ) > P(X = x \condon Y=2)\\
2 & \text{ otherwise.} 
\end{cases}
\eanum

\paragraph*{Optimal classifier for normal and log-normal
  distributions.}  

We now consider the special case where the class-conditionals follow a
particular distribution. Let us start with the normally distributed
case. We assume that both class-conditionals are normal distributions
with means $\mu_1$, $\mu_2$ and equal variance $\sigma^2$,  and we
denote their corresponding probability density functions (PDFs) by
$\varphi_{\mu_1,\sigma}$ and $\varphi_{\mu_2,\sigma}$. Under the
additional assumption that  both classes have equal weights $\pi_1 =
\pi_2 = 0.5$, the  cumulative distribution  function (CDF) of the
input (marginal distribution of $X$)  is given as 
\banum
\label{eq-gaussian}
&\Omega(x) := 0.5 \cdot \Big( \Phi(\frac{x - \mu_1}{\sigma}) +
\Phi(\frac{x - \mu_2}{\sigma} )\Big), 
\eanum
where $\Phi$ denotes the CDF of the standard normal distribution. For
$t \in \R$, we introduce the step function classifier with threshold
$t$ by 
\banum \label{eq-step}
f_t(x) := \begin{cases} 1 & \text{ if } x \leq t\\
2 & \text{ otherwise.} 
\end{cases}
\eanum
In the special case where the threshold $t$ coincides with the median of
the marginal distribution of $X$, we call the resulting step function
classifier the {\em median classifier. }

\begin{propositionnn}[Median classifier is optimal for normal model]
  If the input distribution is given by Equation~\eqref{eq-gaussian}, then 
the optimal classifier $f_{opt}$ coincides with the median
classifier. 
\end{propositionnn}
{\em Proof.}
Because both classes have the same weight of 0.5, the
Bayes classifier is given by $f_{opt}$ as in Equation~\eqref{eq-fopt}. 
For any choice of $\mu_1$, $\mu_2$ and $\sigma$, the class-conditional
PDFs $\varphi_{\mu_1, \sigma}$ and $\varphi_{\mu_2,\sigma}$  intersect
exactly once, namely at $t^* = (\mu_1 + \mu_2) / 2$. By definition of
$f_{opt}$, the optimal classifier $f_{opt}$ is then the step function
classifier with threshold $t^*$. We now compute the value of the CDF at 
$t^*$: 

\ba
\Omega(t^*) & = 0.5 \cdot 
\Big( \Phi(\frac{t^* - \mu_1}{\sigma}) + \Phi(\frac{t^* - \mu_2}{\sigma} )\Big)\\
&= 0.5 \cdot  
\Big( \Phi(\frac{\mu_2 - \mu_1}{2\sigma}) + \Phi(\frac{\mu_1 - \mu_2}{2\sigma})\Big)\\
&= 0.5 \cdot  
\Big(\Phi(\frac{\mu_2 - \mu_1}{2}) + (1 - \Phi(\frac{\mu_2 - \mu_1}{2})\Big)\\
& = 0.5. 
\ea
Here, the second last equality comes from the fact that the normal
distribution is symmetric about 0.  This calculation shows that the
optimal threshold $t^*$ indeed coincides with the median of the input
distribution, which is what we wanted to prove.~\hfill$\Box$

It is easy to see that this proof can be generalized to more general
types of symmetric probability distributions. It is, however, even
possible to prove an analogous statement for log-normal distributions,
which are not symmetric themselves. We introduce the notation
$\lambda_{\mu,\sigma}$ for the PDF of a
log-normal distribution, and $\Lambda_{\mu,\sigma}$ for the
corresponding CDF. These functions are defined as
\ba
&\lambda_{\mu,\sigma}(x) :=  
\frac{1}{x \sigma \sqrt{2\pi}}
\exp\Big(- \frac{(\log x  -\mu)^2 }{2 \sigma^2} \Big) 
&& \text{ and } &&\Lambda_{\mu,\sigma}(x) := \Phi\Big(     
\frac{\log x  -\mu}{\sigma}
\Big).
\ea

Consider the case where the class-conditional distributions
are log-normal distributions with same scale parameter $\sigma$ but
different location parameters $\mu_1$ and $\mu_2$, and 
assume that both classes have the same weights $\pi_1 = \pi_2 =
0.5$. Then the PDF and CDF of the input distribution (marginal
distribution of $X$) are given as
\banum
& g(x) = 0.5 \cdot \;(\; \lambda_{\mu_1, \sigma}(x) + \lambda_{\mu_2,
  \sigma}(x) \;) \nonumber \\
& G(x) = 0.5 \cdot \;(\; \Lambda_{\mu_1, \sigma}(x) + \Lambda_{\mu_2,
  \sigma}(x) \;). \label{eq-log-normal}
\eanum
\begin{propositionnn}[Median classifier is optimal for log-normal
  model] 
  If the input distribution is given by Equation~\eqref{eq-log-normal}, then 
the optimal classifier $f_{opt}$ coincides with the median
classifier. 
\end{propositionnn}
{\em Proof.} %
The proof is analogous to the previous one. 
For any choice of $\mu_1$, $\mu_2$ and $\sigma$, the densities
$\lambda_{\mu_1, \sigma}$ and $\lambda_{\mu_2, \sigma}$ intersect
exactly once. To see this, we solve the equation $\lambda_{\mu_1,
  \sigma}(t^*) = \lambda_{\mu_2, \sigma}(t^*)$, which leads to the
unique solution $t^*= \exp( (\mu_1 + \mu_2)
/ 2)$. 
The input CDF at this value can be computed as 
\ba
G(t^*) & = 0.5 \Big( \Lambda_{\mu_1,
  \sigma}(t^*) +  \Lambda_{\mu_2, \sigma}(t^*) \Big) \\
& =  0.5\Big( \Phi( \frac{\mu_2 - \mu_1}{2\sigma} )+  \Phi( \frac{\mu_1 - \mu_2}{2\sigma} ) \Big)\\
& = 0.5. 
\ea
The last step follows as above by the symmetry of the normal cdf. 
\hfill$\Box$


\section{\textbf{\appendixC}}
\label{supp:reanalysisFormulas}

We use typically reported results from studies on unconscious priming
to estimate the direct and indirect task sensitivities,
$d'_\text{estimated,direct}$ and $d'_\text{estimated,indirect}$.
First, we recapitulate the basic model assumptions of a standard
repeated measures ANOVA and introduce the notation. We then derive 
estimators for the sensitivity and standard error in both tasks using
only the typically reported results. Finally, we compute the
difference between direct vs. indirect task sensitivities and
construct a confidence interval around that difference in order to
test for an \ITA{}. 


\subsection{\textbf{Model assumptions}}
\label{supp:preliminaries}

Our reanalysis of both tasks is based on the standard model of
repeated measures ANOVA and paired $t$ test
\cite{Winer_etal_91,Maxwell_Delaney_00,rouder2018power} as employed in
all reanalyzed studies. In this model $N$ participants perform $M$
trials in each condition. In the specific setting we consider, there
are only $2$ conditions. In the direct task, this corresponds to
trials where the masked stimulus is from either of two categories, A
vs. B. In the indirect task, the two conditions are typically
congruent (A-A, B-B) vs. incongruent (A-B, B-A). In each trial of a
given task, $Y_{ijk}$ denotes the response from participant $i$
($1,...,N$) in condition $j$ ($1$ or $2$) in trial $k$ ($1,...,M$),
where we assume a balanced design such that the total number of trials
$K$ is split evenly into the two conditions for $M=K/2$ trials per
condition.

In the indirect task, responses $Y_{ijk}$ are the indirect measures
(e.g., RTs). In the direct task, it is plausible to assume that
responses $Y_{ijk}$ represent participants' internal evidence about
the masked stimuli (some neural activity indicating whether the
participant saw a masked stimulus from category A or from B). Based on
this noisy internal evidence, participants make an internal
classification and guess in each trial to which category the stimulus
belonged. 

The standard model decomposes participants' responses
$Y_{ijk}$ into five components:
\[
  Y_{ijk} = \mu + p_i + c_j + (p\times c)_{ij} + \epsilon_{ijk}.
\]

To facilitate understanding, we now describe the model for the example
of congruency effects on RTs in the indirect task; but the same
notation applies to other indirect measures and to the direct task as
well. RTs have a grand mean $\mu$. Some participants have faster RTs
than others which is captured in participants' effects $p_i$. The
congruency condition has an effect $c_j$ on RTs. While $c_1$ is
negative leading to faster RTs in congruent trials, $c_2$ is positive
reflecting slower RTs in the incongruent conditions. Participants
differ in the extent to which the congruency conditions affect them
captured in $(p\times c)_{ij}$ so that some participants have a larger
congruency effect than others. The variability in the individual
effects is captured by this term's variance, $\Var[(p\times c)_{ij}] =
\sigma^2_{p\times c}$. Additionally, there is trial-by-trial noise
$\epsilon_{ijk}$ from neuromuscular noise and measurement error
leading to different responses in each trial. This trial-by-trial
measurement error is assumed by the standard models to have a constant
variance (homogeneity) across participants and conditions,
$\Var[\epsilon_{ijk}] = \sigma^2_{\epsilon}$. The congruency effect
$c_j$ is a fixed effect while participant and interaction effects
($p_i$ and $(p\times c)_{ij}$) are random effects because they depend
on the drawn sample of participants. Random effects and trial-by-trial
noise are assumed to be normally distributed with an expected value of
zero and their corresponding variance.

\subsubsection*{Raw effects and sensitivities}  Each participant $i$
has an individual expected congruency effect, $\Delta_i$, which
theoretically would be obtained by sampling infinitely many trials.
The expected RT difference across participants is denoted by
$\Delta$.
\begin{align*}
        \Delta_i &= \left(c_2 + (p\times c)_{i2}\right) - \left(c_1 + (p\times c)_{i1}\right)\\
        \Delta &= c_2-c_1
\end{align*}
In a typical experiment, the individual congruency effects are
estimated by the observed mean difference between conditions. For the
$i$-th participant, this estimate is $\hat{\Delta}_i$ and averaged
across participants this is $\hat{\Delta}$.
\begin{align*}
        \hat{\Delta}_i &= \bar{Y}_{i2\cdot} - \bar{Y}_{i1\cdot} = \frac{1}
        {M}\sum_{k=1}^M Y_{i2k} - \frac{1} {M}\sum_{k=1}^M Y_{i1k}\\
        \hat{\Delta} &= \frac{1}{N} \sum_{i=1}^N \hat{\Delta}_i
\end{align*}

A participant's true sensitivity $d'_{\text{true},i}$ is the
normalized effect---normalized by the trial-by-trial error standard
deviation $\sigma_\epsilon$. This quantity indicates, similar to a
signal to noise ratio, how well a participant's RTs are separable and
therefore to which degree the masked stimuli were processed, cf.
Figure~\ref{fig:toyExample}a. The expectation across participants is
the true sensitivity $d'_\text{true}$ indicating how well the RTs of a
prototypical participant are separated.
\begin{align*}
        d'_{\text{true},i} & = \frac{\Delta_i}{\sigma_\epsilon}\\
        d'_\text{true} &= \frac{\Delta}{\sigma_{\epsilon}}.
\end{align*}

In the direct task, $d'_\text{true}$ is typically measured by the
sensitivity index $d'$ averaged across participants. Participants'
individual $d'_i$ are calculated from hit rate, $\text{HR}$
(\%-correct guesses for masked stimuli from category A), and false
alarm rate, $\text{FA}$ (\%-incorrect guesses for masked stimuli from
category B), where $\Phi^{-1}$ is the inverse cumulative density
function of the normal distribution.
\begin{align*}
    d'_i &= \Phi^{-1}(\text{HR}) - \Phi^{-1}(\text{FA})\\
    d' &= \frac{1}{N}\sum_i d'_i.
\end{align*}
Note that the empirical literature often uses the notation $d'$
without a clear distinction between estimated vs. true value and
individual vs. average effects. Because we need to be more precise in
our derivations: We denote the true value of an individual participant
by $d'_{\text{true},i}$ and the sensitivity index, which is an
estimate for the true value, by $d'_i$. We denote the true sensitivity
across participants by $d'_\text{true}$. In the direct task, this is
estimated by the average across $d'_i$ values denoted by $d'$. We will
also label this averaged estimate $d'_\text{estimated,indirect}$.

\subsubsection*{Two variance sources: true effect (between-) vs. trial-by-trial error (within-subject) variance}

Participants differ in their true congruency effect. The variance of
these true inter-individual differences can be derived from the model
variances using the standard assumptions (1) $(p\times c)_{ij} \sim
\mathcal{N}(0,\sigma^2_{p\times c})$, (2) $\Var[c_1] = \Var[c_2] = 0$,
and (3) $(p\times c)_{i1} = -(p\times c)_{i2}$. We denote this true
effect variance as $\sigma^2_{\text{effect}}$: 
\begin{align*} 
        \sigma^2_{\text{effect}} &= \Var[\Delta_i] = \Var[[c_2 + (p\times c)_{i2}] - [c_1 + (p\times c)_{i1}]]
        \\
        &= \Var[(p\times c)_{i2} - (p\times c)_{i1}] = \Var[2(p\times c)_{i2}]
        \\
        &= 4\sigma^2_{p\times c}.
\end{align*} 

The variance of the actually observed congruency effects is
conceptually different from the variance of the true effects. We
denote the variance of the observed congruency effects as
$\sigma^2_{\hat{\Delta}_i}$. The observed congruency effects vary more
because they are not only affected by true inter-individual difference
but also by trial-by-trial measurement errors: 
\begin{align*} 
    \sigma^2_{\hat{\Delta}_i} &= \Var[\hat{\Delta}_i] \\ 
    &= \Var\left[\bar{Y}_{i2\cdot} - \bar{Y}_{i1\cdot}\right]\\ &= \Var\left[\frac{1}{M} \left(\sum_{k=1}^M \mu + p_i + c_2 + (p\times c)_{i2} + \epsilon_{i2k}\right) \right. \\
    & ~~~~~~~~~~~~~~~~~~~ - \left. \frac{1}{M} \left(\sum_{k=1}^M \mu + p_i + c_1 + (p\times c)_{i1} + \epsilon_{i1k}\right)\right] \\ 
    &= \Var\left[[c_2 + (p\times c)_
    {i2}] - [c_1 + (p\times c)_{i1}] + \frac{1}{M} \left(\sum_{k=1}^M \epsilon_
    {i2k}\right) - \frac{1}{M} \left(\sum_{k=1}^M \epsilon_{i1k}\right)\right]\\
    &= \Var\left[\Delta_i +
    \frac{1}{M} \left(\sum_{k=1}^M \epsilon_{i2k}\right) - \frac{1}{M}
    \left(\sum_{k=1}^M \epsilon_{i1k}\right)\right]\\ 
    &= \Var[\Delta_i] + \Var\left[\frac{1}{M} \sum_{k=1}^M \epsilon_{i2k}\right] + \Var\left[\frac{1}{M}
    \sum_{k=1}^M \epsilon_{i1k}\right]\\ 
    &= \sigma^2_{\text{effect}} + \frac{2}{M}\sigma^2_\epsilon \\
    &= \sigma^2_{\text{effect}} + \frac{4}{K}\sigma^2_\epsilon.
\end{align*}
This has an implication for the variance of average congruency
effects, $\hat \Delta = \frac{1}{N}\sum_i \hat \Delta_i$. These
observed, average congruency effects vary due to two variance sources,
the true inter-individual differences and trial-by-trial measurement
error. 
\[
    \hat{\Delta} \sim \mathcal{N}\left(\Delta, 
    \frac{\sigma^2_{ \text{effect}} + \frac{4}{K}\sigma^2_\epsilon}{N}
    \right).
 \]

We will later have to estimate $\sigma^2_\epsilon$ from a given
$\sigma^2_{\hat{\Delta}_i}$. To achieve this, we must disentangle
$\sigma^2_{\text{effect}}$ and $\sigma^2_\epsilon$. We do so by
defining the ratio $q^2$ between these two sources of variance: 
\[
        q^2 = \frac{\sigma^2_{\text{effect}}}{\sigma^2_\epsilon}.
\]
This parameter tells us how much of the observed variability comes
from true differences vs. noise. If $q^2=0$ then all participants
would have the same true congruency effect and observed differences
are only due to trial-by-trial error. If $q^2$ is large then there is
relatively small trial-by-trial error variance and observed
differences between participants stem from reliable, true differences
between participants. Crucially, note that $q^2$ is also the variance
of true, individual sensitivities. Thus, the square root of this
ratio, $q$, is the standard deviation of true, individual
sensitivities. 
\begin{align*}
     \text{\textit Var}[d'_{\text{true},i}] = Var\left[ \frac{\Delta_i}{\sigma_\epsilon}  \right] 
     = \frac{\sigma^2_\text{effect}}{\sigma^2_\epsilon} 
     = q^2 \text{~~~~corresponding to~~~~} SD[d'_{\text{true},i}] = q
\end{align*}

We derive a reasonable value to use for our setting in Appendix
\ref{supp:roleOfq}, which is $q^2=0.0225$. This means that we will
assume that participants' sensitivities $d'_{\text{true},i}$ vary
around some true value $d'_\text{true}$ with a standard deviation of
$q = 0.15$.

\subsubsection*{Relationship between sensitivity and accuracy}

As we have already mentioned, some published studies report $d'$
values, whereas other studies report \%-correct values in the direct
task. Because we would like to be able to work with either of them, we
now discuss the relationship that can transform \%-correct values into
$d'$ values and vice versa. 

Recall that $d'_{\text{true},i}$ denotes the true sensitivity of
participant $i$, and let us introduce the notation $\pi_i$ for the
true probability of a correct classification of a masked stimulus by
participant $i$. We now make the assumption of a neutral criterion in
the direct task, that is, participants are not inclined to guess one
category of the masked stimuli (A or B) more often than the other.
Under this assumption, the true relationship is $d'_{\text{true},i} =
2\Phi^{-1}(\pi_i)$ where $\Phi^{-1}$ is the inverse cumulative normal
distribution \cite{Macmillan_Creelman_90}. To simplify our later
analysis, we now introduce the linear approximation $h(x) = 5(x - 0.5)
\approx 2\Phi^{-1}(x)$. This approximation works remarkably well in
the regime of sensitivities being close to zero:
\[
    \text{given }\pi_i\text{, we approximate }  d'_{\text{true},i} \approx h(\pi_i) = 5(\pi_i-0.5)
\]
\[
    \text{given }d'_{\text{true},i}\text{, we approximate }\pi_i \approx h^{-1}(d'_{\text{true},i}) = \frac{1}{5}d'_{\text{true},i}+0.5 
\]
For example, an accuracy of 54\%-correct is approximately translated
into the sensitivity $d'_{\text{true},i} \approx 5\cdot(0.54-0.5) =
0.2$. This is very close to the exact translation, $d'_{\text{true},i}
= 2\Phi^{-1}(\pi_i) = 0.201$. Table \ref{tab:approximation} shows that
this approximation provides a very tight fit in the range of $\pi_i
\in [0.4;0.6]$ or, equivalently, $d'_{\text{true},i} \in [-0.5; 0.5]$.
Larger values, that is, an accuracy above 60\%-correct, would be at
odds with the experimental setting in which direct task performance is
assumed to be close to chance ($d'_{\text{true},i}$ close to 0 and
$\pi_i$ close to 0.5).

\begin{table}[!ht]
\centering
\caption{Relation between the true accuracy (first column), the
approximation of the sensitivity (second column) and the true
sensitivity (third column). Note, that for $\pi_i$ in the range of
$[0.5,0.6]$ and $D_i$ in the range of $[0,0.5]$ (first six rows in the
table) there is a very tight fit between $h(\pi_i)$ and
$d'_{\text{true},i}$. Negative values of $d'_{\text{true},i}$ follow
symmetrically.\newline}
\begin{tabular}{rrr}
  \hline
$\pi_i$ & $h(\pi_i)$ & $d'_{\text{true},i}$ \\ 
  \hline
0.50 & 0.000 & 0.000 \\ 
  0.52 & 0.100 & 0.100 \\ 
  0.54 & 0.200 & 0.201 \\ 
  0.56 & 0.300 & 0.302 \\ 
  0.58 & 0.400 & 0.404 \\ 
  0.60 & 0.500 & 0.507 \\ 
  0.62 & 0.600 & 0.611 \\ 
  0.64 & 0.700 & 0.717 \\ 
  0.66 & 0.800 & 0.825 \\ 
  0.68 & 0.900 & 0.935 \\ 
  0.70 & 1.000 & 1.049 \\ 
   \hline
\end{tabular}
\label{tab:approximation}
 \end{table}
%


\subsection[Estimating sensitivity from mean sensitivity index]{\textbf{Estimated sensitivity, \texorpdfstring{$d'_\text{estimated,direct}$}{}, from mean sensitivity index \texorpdfstring{$d'$}{}}}
\label{supp:Dfromsensitivity}

We want to estimate the sensitivity and corresponding standard error
from the typically reported direct task results. Usually, the average
across individual sensitivity indices is reported as $d'$. This
sensitivity index is already an estimate of the true sensitivity and
we take it as it is \cite{Macmillan_Creelman_90}, 
\begin{align*}
    d'_\text{estimated,direct} = d'. \numberthis
     \label{eqn:DfromdprimeUnocrrected}
\end{align*}
The standard error of $d'$ is composed of two variances, one due to
systematic variation between individuals' true sensitivities
($d'_\text{true,i}$) and the other due to non-systematic measurement
error ($\epsilon_{d'_i}$). We use two simplifications: (a)~We neglect
dependencies between them because the variance of random error
$\Var[\epsilon_{d'_i}]$ does not change substantially for different
sensitivity values in the relevant range, $D_i^\text{dir} \in [-0.5,
0.5]$; (b)~We apply the approximation function $h$ that relates $d'_i$
to $\hat\pi_i$. This allows us to use the variance of the binomially
distributed accuracies $\hat\pi_i$ from $K$ trials,
$\Var[\epsilon_{\hat \pi_i}] = \pi_i(1-\pi_i)/K$, and relate them back
to the variance of $d'_i$, which leads to $\Var[\epsilon_{d'_i}]
\approx 5^2\Var[\epsilon_{\hat \pi_i}]$. 
\begin{align*}
        SE_\text{direct} &= \sqrt{\Var\left[d'\right] } \\
        &= \sqrt{\Var\left[\frac{1}{N} \sum_i d'_i\right]} = \frac{1}{\sqrt{N}}\sqrt{\Var[d'_i]} 
        = \frac{1}{\sqrt{N}}\sqrt{\Var[d'_\text{true,i} + \epsilon_{d'_i}]}\\
        &\overset{(a)}{\approx} \frac{1}{\sqrt{N}}\sqrt{\Var[d'_\text{true,i}] + \Var[\epsilon_{d'_i}]}
        \overset{(b)}{\approx} \frac{1}{\sqrt{N}}\sqrt{\Var[d'_\text{true,i}] + 5^2\Var[\epsilon_{\hat \pi_i}]}\\
        & \overset{}{=} \underbrace{\frac{1}{\sqrt{N}}}_\text{average}\sqrt{\underbrace{q^2}_{\text{between subject variance}} + ~5^2~ \underbrace{\frac{\left(\frac{1}{5}d'+0.5\right)\left(1-\left(\frac{1}{5}d'+0.5\right)\right)}{K}}_\text{non-systematic error of $\hat \pi_i$}}         \numberthis \label{eqn:SEfromdprime}.
\end{align*}
Without simplifications (a) and (b), one could construct an exact
estimator. Exact calculations from \citeA{Miller_96} show that $d'$
slightly overestimates the true sensitivity $d'_\text{true}$ but that
this bias is so small that the estimator can be considered
approximately unbiased when typical sample sizes as in our context are
used. On the other hand, our simplifications allow us to find a closed
form solution that is simple to compute. Our estimators are well
aligned with the true values, which we have shown by validating
simulations in Appendix \ref{supp:validation}.


\subsection[Estimating sensitivity from mean accuracy]{Estimated sensitivity, \texorpdfstring{$d'_\text{estimated,direct}$}{}, from mean accuracy \texorpdfstring{$\hat{\pi}$}{}}
\label{supp:Dfromacc}

Instead of $d'$, some studies report the average classification
accuracy $\hat{\pi}$ (\%-correct) for the direct task. We estimate the
sensitivity~$d'_\text{estimated,direct}$ from the mean accuracy
$\hat{\pi}$ by a plug-in estimator \cite{Macmillan_Creelman_90},
\[
        d'_\text{estimated,direct} = 2\Phi^{-1}(\hat{\pi}) \approx 5\cdot(\hat\pi - 0.5)
        \numberthis \label{eqn:Dfrompi}
\]
where $\Phi^{-1}$ is the inverse cumulative normal distribution.
Exploiting the linearity of approximation $h$ in (*), we can derive
that this estimator is approximately unbiased:
\begin{align*} \text{E}[d'_\text{estimated,direct}]
        &= \text{E}[2\Phi^{-1}(\hat{\pi})] \approx \text{E}[h(\hat{\pi})] \overset{(*)}{=} h[\text{E}(\hat{\pi})] = h(\pi) \approx 2\Phi^{-1}(\pi) = d'_\text{true}.
\end{align*}
Next, the standard error can be derived in the same fashion as for
reported $d'$ values so that we obtain:
\begin{align*}
    SE_\text{direct} = \frac{1}{\sqrt{N}}\sqrt{q^2 + 5^2\frac{\hat 
    \pi \left(1-\hat \pi \right)}{K}}. \numberthis \label{eqn:SEfrompi}
\end{align*}


\subsection[Estimating sensitivity from t and F values]{Estimated sensitivity, \texorpdfstring{$d'_\text{estimated,indirect}$}{}, from \texorpdfstring{$t$}{} and
\texorpdfstring{$F$}{} values}
\label{supp:Dfromt}

Now let us move to estimating sensitivities from $t$ values in the
indirect task. We will show that an unbiased estimator is obtained
from multiplying the $t$ value by the constant $c_{N,K,q^2}$:
\[
        d'_\text{estimated,indirect} = t\cdot c_{N,K,q^2} ~~~~\text{with}~~~~~ c_{N,K,q^2}
        =
        \sqrt{\frac{q^2 + \frac{4}{K}}{N}} \sqrt{\frac{2}{N-1}} ~~\frac{\Gamma\left(\frac{N-1}{2}\right)}{
        \Gamma\left( \frac{N-2}{2} \right)},
        \numberthis \label{eqn:Dfromt}
\]
where $\Gamma$ is the gamma distribution.

We start by considering how the $t$ value in our setting is computed
from the observed congruency effect:
\[
        t = \frac{ \hat{\Delta} }{\hat{\sigma}_{\hat{\Delta}_i} } 
        \sqrt{N}
\]
We know that $\hat{\Delta} \sim \mathcal{N}\left(\Delta, (\sigma^2_{
\text{effect}} + \frac{4}{K}\sigma^2_\epsilon)/N \right)$ from above.
Now we introduce independent random variables $Z \sim
\mathcal{N}\left(0, 1\right)$ and $V \sim \chi^2\left (N-1\right)$ and
rearrange  $t$: 
\begin{align*}
    t &= \frac{Z\sqrt{\left(\sigma^2_{\text{effect}} + \frac{4}{K}\sigma^2_\epsilon\right)/N} + \Delta }{ \sqrt{ \left(\sigma^2_{\text{effect}} + \frac{4}{K}\sigma^2_\epsilon\right) }\sqrt{\frac{V}{N-1}} }\sqrt{N} \\ 
  &= \left(Z+\Delta\sqrt{\frac{N}{\left(\sigma^2_{\text{effect}} + \frac{4}{K}\sigma^2_\epsilon\right)}}\right)\frac{\sqrt{N-1}}{\sqrt{V}}.
\end{align*}
We now use $\sigma^2_{\text{effect}} = q^2 \sigma^2_\epsilon$ (also
from above) to isolate $\sigma^2_\epsilon$ and obtain
$d'_{true}$.  
\begin{align*} 
        t &= \left(Z+\Delta\sqrt{\frac{N}{\left(q^2\sigma^2_\epsilon + \frac{4}{K}\sigma^2_\epsilon\right)}}\right)\frac{\sqrt{N-1}}{\sqrt{V}} \\
        &= \left(Z+\frac{\Delta}{\sigma_\epsilon}\sqrt{\left(\frac{N}{q^2 + \frac{4}{K}}\right)}\right)\frac{\sqrt{N-1}}{\sqrt{V}} \\
        &= \left(Z+d'_{true}\sqrt{\left(\frac{N}{q^2 + \frac{4}{K}}\right)}\right)\frac{\sqrt{N-1}}{\sqrt{V}} \\
\end{align*}
As a result, $t$ follows a $t$ distribution with degrees of freedom
$df = N-1$ and non-centrality parameter $\delta =
d'_{true}\sqrt{\frac{N}{q^2 + \frac{4}{K}}}$. From
\citeA{hogben1961moments} and \citeA{hedges1981distribution}, we know
the expected $t$ value to be
\begin{align*}
  \text{E}[t] 
  &= \delta \sqrt{\frac{N-1}{2}}\frac{\Gamma(\frac{N-2}{2})}{\Gamma(\frac{N-1}{2})} \\
  &= d'_{true} \cdot \sqrt{\frac{N}{q^2 + \frac{4}{K}}} \sqrt{\frac{N-1}{2}}\frac{\Gamma(\frac{N-2}{2})}{\Gamma(\frac{N-1}{2})} \\
  &= d'_\text{true} \cdot c^{-1}_{N,K,q^2}.
\end{align*}
In consequence, an unbiased estimator
of $d'_\text{true}$ is $d'_\text{estimated, indirect} = t\cdot
c_{N,K,q^2}$.

As for the expected value, the variance of $t$ values is also given by
the properties of a non-central $t$ distribution. Multiplying this
variance by the constant $c_ {N,K,q^2}$ yields the variance of our
estimated sensitivity $ d'_\text{estimated, indirect}$. Since this
depends on the non-centrality parameter, we use the plugin estimator
\begin{align*}
  \hat{\delta} &= d'_\text{estimated, indirect}\sqrt{\frac{N}{q^2 + \frac{4}{K}}} \\
  &= t \cdot c_{N,K,q^2} \sqrt{\frac{N}{q^2 + \frac{4}{K}}} \\
  &= t \cdot \sqrt{\frac{2}{N-1}} \frac{\Gamma\left(\frac{N-1}{2}\right)}{\Gamma\left( \frac{N-2}{2} \right)}
\end{align*}
The standard error being its positive square root
follows accordingly:
\begin{align*}
    SE_\text{direct}
    &= \sqrt{\Var[c_{N,K,q^2} \cdot t]} 
    = c_{N,K,q^2}\sqrt{\Var[t]} \\
    &= c_{N,K,q^2}\sqrt{\frac{1+\hat\delta^2}{N-3} - \frac{\hat\delta^2(N-1)\Gamma\left( \frac{N-2}{2} \right)^2}{2\Gamma\left(\frac{N-1}{2} \right)^2}} \\
    &= c_{N,K,q^2}\sqrt{\left(1+\frac{2t^2}{N-1}\left(\frac{\Gamma(\frac{N-1}{2})}{\Gamma(\frac{N-2}{2})}\right)^2 \right) \left(\frac{N-1}{N-3}\right) - t^2}
    \numberthis \label{eqn:SEfromt}
\end{align*}
With this, we can estimate the sensitivity and its standard error from
a given $t$ value in a repeated measures design.

Note that this approach can be applied identically to
reported $F$ values instead of $t$ values. The reason is that in
repeated measures ANOVA settings with two conditions the equality $|t|
= \sqrt{F}$ holds. The main argument can be derived in the following
equations using the standard definitions for the explained (SSE) and
residual summed squares (SSR), see
\citeA{Winer_etal_91,Maxwell_Delaney_00}:

\begin{align*}
    t^2 &= \left( \frac{ \hat{\Delta} } { \hat{\sigma}_{\hat{\Delta}_i} } \cdot  \sqrt{N} \right)^2 
        = \frac{4 \cdot N \cdot ( \hat{\Delta}/2 )^2 }{  \frac{1}{N-1} \sum_{i=1}^{N} \left(\hat{\Delta}_i - \hat{\Delta} \right)^2  } 
        = \frac{2 \cdot N \cdot ( \hat{\Delta}/2 )^2 }{  \frac{1}{N-1} \sum_{i=1}^{N} \left(\frac{\hat{\Delta}_i - \hat{\Delta}}{2} \right)^2  } \\
        &= \frac{2 \cdot N \cdot ( \hat{\Delta}/2 )^2 }{  2\sum_{i=1}^{N} \left(\frac{\hat{\Delta}_i - \hat{\Delta}}{2} \right)^2/(2N-2)  } = \frac{\text{SSE}/df_\text{E}}{\text{SSR}/df_\text{R}} 
        = F.
\end{align*}

Finally, note that this reanalysis for the indirect task can be
extended to unbalanced settings in which the total number of trials
$K$ is not equally distributed to the two conditions for $M=K/2$
trials per condition but instead to $M_1$ and $M_2$ trials per
condition $j=1$ and $j=2$, respectively. In these situations, one can
analogously show that $\hat{\Delta} \sim \mathcal{N}\left(\Delta,
(\sigma^2_{ \text{effect}} + \frac{M_1 + M_2}{M_1
M_2}\sigma^2_\epsilon)/N \right)$. Following the same steps as above,
one would obtain an alternative constant that now depends on the split
$M_1$ versus $M_2$ instead of only $K$. 
\begin{align*}
c_{N,M_1,M_2,q^2} = \sqrt{\frac{q^2 + \frac{M_1+M_2}{M_1 M_2}}{N}} \sqrt{\frac{2}{N-1}} \frac{\Gamma\left(\frac{N-1}{2}\right)}{\Gamma\left( \frac{N-2}{2} \right)}.
\end{align*}
As a sanity check, set $M_1=M_2=K/2$ and find $c_{N,M_1,M_2,q^2} =
c_{N,K,q^2}$.


\subsection{Confidence intervals for the difference in sensitivities}
\label{supp:ci}

Based on the previous estimators, we now need to test for a
significant difference between sensitivities in direct vs. indirect
tasks. For this purpose we construct a 95\% confidence interval around
the difference $d'_\text{difference}$ while taking the standard error
$SE_\text{difference}$ of the estimated difference into account:
\begin{align*}
        d'_\text{difference} &= d'_\text{estimated,indirect} - d'_\text{estimated,direct}  \numberthis \label{eqn:Ddiff} \\
        SE_\text{difference} &= \sqrt{(SE_\text{direct})^2 + (SE_\text{indirect})^2}  \numberthis \label{eqn:SEdiff} \\
        95\%\text{~CI} &= 
        \left[d'_\text{difference} \pm z_{0.975} \cdot SE_\text{difference} \right]
        \numberthis \label{eqn:ci}, 
\end{align*}

where $z_{0.975}=1.96$ is the 97.5\% quantile of the normal
distribution.  If zero is included in the confidence interval, $0 \in
95\%\text{~CI}$, then there is not sufficient evidence for an \ITA{}
because the observed difference can be explained by measurement error
in a situation where the true direct and indirect task sensitivities
are equal. Only if the confidence interval lies above zero, that is
$95\%\text{~CI}=[a,b]$ and $a>0$, there is evidence for the presence
of an \ITA{}. 

Note that in this test we use quantiles $z_\alpha$ of the normal
distribution and not quantiles of the $t$ distribution. Using the $t$
distribution would require to estimate the degrees of freedom, which
is unnecessarily complicated for our approach. We use the quantiles of
the normal distribution which leads to a more liberal test increasing
the likelihood of confirming an \ITA{} and following the
benefit-of-the-doubt approach (see General Discussion).


\section{\textbf{\appendixD}}
\label{supp:roleOfq}

As we have seen in the reanalysis of direct and indirect task
sensitivities, we need to know one parameter: $q^2$, a ratio of
systematic vs. noise variance. This is not an artifact of our
reanalysis but unavoidable.

\subsection*{What Does the Parameter $q^2$ Mean?}

To see what this parameter means and why we need to estimate it,
consider estimating the indirect task sensitivity
$d'_\text{estimated,indirect}$ from $t$ values. A $t$ value is
computed by dividing an observed effect by its standard error, $t =
\bar{x}/SE$. In the indirect task, $\bar{x}$ may be the average
congruency effect and $SE$ the estimated standard error of congruency
effects across participants. This standard error is influenced by two
sources of variability: variance due to inter-individual differences
in true congruency effects across participants
($\sigma^2_{\text{effect}}$) and variance due to trial-by-trial
measurement error ($\sigma^2_{\epsilon}$). We want to isolate the
latter variance, $\sigma^2_{\epsilon}$, because we want to estimate
the underlying sensitivity $d'_\text{true} = \Delta/\sigma_{\epsilon}$
from the $t$ value. Thus, we need to distinguish the two sources of
variability. We do so by defining the ratio $q^2$:
\[
    q^2 = \frac{\sigma^2_{\text{effect}}}{\sigma^2_\epsilon}.
\]

Note that this parameter is equal to the variance of individual true
sensitivities, $q^2 = Var[d'_{\text{true},i}]$, see
Supplement~\ref{supp:preliminaries}. Therefore, it might be more
intuitive to consider the un-squared parameter, which is the standard
deviation of participants' true sensitivities, $q =
\text{SD}[d'_{\text{true},i}]$.

\subsection*{Literature Review to Determine $q^2$ (Following Benefit-Of-The-Doubt Approach)}
\noindent

To estimate $q^2$, we consider multiple studies that either provide
estimates or make explicit assumptions. All these studies yield a
specific value, see our summary in Table~\ref{qtable}, columns $q^2$
and $q$. For our reanalysis, we will use the largest plausible value,
$q^2 = 0.0225$. Thus, we follow the benefit-of-the-doubt approach
giving a previously established \ITA{} the best chance to be confirmed
in our reanalysis.

\begin{table}[!ht]
\centering
\caption{We repeated our reanalysis of the indirect task sensitivity from \protect\citeA{Dehaene_etal_98} (last column) based on the $q^2$ values from different studies. Larger values of $q^2$ increase the estimated, indirect task sensitivity. We took the largest plausible value our reanalysis method.}
\label{qtable}
\begingroup\footnotesize
\begin{tabular}{lccc}
  \toprule                                                                                     
 \multirow{2}{*}{Study} & \multirow{2}{*}{\textbf{$q^2$}} & \multirow{2}{*}{$q$} &Reanalysis of Dehaene et al. (1998) \\                                                                       
 &  &  & $d'_\text{estimated, indirect}$ \\                                                    
 \cmidrule{1-4} 
  ten Brinke et al. (2014) & \textbf{0.0020} & 0.04 & 0.16 \\                    
  Our example study & \textbf{0.0074} & 0.09 & 0.20 \\                                         
  Rouder \& Haaf (2018) & \textbf{0.0087} & 0.09 & 0.21 \\       
  Miller \& Ulrich (2013) & \textbf{0.0121} & 0.11 & 0.23 \\                               
  Jensen (2002) & \textbf{0.0142} & 0.12 & 0.25 \\                                             
  Ribeiro et al. (2016) & \textbf{0.0214} & 0.15 & 0.28 \\                                     
   \hline                                                                                      
Our assumption & \textbf{0.0225} & 0.15 & 0.29 \\                                              
   \bottomrule                                                                                 
\end{tabular}
\endgroup
\end{table}

First, we estimated $q^2$ from the data of \citeA{tenBrinke_14}. This
yielded $\hat{\sigma}_\text{effect}^2 = (6.5~\text{ms})^2$ and
$\hat{\sigma}_\epsilon^2 = (144~\text{ms})^2$, which translates into
an estimated ratio of
$\hat{q}^2 = (6.5~\text{ms})^2 / (144~\text{ms})^2 = 0.0020$.

Our replication based on \citeA{Dehaene_etal_98} produced estimates
for the variances of $\hat{\sigma}_\text{effect}^2 =
(6.7~\text{ms})^2$ and $\hat{\sigma}_\epsilon^2 = (78~\text{ms})^2$
translating into an estimated ratio of $\hat{q}^2 = 0.0074$.

Similarly, \citeA[p.~21]{rouder2018power} discuss the relation between
the two sources of variance in psychophysics. Their formulas are
identical to ours when changing the notation from
$\sigma^2_{\text{effect}}$ to $\sigma^2_\beta$ and
$\sigma^2_{\epsilon}$ to $\sigma^2$. They argue that reasonable values
are $\sigma_{\text{effect}} = 28$~ms and $\sigma_{\epsilon} = 300$,
which leads to $q^2 = \sigma^2_{\text{effect}}/\sigma^2_{\epsilon} =
0.0087$. 

Other studies did not discuss the ratio between the two variances,
$\sigma^2_{\text{effect}}$ and $\sigma^2_{\epsilon}$, but only the
trial-by-trial error variability $\sigma^2_{\epsilon}$. We can combine
this with \citeA{Dehaene_etal_98} reporting the observed standard
deviation of RT effects to be $13.5$~ms. This variability is
constituted by $\hat{\sigma}^2_{\hat{\Delta}_i} =
\hat\sigma^2_{\text{effect}} + \frac{4}{K}\hat\sigma^2_\epsilon =
(13.5~\text{ms})^2$. By knowing $\hat\sigma^2_{\epsilon}$ and the
number of trials, $K$, we can rearrange the formula and estimate
$\hat\sigma^2_{\text{effect}}$ and thereupon $\hat{q}^2$.

\citeA[p. 846, in their Table~3]{Miller_Ulrich_13} suggested
$\sigma_\epsilon = 96$~ms in a binary forced-choice task (without
masked stimuli): Their error term $E_k$ with variance $Var[E_k] =
91.5$ corresponds to the mean noise across 100 trials, see their
Table~15. From this, we obtained $\sigma_\epsilon = \sqrt{Var[E_k]
\cdot 100} = 96$~ms, as noted above. Combining this with Dehaene et
al.'s results yields $\sigma_{\text{effect}} = 10.5$~ms and thereupon
$q^2 = 0.0121$.

\citeA[p. 877, Table 7 for task ``Hick SS 2'']{jensen1992importance} 
reported an average estimate of $\hat{\sigma}_\epsilon = 91$~ms
measured in $N=863$ nine to twelve year olds yielding $q^2 = 0.014$.
\citeA{ribeiro2016spontaneous} report $\hat{\sigma}_\epsilon = 79$~ms
in a speeded binary choice task without priming suggesting $q^2 =
0.021$. Even though the specific tasks and populations from these last
two studies do not match Dehaene et al.'s setting exactly, it is
plausible  that variances are by and large comparable.

Given this range of parameter values, we use a value that is larger
than any $q^2$ value reported in these studies:
\[
    q^2 = 0.0225 \text{~~~~corresponding to~~~~} q = SD[d'_{\text{true},i}] = 0.15.
\]

By choosing this upper bound on $q^2$, we follow the
benefit-of-the-doubt approach because large values of $q^2$ favor the
\ITA{} hypothesis in our reanalysis attributing more variance to
$\sigma^2_{ \text{effect}}$ and less to $\sigma^2_\epsilon$. This, in
turn, increases our estimate of $d'_\text{true} = \Delta /
\sigma_\epsilon$. For example, see how larger values of $q^2$ increase
the estimated indirect task sensitivity from \citeA{Dehaene_etal_98}
in the last column of Table~\ref{qtable}. Hence, overestimating $q^2$
leads to an overestimation of the indirect task sensitivity increasing
the chances of confirming an \ITA{}.

\subsection*{How Would our Reanalysis Look Like With a Different $q^2$?}
\noindent

\begin{figure}[!ht]
\centering
\includegraphics[ width = \textwidth, height = 19.25cm]{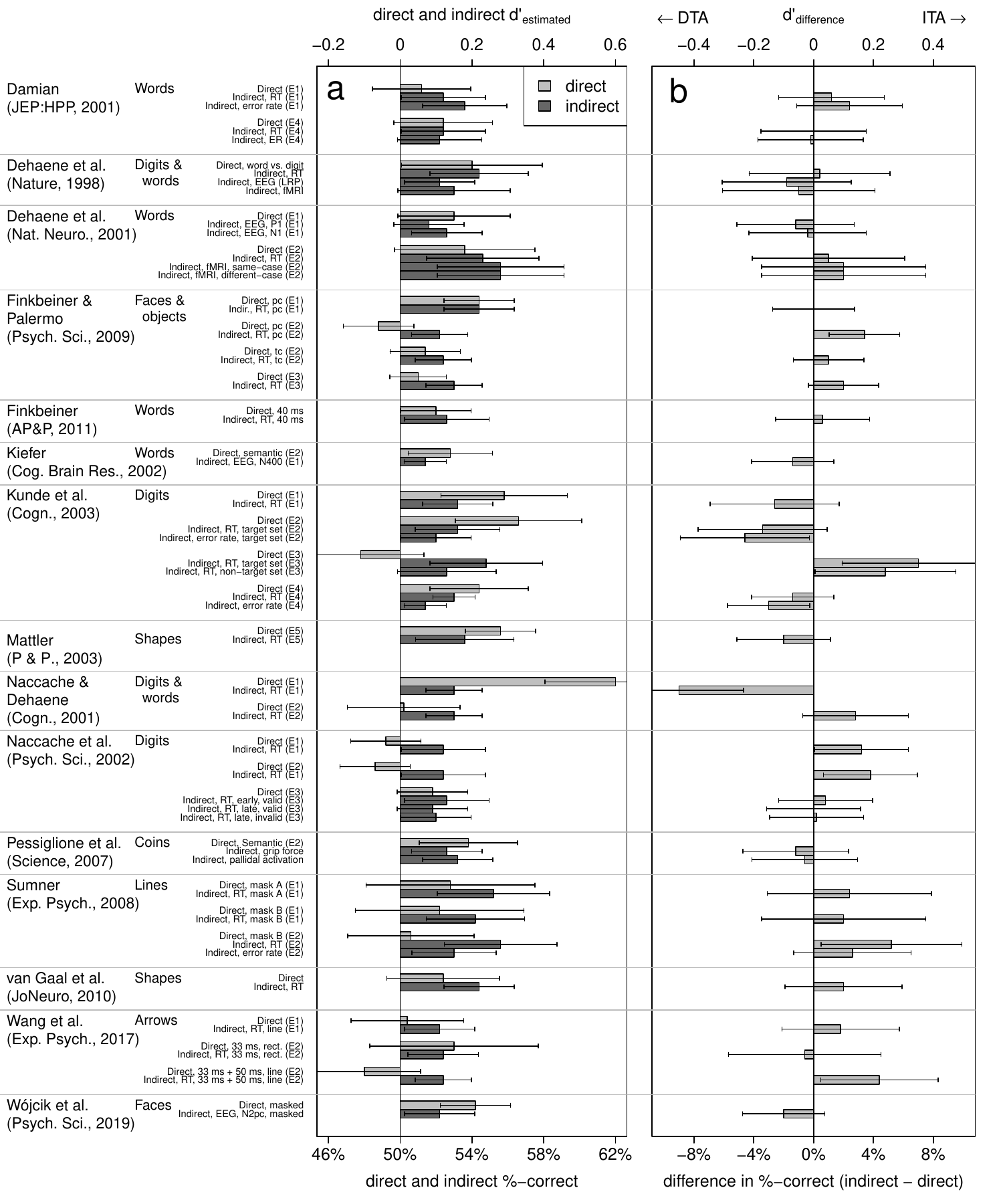}

\caption{\footnotesize \textbf{Reanalysis with $q^2 = 0.01$.}  Same as
Figure~\ref{fig:reanalysis} assuming that the standard deviation of
true sensitivities across participants is
$\text{SD}[d'_{\text{true},i}] = q = 0.1$. This assumption matches the
results of our replication and is therefore more realistic but also
more strict in dismissing results of an indirect task advantage
(\ITA). Here, only 7 reanalyzed \ITAs{} are confirmed while 3 results
yield the opposite result of a larger sensitivity in the direct task
(direct task advantage [DTA]). Error bars represent 95\%-confidence
intervals. }

\label{fig:reanalysisSmaller}
\end{figure}

\begin{figure}[!ht]
\centering
\includegraphics[ width = \textwidth, height = 19.25cm]{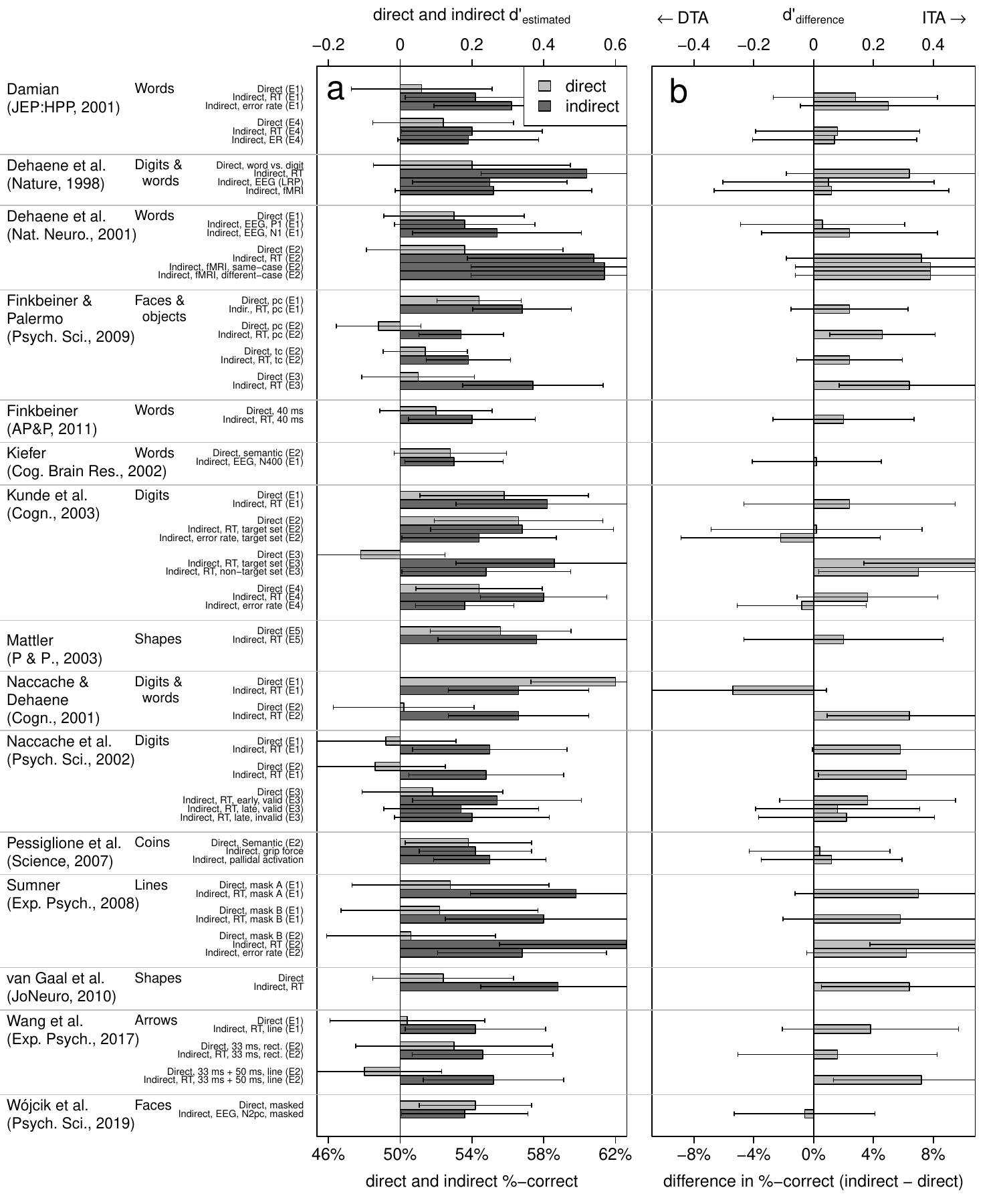}
\caption{\footnotesize
\textbf{Reanalysis with $q^2 = 0.09$.}
Same as Figure~\ref{fig:reanalysis} assuming that the standard
deviation of true sensitivities across participants is
$\text{SD}[d'_{\text{true},i}] = q = 0.3$. With this or even larger
$q^2$, reanalyzed sensitivities tend to become clearly larger in the
indirect compared to the direct task. However, this assumption is
clearly unrealistic. First, in the direct task, this would mean that a
substantial percentage of participants had a true sensitivity of
$d'_{\text{true},i} = 0.5$ or higher indicating that they could
discriminate the masked stimuli better than 60\%-correct. In the
indirect task, an unrealistic implication of this assumption is that,
in the study of \protect\citeA{Dehaene_etal_98}, trial-by-trial
reaction times (RTs) would be estimated to vary with a standard
deviation of only $\pm43$~ms (within-subject variance
$\sigma_\epsilon^2 = 43^2$) even though RTs typically vary more than
$\pm80$~ms from trial to trial, see Appendix
\protect\ref{supp:roleOfq}.
}
\label{fig:reanalysisLarger}
\end{figure}

We have repeated our literature reanalysis from Figure~5 with
different parameter values. In Figure~\ref{fig:reanalysisSmaller}, we
show a more realistic reanalysis with $q^2 = 0.01$. Here, the picture
resembles a null effect. In contrast, we show an overly optimistic
reanalysis with $q^2=0.09$ in Figure~\ref{fig:reanalysisLarger}, in
which an \ITA{} starts to emerge for many studies. However, even in
this case there is no conclusive evidence for an \ITA{} in most
studies because confidence intervals for the sensitivity difference
still include 0.

Note that, when assuming large $q^2$ values as in
Figure~\ref{fig:reanalysis} and \ref{fig:reanalysisLarger}, one cannot
take the reanalysis result as \emph{evidence for} an \ITA{}. This is
because large $q^2$-values bias our reanalysis in favor of finding an
\ITA{}. Only when we nevertheless do not find an \ITA{}, these results
can be meaningfully interpreted as \emph{evidence against} an \ITA{}.
In order to establish evidence for an \ITA{}, one would have to use
smaller values for $q^2$ or, better yet, use the trial-by-trial data
so that no assumption on $q^2$ is necessary. Otherwise, an apparent
\ITA{} result may only be due to the bias introduced by a large $q^2$.

We provide an online tool to perform our reanalysis with different
values of $q^2$ at
\url{http://www.ecogsci.cs.uni-tuebingen.de/ITAcalculator/}. There, we
suggest three different values for $q^2$: To establish a lack of
evidence for an \ITA{}, we suggest $q^2=0.0225$ as in our reanalysis
proper. This assumption rarely rejects evidence for an \ITA{} if there
is any. On the other hand, to establish evidence for an \ITA{}, we
instead suggest $q^2 = 0.0025$, which is more restrictive. Only when
an \ITA{} is established with a relatively small $q^2$ like this, we
can be sure that it is a genuine \ITA{} instead of being produced only
due to a lenient assumption on $q^2$. Lastly, we suggest an
intermediate value of $q^2 = 0.01$, which is suitable for an
exploratory reanalysis. Note that depending on the exact experimental
setup, different values of $q^2$ may be appropriate.

\subsection*{Overall Summary Regarding our Choice of $q^2$}
\noindent

Taken together, our replication, our simulations, and the literature
review suggests that $q^2$ is clearly below $0.0225$. We adopted this
upper bound as our assumption because it increases the chances of
finding an \ITA{}, thereby, following the benefit-of-the-doubt
approach. We use this assumption to show that evidence for an \ITA{}
is missing in many studies. To establish evidence for an \ITA{}, the
reanalysis would have to use smaller values to rule out the
possibility that an \ITA{} was only the product of the overestimation
bias coming from a too large $q^2$.

Up to now, we have only discussed behavioral data (RTs) but we applied
our reanalysis method also to EEG and fMRI data.  The justification
for this is that the relative noise level is even larger in
single-trial event related potentials (ERPs) and
blood-oxygen-level-dependent signals (BOLD signals) because they
incorporate much more noise \cite{stahl2010modeling}. Thus, the ratio
between effect vs. noise variance in these measures will be even
smaller, again, justifying our choice of $q^2 = 0.0225$.

\newpage

\FloatBarrier


\section{\textbf{\appendixE}}
\label{supp:reportedResults}

For each study, we give an overview of the study's structure, indicate
in a table which values we extracted and explain our decisions for in-
and exclusion of particular results. We only use results that follow
the standard reasoning, claim an \ITA{} and fit into our reanalysis
method. We include quotes from the reanalyzed studies indicating their
adherence to the standard reasoning. We use the following two
abbreviations:

\begin{enumerate}
\item[\textbf{NR}] \textbf{N}ot \textbf{r}eanalyzable: Reported
statistics do not match our reanalysis method. For example when the
congruency factor has more than two levels (congruent, incongruent,
and neutral) or when there are additional between-subject factors.
\item[\textbf{NIE}] \textbf{N}o \textbf{i}ndirect \textbf{e}ffect: The
study attempted to find an \ITA{} but failed due to a non-significant
indirect task result. In such cases, the studies usually abort the
standard reasoning, such that these cases are not relevant for us. 
\end{enumerate}

We report the number $N$ of participants, the total number of trials
$K$, and the reported statistic of the original study. Additionally,
we report the sensitivities and standard errors according to our
reanalysis. These are the values from Figure~\ref{fig:reanalysis}a. We
then report the differences in sensitivities and their standard
errors; here the difference is always taken between the current row's
indirect task compared to the previously reported direct task. These
results are presented in Figure~\ref{fig:reanalysis}b. We abbreviate
Experiment 1 by E1, etc. 

We also mark studies that excluded participants with good direct-task
performance by adding the label \textbf{Regression to the mean} (see
Discussion on why this is problematic). We still reanalyzed the
reported results, although the exclusion introduced a bias for which
our reanalysis method does not correct. This bias is liberal and
favors finding an \ITA{}. Thus, we follow the benefit-of-the-doubt
approach.

\subsection{15 Reanalyzed Studies}

\subsubsection[Damian (2001)]{\citeA{Damian_01}} 
The study reports four experiments but concludes an \ITA{} only in
Experiment~1 and 4. Experiments~2 and 3 were NIE. 

Standard Reasoning: ``Two control experiments investigated
participants’ ability to consciously perceive the masked primes. It
was shown that performance was at chance level on both
presence-absence judgments and on a number vs. random letter string
discrimination task when the temporal characteristics of a trial were
identical to those of the main experiment. Thus, the congruity effect
described above must indeed have occurred outside of the participants’
awareness'' (p.~1).

\begin{center}
\begingroup
\scriptsize
\renewcommand{\arraystretch}{0.7} 
\begin{tabular}{lrrlrr}
  \toprule 
  & \multicolumn{3}{c}{Original data} & \multicolumn{2}{c}{Our reanalysis (Figure 5)} \\
 \cmidrule(l){2-4} \cmidrule(l){5-6} 
  &  $N$  &  $K$  & Statistic & $d'_\text{estimated} \pm SE$ & $d'_\text{diff}\pm SE_\text{diff}$ \\ 
 \cmidrule(l){1-6}
 Direct (E1) & 16 & 96 & $d' = 0.064$ & $0.06\pm0.07$ &  \\ 
  Indirect, RT (E1) & 16 & 120 & $F(1,15) = 6.15$ & $0.14\pm0.07$ & $0.07\pm0.10$ \\ 
  Indirect, error rate (E1) & 16 & 120 & $F(1,15) = 13.8$ & $0.21\pm0.07$ & $0.14\pm0.10$ \\ 
   \cmidrule(l){1-6} 
Direct (E4) & 16 & 96 & $d' = 0.117$ & $0.12\pm0.07$ &  \\ 
  Indirect, RT (E4) & 16 & 120 & $F(1,15) = 5.67$ & $0.13\pm0.07$ & $0.02\pm0.10$ \\ 
  Indirect, ER (E4) & 16 & 120 & $F(1,15) = 5$ & $0.13\pm0.07$ & $0.01\pm0.10$ \\ 
   \bottomrule 
\end{tabular}
\endgroup
\end{center}

\subsubsection[Dehaene et al. (1998)]{\citeA{Dehaene_etal_98}}
The study reported two direct tasks and three indirect tasks. From the
two direct tasks, we consider only the second direct task (word vs.
digit discrimination) because it fits the neutral criterion assumption
and it also shows lower sensitivity ($d'=0.2$ in the first and $d' =
0.3$ in the second task). This way, we favor confirming the \ITA{}
hypothesis. For the first indirect measure, we computed the $t$ value
from the given estimates for the congruency effect ($M=24$~ms and
$SD=13.5$). For the second indirect measure, the statistic ($t(11)<3$)
is taken from Figure 4, where the covert activation reflects
processing of the prime as opposed processing of the target in the
overt activation. For the third indirect measure, we only considered
the congruency effect on fMRI the results are provided in Figure 5.

Standard Reasoning: ``Under these conditions, even when subjects
focused their attention on the prime, they could neither reliably
report its presence or absence nor discriminate it from a nonsense
string (Table 1). Nevertheless, we show here that the prime is
processed to a high cognitive level [by demonstrating a priming
effect].''

\begin{center}
\begingroup
\scriptsize
\renewcommand{\arraystretch}{0.7} 
\begin{tabular}{lrrlrr}
  \toprule 
  & \multicolumn{3}{c}{Original data} & \multicolumn{2}{c}{Our reanalysis (Figure 5)} \\
 \cmidrule(l){2-4} \cmidrule(l){5-6} 
  &  $N$  &  $K$  & Statistic & $d'_\text{estimated} \pm SE$ & $d'_\text{diff}\pm SE_\text{diff}$ \\ 
 \cmidrule(l){1-6}
 Direct, word vs. digit & 7 & 112 & $d' = 0.2$ & $0.20\pm0.11$ &  \\ 
  Indirect, RT & 12 & 512 & $t(11) = 6.16$ & $0.29\pm0.09$ & $0.09\pm0.14$ \\ 
  Indirect, EEG (LRP) & 12 & 512 & $t(11) = 3$ & $0.14\pm0.06$ & $-0.06\pm0.12$ \\ 
  Indirect, fMRI & 9 & 128 & $F(1,8) = 6.23$ & $0.17\pm0.10$ & $-0.03\pm0.14$ \\ 
   \bottomrule 
\end{tabular}
\endgroup
\end{center}

\subsubsection[Dehaene et al. (2001)]{\texorpdfstring{\protect\citeA{Dehaene_etal_01}}{}}
The study reports two experiments. In E1, multiple measures assessed
the visibility of the masked stimulus and we chose the reported binary
forced-choice task (no stimulus vs. masked stimulus) because it is the
most relevant result. In this experiment, the \ITA{} refers to the
absence vs. presence of the masked stimuli. The fMRI results in E1
were NR. In E2, the \ITA{} referred to the congruency effect of
repeated (congruent, either in same or in different case) vs.
different words (incongruent). 

Standard Reasoning: ``Behaviorally, participants again denied seeing
the primes and were unable to select them in a two-alternative
forced-choice test [...]. However, case-independent repetition priming
was observed in response times recorded during imaging [...]''
(p.~755) and ``As this phenomenon depends only on the identity of the
masked prime, specific information about word identity must have been
extracted and encoded unconsciously [...]'' (p.~756).

\begin{center}
\begingroup
\scriptsize
\renewcommand{\arraystretch}{0.7} 
\begin{tabular}{lrrlrr}
  \toprule 
  & \multicolumn{3}{c}{Original data} & \multicolumn{2}{c}{Our reanalysis (Figure 5)} \\
 \cmidrule(l){2-4} \cmidrule(l){5-6} 
  &  $N$  &  $K$  & Statistic & $d'_\text{estimated} \pm SE$ & $d'_\text{diff}\pm SE_\text{diff}$ \\ 
 \cmidrule(l){1-6}
 Direct (E1) & 27 & 36 & 52.9\%-correct & $0.15\pm0.09$ &  \\ 
  Indirect, EEG, P1 (E1) & 12 & 300 & $t(11) = 2.04$ & $0.10\pm0.06$ & $-0.04\pm0.10$ \\ 
  Indirect, EEG, N1 (E1) & 12 & 300 & $F(1,11) = 9.79$ & $0.16\pm0.07$ & $0.01\pm0.11$ \\ 
   \cmidrule(l){1-6} 
Direct (E2) & 10 & 64 & 53.6\%-correct & $0.18\pm0.11$ &  \\ 
  Indirect, RT (E2) & 10 & 480 & $F(1,9) = 36$ & $0.30\pm0.10$ & $0.12\pm0.15$ \\ 
  Indirect, fMRI, same-case (E2) & 10 & 240 & $t(9) = 1.98$ & $0.34\pm0.11$ & $0.16\pm0.16$ \\ 
  Indirect, fMRI, different-case (E2) & 10 & 240 & $t(9) = 2.68$ & $0.34\pm0.11$ & $0.16\pm0.16$ \\ 
   \bottomrule 
\end{tabular}
\endgroup
\end{center}

\subsubsection[Finkbeiner and Palermo (2009)]{\texorpdfstring{\protect\citeA{Finkbeiner_Palermo_09}}{}}
The study reported four experiments. Prime and target stimuli were
presented in different locations to the participants. In half of the
trials the prime location was cued (pc) and in the other half it was
the target location (tc). We excluded the target cued condition in E1
because it was NIE. In E3, multiple within-subject factors were tested
but since those do not change the reported $F$ value of the congruency
effect we could nevertheless reanalyze it. E4 did not follow the
standard reasoning.

\begin{center}
\begingroup
\scriptsize
\renewcommand{\arraystretch}{0.7} 
\begin{tabular}{lrrlrr}
  \toprule 
  & \multicolumn{3}{c}{Original data} & \multicolumn{2}{c}{Our reanalysis (Figure 5)} \\
 \cmidrule(l){2-4} \cmidrule(l){5-6} 
  &  $N$  &  $K$  & Statistic & $d'_\text{estimated} \pm SE$ & $d'_\text{diff}\pm SE_\text{diff}$ \\ 
 \cmidrule(l){1-6}
 Direct, pc (E1) & 40 & 80 & $d' = 0.22$ & $0.22\pm0.05$ &  \\ 
  Indirect, RT, pc (E1) & 40 & 80 & $F(1,39) = 33.94$ & $0.24\pm0.05$ & $0.02\pm0.07$ \\ 
   \cmidrule(l){1-6} 
Direct, pc (E2) & 40 & 80 & $d' = -0.06$ & $-0.06\pm0.05$ &  \\ 
  Indirect, RT, pc (E2) & 40 & 80 & $F(1,39) = 8.5$ & $0.12\pm0.05$ & $0.18\pm0.07$ \\ 
   \cmidrule(l){1-6} 
Direct, tc (E2) & 40 & 80 & $d' = 0.07$ & $0.07\pm0.05$ &  \\ 
  Indirect, RT, tc (E2) & 40 & 80 & $F(1,39) = 10.6$ & $0.14\pm0.05$ & $0.07\pm0.07$ \\ 
   \cmidrule(l){1-6} 
Direct (E3) & 20 & 240 & $d' = 0.05$ & $0.05\pm0.05$ &  \\ 
  Indirect, RT (E3) & 20 & 720 & $F(1,19) = 31.37$ & $0.20\pm0.05$ & $0.15\pm0.07$ \\ 
   \bottomrule 
\end{tabular}
\endgroup
\end{center}

\subsubsection[Finkbeiner (2011)]{\texorpdfstring{\protect\citeA[Regression to the
mean]{finkbeiner2011subliminal}}{}} 
The study presented trials in two conditions, one with a short
($40$~ms) and one with a long ($50$~ms) prime presentation duration.
An \ITA{} was concluded only for the short duration and with respect
to the semantic content (not color).

Standard Reasoning: ``In contrast, 16 of the 21 subjects were judged
to be at chance with the 40-ms primes. Following Rouder et al. (2007),
the RTs for the 17 subject-by-prime-duration combinations for which
subliminality was confirmed were entered into a paired-samples $t$
test (two-tailed) to determine whether subliminal priming had
occurred'' (p.~1260).

\begin{center}
\begingroup
\scriptsize
\renewcommand{\arraystretch}{0.7} 
\begin{tabular}{lrrlrr}
  \toprule 
  & \multicolumn{3}{c}{Original data} & \multicolumn{2}{c}{Our reanalysis (Figure 5)} \\
 \cmidrule(l){2-4} \cmidrule(l){5-6} 
  &  $N$  &  $K$  & Statistic & $d'_\text{estimated} \pm SE$ & $d'_\text{diff}\pm SE_\text{diff}$ \\ 
 \cmidrule(l){1-6}
 Direct, 40 ms & 21 & 120 & $d' = 0.098$ & $0.10\pm0.06$ &  \\ 
  Indirect, RT, 40 ms & 21 & 80 & $t(20) = 2.5$ & $0.14\pm0.06$ & $0.04\pm0.09$ \\ 
   \bottomrule 
\end{tabular}
\endgroup
\end{center}

\subsubsection[Kiefer 2002]{\texorpdfstring{\protect\citeA{Kiefer_02}}{}}
The study reported two experiments. E1 reported the indirect task
results and E2 reported the direct task results. In E1, indirect
effects on RT, error rates and some EEG components were NR because the
reported statistics combine masked and unmasked conditions (for
unmasked conditions, they claimed no \ITA{}) except for the N400
component in EEG. In E2, there were multiple direct tasks (see their
Table 1). We chose the direct task on semantic judgment because the
indirect task's congruency effect was an effect from semantic
relatedness too. 

Standard Reasoning: ``Average $d'$ measures in all tasks and context
conditions did not deviate significantly from zero demonstrating that
masked words were not identified'' (p.~36).

\begin{center}
\begingroup
\scriptsize
\renewcommand{\arraystretch}{0.7} 
\begin{tabular}{lrrlrr}
  \toprule 
  & \multicolumn{3}{c}{Original data} & \multicolumn{2}{c}{Our reanalysis (Figure 5)} \\
 \cmidrule(l){2-4} \cmidrule(l){5-6} 
  &  $N$  &  $K$  & Statistic & $d'_\text{estimated} \pm SE$ & $d'_\text{diff}\pm SE_\text{diff}$ \\ 
 \cmidrule(l){1-6}
 Direct, semantic (E2) & 24 & 80 & $d' = 0.14$ & $0.14\pm0.06$ &  \\ 
  Indirect, EEG, N400 (E1) & 24 & 320 & $F(1,23) = 5.48$ & $0.09\pm0.04$ & $-0.05\pm0.08$ \\ 
   \bottomrule 
\end{tabular}
\endgroup
\end{center}

\subsubsection[Kunde et al. (2003)]{\texorpdfstring{\protect\citeA{kunde2003conscious}}{}}
The study reported four experiments. In E1, there were multiple direct
task measures from which we chose the one that fit our model
assumptions of a neutral criterion (the identification rate is not
comparable by our method). Also in E1, we chose not to consider
sub-analyses of the indirect effects because they are essentially
repetitions of the same comparison. In E2, we did not consider the
non-target set condition and in E3 we did not consider the error rate
analysis as they were NIE. In E1-E3, trials with neutral primes were
not considered for calculating the priming effect. 

Standard Reasoning: ``The identification rate for the prime numbers
was 2.2\% (the chance level is 6.25\% as each prime is presented four
times in the 64 test trials). Thus, the primes were indeed
unidentifiable, as is usually found under the experimental conditions
that we adopted (Damian, 2001; Dehaene et al., 1998; Koechlin et al.,
1999; Naccache \& Dehaene, 2001)'' (p.~230).

\begin{center}
\begingroup
\scriptsize
\renewcommand{\arraystretch}{0.7} 
\begin{tabular}{lrrlrr}
  \toprule 
  & \multicolumn{3}{c}{Original data} & \multicolumn{2}{c}{Our reanalysis (Figure 5)} \\
 \cmidrule(l){2-4} \cmidrule(l){5-6} 
  &  $N$  &  $K$  & Statistic & $d'_\text{estimated} \pm SE$ & $d'_\text{diff}\pm SE_\text{diff}$ \\ 
 \cmidrule(l){1-6}
 Direct (E1) & 12 & 64 & $d' = 0.29$ & $0.29\pm0.10$ &  \\ 
  Indirect, RT (E1) & 12 & 1152 & $F(1,11) = 25.17$ & $0.22\pm0.07$ & $-0.07\pm0.12$ \\ 
   \cmidrule(l){1-6} 
Direct (E2) & 12 & 64 & $d' = 0.33$ & $0.33\pm0.10$ &  \\ 
  Indirect, RT, target set (E2) & 12 & 288 & $F(1,11) = 15.24$ & $0.20\pm0.07$ & $-0.13\pm0.12$ \\ 
  Indirect, error rate, target set (E2) & 12 & 288 & $F(1,11) = 6.35$ & $0.13\pm0.06$ & $-0.20\pm0.12$ \\ 
   \cmidrule(l){1-6} 
Direct (E3) & 12 & 64 & $d' = -0.11$ & $-0.11\pm0.10$ &  \\ 
  Indirect, RT, target set (E3) & 12 & 144 & $F(1,11) = 21.67$ & $0.28\pm0.09$ & $0.39\pm0.14$ \\ 
  Indirect, RT, non-target set (E3) & 12 & 144 & $F(1,11) = 6.58$ & $0.15\pm0.08$ & $0.26\pm0.13$ \\ 
   \cmidrule(l){1-6} 
Direct (E4) & 24 & 64 & $d' = 0.22$ & $0.22\pm0.07$ &  \\ 
  Indirect, RT (E4) & 24 & 1152 & $F(1,23) = 43.2$ & $0.21\pm0.05$ & $-0.01\pm0.08$ \\ 
  Indirect, error rate (E4) & 24 & 1152 & $F(1,23) = 9.17$ & $0.10\pm0.04$ & $-0.12\pm0.08$ \\ 
   \bottomrule 
\end{tabular}
\endgroup
\end{center}

\subsubsection[Mattler (2003)]{\texorpdfstring{\protect\citeA{mattler2003priming}}{}}
The study reports five experiments. Only Experiments~3 and~5 are
considered to be evidence for unconscious priming. Experiment~3
suffers severely from regression to the mean and is therefore not
reanalyzed. 

Standard Reasoning: ``We might assume that performance at chance level
indexes absence of all conscious information. This assumption was made
in a number of studies (e.g., Dehaene et al., 1998; Klotz \& Neumann,
1999; Neumann \& Klotz, 1994; Vorberg et al., in press). In the
present study, evidence for priming without awareness comes from
Experiment 3 and Experiment 5, in which participants showed
substantial non-motor priming effects although they could not
discriminate primes better than chance'' (p.~184) 

\begin{center}
\begingroup
\scriptsize
\renewcommand{\arraystretch}{0.7} 
\begin{tabular}{lrrlrr}
  \toprule 
  & \multicolumn{3}{c}{Original data} & \multicolumn{2}{c}{Our reanalysis (Figure 5)} \\
 \cmidrule(l){2-4} \cmidrule(l){5-6} 
  &  $N$  &  $K$  & Statistic & $d'_\text{estimated} \pm SE$ & $d'_\text{diff}\pm SE_\text{diff}$ \\ 
 \cmidrule(l){1-6}
 Direct (E5) & 11 & 320 & $d' = 0.28$ & $0.28\pm0.06$ &  \\ 
  Indirect, RT (E5) & 11 & 320 & $F(1,10) = 18.5$ & $0.22\pm0.08$ & $-0.06\pm0.10$ \\ 
   \bottomrule 
\end{tabular}
\endgroup
\end{center}

\subsubsection[Naccache and Dehaene (2001)]{\texorpdfstring{\protect\citeA{Naccache_Dehaene_01_novelStimuli}}{}}
The study reports two experiments. For the direct tasks in both
experiments, the authors additionally conducted the Greenwald method
\cite{Greenwald_etal_96,Draine_Greenwald_98} which, however, has been
criticized before
\cite{Dosher_98,Klauer_etal_98_comment,Miller_00,Merikle_Reingold_98}.
Therefore, we only considered typical results as in all other studies.
We considered only the main congruency effects on RT and no further
subanalyses because the reported direct task would not have been
comparable. In both experiments, an old and a new stimulus set were
used. In E1, we only reanalyzed the RT effect based on the old
stimulus set because the direct task sensitivity was estimated only
for the old set. In E2, we reanalyzed the RT effect for the mixed,
both new and old, stimulus set because the direct task sensitivity was
estimated for this mixed set, too.

Standard Reasoning: ``In this task, subjects performed at chance
level, while priming effects were replicated. This study provides
strong evidence for the unconscious nature of our semantic priming
effects'' (p.~227).

\begin{center}
\begingroup
\scriptsize
\renewcommand{\arraystretch}{0.7} 
\begin{tabular}{lrrlrr}
  \toprule 
  & \multicolumn{3}{c}{Original data} & \multicolumn{2}{c}{Our reanalysis (Figure 5)} \\
 \cmidrule(l){2-4} \cmidrule(l){5-6} 
  &  $N$  &  $K$  & Statistic & $d'_\text{estimated} \pm SE$ & $d'_\text{diff}\pm SE_\text{diff}$ \\ 
 \cmidrule(l){1-6}
 Direct (E1) & 18 & 32 & $d' = 0.6$ & $0.60\pm0.11$ &  \\ 
  Indirect, RT (E1) & 18 & 384 & $F(1,17) = 21.99$ & $0.19\pm0.06$ & $-0.41\pm0.12$ \\ 
   \cmidrule(l){1-6} 
Direct (E2) & 18 & 64 & $d' = 0.01$ & $0.01\pm0.08$ &  \\ 
  Indirect, RT (E2) & 18 & 384 & $F(1,17) = 21.62$ & $0.19\pm0.06$ & $0.18\pm0.10$ \\ 
   \bottomrule 
\end{tabular}
\endgroup
\end{center}

\subsubsection[Naccache et al. (2002)]{\texorpdfstring{\protect\citeA{Naccache_etal_02}}{}} 
The study reported three experiments. We did not consider the
subanalyses for cued trials as the standard reasoning only related to
the congruency effects. Note that we only counted the number of
``critical'' trials which were used in their analysis.

\begin{center}
\begingroup
\scriptsize
\renewcommand{\arraystretch}{0.7} 
\begin{tabular}{lrrlrr}
  \toprule 
  & \multicolumn{3}{c}{Original data} & \multicolumn{2}{c}{Our reanalysis (Figure 5)} \\
 \cmidrule(l){2-4} \cmidrule(l){5-6} 
  &  $N$  &  $K$  & Statistic & $d'_\text{estimated} \pm SE$ & $d'_\text{diff}\pm SE_\text{diff}$ \\ 
 \cmidrule(l){1-6}
 Direct (E1) & 12 & 240 & $d' = -0.04$ & $-0.04\pm0.06$ &  \\ 
  Indirect, RT (E1) & 12 & 240 & $F(1,11) = 7.88$ & $0.15\pm0.07$ & $0.19\pm0.09$ \\ 
   \cmidrule(l){1-6} 
Direct (E2) & 12 & 240 & $d' = -0.07$ & $-0.07\pm0.06$ &  \\ 
  Indirect, RT (E1) & 12 & 240 & $F(1,11) = 7.32$ & $0.14\pm0.07$ & $0.21\pm0.09$ \\ 
   \cmidrule(l){1-6} 
Direct (E3) & 12 & 240 & $d' = 0.09$ & $0.09\pm0.06$ &  \\ 
  Indirect, RT, early, valid (E3) & 12 & 240 & $F(1,11) = 9.23$ & $0.16\pm0.07$ & $0.07\pm0.09$ \\ 
  Indirect, RT, late, valid (E3) & 12 & 240 & $F(1,11) = 3.97$ & $0.11\pm0.06$ & $0.02\pm0.09$ \\ 
  Indirect, RT, late, invalid (E3) & 12 & 240 & $F(1,11) = 5.34$ & $0.12\pm0.07$ & $0.03\pm0.09$ \\ 
   \bottomrule 
\end{tabular}
\endgroup
\end{center}

\subsubsection[Pessiglione et al. (2007)]{\texorpdfstring{\protect\citeA[Regression to the mean]{Pessiglione_etal_07}}{}}
The study deviated from the standard priming paradigm by just showing
masked stimuli (in this case, coins) and no target stimuli.
Presentation duration was varied in three conditions. For the separate
conditions, participants were measured in one direct task and with
three indirect measures. The appendix provided the required
information for our reanalysis. We digitized their Figure S2 to derive
the $t$ values for the two indirect measures grip force and pallidal
activation. The third indirect measure, skin conductance, was NIE.
Even though these results were only reported in the appendix, the
study bases their interpretation on these results. Note, that $N=24$
relates to 24 participant $\times$ stimulus duration conditions in
which the direct task was non-significant at an individual level.

Standard Reasoning: ``Based on the percentage of correct responses,
the analysis could then be restricted to all situations where subjects
guess at chance level about stimulus identity (fig. S2) [by removing
situations with significant direct task results]. Even in these
situations, pallidal activation and hand-grip force were significantly
higher for pounds as compared to pennies [...] '' (p.~906).

\begin{center}
\begingroup
\scriptsize
\renewcommand{\arraystretch}{0.7} 
\begin{tabular}{lrrlrr}
  \toprule 
  & \multicolumn{3}{c}{Original data} & \multicolumn{2}{c}{Our reanalysis (Figure 5)} \\
 \cmidrule(l){2-4} \cmidrule(l){5-6} 
  &  $N$  &  $K$  & Statistic & $d'_\text{estimated} \pm SE$ & $d'_\text{diff}\pm SE_\text{diff}$ \\ 
 \cmidrule(l){1-6}
 Direct, Semantic (E2) & 24 & 60 & $d' = 0.19$ & $0.19\pm0.07$ &  \\ 
  Indirect, grip force & 24 & 90 & $t(23) = 2.92$ & $0.15\pm0.06$ & $-0.04\pm0.09$ \\ 
  Indirect, pallidal activation & 24 & 90 & $t(23) = 3.41$ & $0.17\pm0.06$ & $-0.02\pm0.09$ \\ 
   \bottomrule 
\end{tabular}
\endgroup
\end{center}

\subsubsection[Sumner (2008)]{\texorpdfstring{\protect\citeA[Regression to the mean]{sumner2008mask}}{}}
The study reported two experiments. Both, E1 and E2, had different
mask conditions (A vs. B). Only E1 provided indirect task results such
that we could reanalyze both conditions separately. For E2 we had to
apply our reanalysis to both conditions aggregated. Therefore, we
averaged over the given $d'$ values from both conditions. We did not
consider the subanalyses on the difference and interaction between the
two masks but only the congruency effects as they are taken for the
standard reasoning.

\begin{center}
\begingroup
\scriptsize
\renewcommand{\arraystretch}{0.7} 
\begin{tabular}{lrrlrr}
  \toprule 
  & \multicolumn{3}{c}{Original data} & \multicolumn{2}{c}{Our reanalysis (Figure 5)} \\
 \cmidrule(l){2-4} \cmidrule(l){5-6} 
  &  $N$  &  $K$  & Statistic & $d'_\text{estimated} \pm SE$ & $d'_\text{diff}\pm SE_\text{diff}$ \\ 
 \cmidrule(l){1-6}
 Direct, mask A (E1) & 12 & 40 & $d' = 0.14$ & $0.14\pm0.12$ &  \\ 
  Indirect, RT, mask A (E1) & 12 & 200 & $t(11) = 5.5$ & $0.30\pm0.10$ & $0.16\pm0.15$ \\ 
   \cmidrule(l){1-6} 
Direct, mask B (E1) & 12 & 40 & $d' = 0.11$ & $0.11\pm0.12$ &  \\ 
  Indirect, RT, mask B (E1) & 12 & 200 & $t(11) = 4.5$ & $0.25\pm0.09$ & $0.14\pm0.15$ \\ 
   \cmidrule(l){1-6} 
Direct, mask B (E2) & 12 & 80 & 50.5\%-correct & $0.03\pm0.09$ &  \\ 
  Indirect, RT (E2) & 12 & 400 & $t(11) = 7.4$ & $0.36\pm0.10$ & $0.33\pm0.14$ \\ 
  Indirect, error rate (E2) & 12 & 400 & $t(11) = 4$ & $0.19\pm0.07$ & $0.17\pm0.12$ \\ 
   \bottomrule 
\end{tabular}
\endgroup
\end{center}

\subsubsection[van Gaal et a. (2010)]{\texorpdfstring{\protect\citeA[Regression to the
mean]{van2010unconscious}}{}} 
The study reported one experiment with one direct task and multiple
indirect measures. However, we only considered the indirect effect on
RTs as the fMRI analyses were NR.

Standard Reasoning: ``[...] a, Participants were unable to
discriminate between trials with a strongly masked square or diamond,
as revealed by chance-level performance in a two-choice discrimination
task administered after the main experiment. b, Although strongly
masked no-go signals could not be perceived consciously, they still
triggered inhibitory control processes, as revealed by significantly
longer response times on these trials than on strongly masked go
trials.'' (in Figure 2, p.~4145).

\begin{center}
\begingroup
\scriptsize
\renewcommand{\arraystretch}{0.7} 
\begin{tabular}{lrrlrr}
  \toprule 
  & \multicolumn{3}{c}{Original data} & \multicolumn{2}{c}{Our reanalysis (Figure 5)} \\
 \cmidrule(l){2-4} \cmidrule(l){5-6} 
  &  $N$  &  $K$  & Statistic & $d'_\text{estimated} \pm SE$ & $d'_\text{diff}\pm SE_\text{diff}$ \\ 
 \cmidrule(l){1-6}
 Direct & 20 & 48 & $d' = 0.118$ & $0.12\pm0.09$ &  \\ 
  Indirect, RT & 20 & 240 & $t(19) = 6.24$ & $0.27\pm0.06$ & $0.15\pm0.11$ \\ 
   \bottomrule 
\end{tabular}
\endgroup
\end{center}

\subsubsection[Wang et al. (2017)]{\texorpdfstring{\protect\citeA{wang2017role}}{}}
The study reported two experiments. In E1, there were two outline
conditions, line vs. rectangle. The line condition yielded a negative
congruency effect which we treated similar to a standard (positive)
priming effect. The rectangle condition was NIE. In E2, the rectangle
condition with prime duration of $50$~ms produced a large $d'$ so that
no \ITA{} was claimed. Hence, we only considered the rectangle
condition only for $33$~ms. For the line condition, $33$~ms and
$50$~ms trials were analyzed together since there was no interaction
effect. 

Standard Reasoning: ``The results from the FC task indicated that
similar prime visibility, equivalent to chance level, was obtained in
the two preposed object type conditions. This finding confirmed that
primes were processed subliminally in the primary task'' (p.~425).

\begin{center}
\begingroup
\scriptsize
\renewcommand{\arraystretch}{0.7} 
\begin{tabular}{lrrlrr}
  \toprule 
  & \multicolumn{3}{c}{Original data} & \multicolumn{2}{c}{Our reanalysis (Figure 5)} \\
 \cmidrule(l){2-4} \cmidrule(l){5-6} 
  &  $N$  &  $K$  & Statistic & $d'_\text{estimated} \pm SE$ & $d'_\text{diff}\pm SE_\text{diff}$ \\ 
 \cmidrule(l){1-6}
 Direct (E1) & 15 & 64 & $d' = 0.02$ & $0.02\pm0.09$ &  \\ 
  Indirect, RT, line (E1) & 15 & 208 & $F(1,14) = 6.86$ & $0.13\pm0.06$ & $0.11\pm0.11$ \\ 
   \cmidrule(l){1-6} 
Direct, 33 ms, rect. (E2) & 15 & 32 & $d' = 0.15$ & $0.15\pm0.12$ &  \\ 
  Indirect, RT, 33 ms, rect. (E2) & 15 & 208 & $F(1,14) = 8.15$ & $0.14\pm0.06$ & $-0.01\pm0.14$ \\ 
   \cmidrule(l){1-6} 
Direct, 33 ms + 50 ms, line (E2) & 15 & 64 & $d' = -0.103$ & $-0.10\pm0.09$ &  \\ 
  Indirect, RT, 33 ms + 50 ms, line (E2) & 15 & 416 & $F(1,14) = 11.47$ & $0.15\pm0.06$ & $0.25\pm0.11$ \\ 
   \bottomrule 
\end{tabular}
\endgroup
\end{center}

\subsubsection[W\'ojcik et al. (2019)]{\texorpdfstring{\protect\citeA{wojcik2019unconscious}}{}}
The study reported one experiment with masked and unmasked conditions.
We only considered the masked condition for which an \ITA{} was
claimed but not the unmasked condition. In the direct task, we had to
compute average $d'$ from the openly accessible material. In the
indirect task, EEG components were measured. For EEG preprocessing,
some trials had to be rejected leading to an average of 131 trials. We
assumed that rejection rate was approximately equal in the two
indirect task conditions.

Standard Reasoning: ``Analysis of the sensitivity measure $d'$
indicated that faces were not consciously identified in the masked
condition. A clear N2 posterior-contralateral (N2pc) component (a
neural marker of attention shifts) was found in both the masked and
unmasked conditions, revealing that one’s own face automatically
captures attention when processed unconsciously'' (in the abstract,
p.~471).

\begin{center}
\begingroup
\scriptsize
\renewcommand{\arraystretch}{0.7} 
\begin{tabular}{lrrlrr}
  \toprule 
  & \multicolumn{3}{c}{Original data} & \multicolumn{2}{c}{Our reanalysis (Figure 5)} \\
 \cmidrule(l){2-4} \cmidrule(l){5-6} 
  &  $N$  &  $K$  & Statistic & $d'_\text{estimated} \pm SE$ & $d'_\text{diff}\pm SE_\text{diff}$ \\ 
 \cmidrule(l){1-6}
 Direct, masked & 18 & 160 & $d' = 0.211$ & $0.21\pm0.06$ &  \\ 
  Indirect, EEG, N2pc, masked & 18 & 131 & $t(17) = 2.34$ & $0.12\pm0.06$ & $-0.09\pm0.08$ \\ 
   \bottomrule 
\end{tabular}
\endgroup
\end{center}


\section{\textbf{\appendixF}}
\label{supp:bayes}

The main fallacy of the standard reasoning persists independently of
which statistical methods are chosen (significance testing or Bayesian
analysis). It comes from evaluating the two tasks separately instead
of using the appropriate analysis of measuring a difference between
direct vs. indirect task sensitivities. To see why problems occur in
both methods, consider the following simulation demonstrating the cost
of dichotomization.

In multiple runs, we simulate one data set by sampling responses from
$N=12$ participants and $K=256$ trials per participant. Thus, we
sample $K/2 = 128$ observations in each of two conditions based on two
normal distributions that are shifted by $d'_\text{true} = 0.15$
standard deviations (corresponding to a true performance of
53\%-correct; log-normal distributions produce similar results). We
analyze this one data set (a) as in the indirect task and (b) as in
the direct task. We will show that both methods, significance testing
and Bayesian analysis, produce misleading results in favor of the
indirect task even though the \emph{exact same data} is the basis for
both tasks.

\begin{figure}[!ht]
\centering
\includegraphics[width=\textwidth]{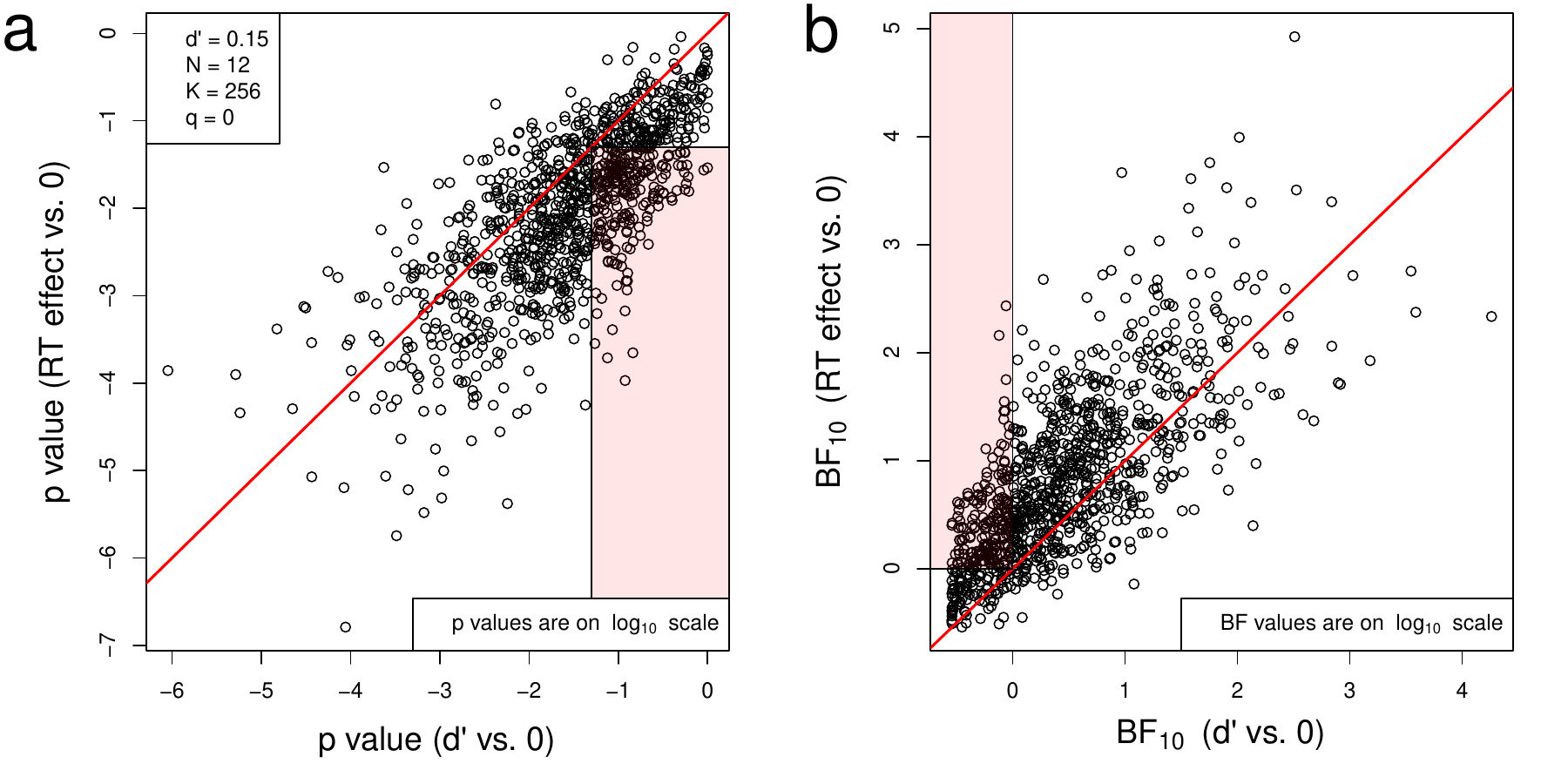}

\caption{\footnotesize \textbf{Cost of Dichotomization in significance
testing and Bayesian analysis.} Each point corresponds to one
simulated data set. We analyzed each data set as in the direct task
($x$ axis) and indirect task ($y$ axes). We find that $p$ values in
(a) as well as Bayes Factors in (b) diverge from the red equality line
indicating more evidence in the indirect task due to the loss of
information from median splitting the data in the direct task. Shaded
regions indicate a misleading pattern of result: (a) a significant
indirect task vs. a non-significant direct task result; (b) a Bayes
Factor supporting the null hypothesis in the direct task vs. a Bayes
Factor supporting the alternative hypothesis in the indirect task.}

\label{fig:costOfDichotomization}
\end{figure}

To mimic the RT effect from the indirect task, we tested the mean
difference between two conditions against 0 (y axes in
Figure~\ref{fig:costOfDichotomization}). To mimic the direct task, we
conducted a median split and tested sensitivity $d'$ against 0 (x axes
in Figure~\ref{fig:costOfDichotomization}). The assumption here is
that participants have access to the same information in both tasks
and were forced to give a binary response (dichotomize) in the direct
task so that the best they could do is to respond according to the
optimal median split criterion. To test against 0, we used a $t$ test
(see Figure~\ref{fig:costOfDichotomization}a) and we computed Bayes
Factor (Figure~\ref{fig:costOfDichotomization}b) using the R package
provided by \citeA{morey2015package}

Inspecting the results in Figure~\ref{fig:costOfDichotomization}, we
find that $p$ values and Bayes Factors diverge from the red equality
line indicating more evidence for an effect in the indirect task
analysis compared to the direct task analysis. This is so because a
median split dichotomization discards information \cite{Cohen_83}
producing larger $p$ values and smaller Bayes Factors in the direct as
compared to the indirect task.

In 23\% of the simulations, there is a non-significant direct task vs.
a significant indirect task result (shaded area in
Figure~\ref{fig:costOfDichotomization}a). This pattern may mislead
researchers into thinking that there is an effect in the indirect task
but none in the direct task.  Note that this is a well--known error:
One cannot take a non-significant result as evidence for the absence
of an effect without a power analysis \cite<see for
example>{vadillo2020unconscious}.

The pattern of results from Bayes Factors is misleading in an even
more severe way. In 20\% of the simulations, we find Bayes Factors
supporting the null hypothesis of no effect in the direct task
($BF_{10}<1$) and simultaneously supporting the alternative hypothesis
in the indirect task ($BF_{10}>1$; on the log scale these are values
below and above 0, see shaded area in
Figure~\ref{fig:costOfDichotomization}b). We even found some
simulations, in which there is substantial evidence for the
\emph{null} hypothesis in the direct task ($BF_{10}<1/3$) and
substantial evidence for the \emph{alternative} hypothesis in the
indirect task ($BF_{10}>3$). That is, if we ignored the main fallacy
of the standard reasoning and followed the Bayesian analysis naively,
we would conclude a difference in the two tasks even though the
analyses in both tasks is based on the exact same data!

Analyzing the simulated data separately---computing mean difference in
the indirect task and sensitivity in the direct task---produces
misleading patterns of results. This problem occurs independent of the
statistical methods used, significance testing or Bayes analysis, and
even if the exact same data underlies both tasks. In a real
experiment, direct and indirect tasks would not be based on the exact
same data but on two samples, which produces additional measurement
error. But in our idealized simulation here, there is no additional
sampling error because both tasks are based on the same sample. Hence,
no difference between the two tasks should be found. Accordingly, the
appropriate analysis based on the sensitivity comparison would find
exactly $d'_{\text{estimated, indirect}} - d'_\text{estimated, direct}
= 0$ correctly identifying no difference between the two tasks and
solving this problem.


\newpage
\section{\textbf{\appendixG}}
\label{supp:symbols}

\begin{table}[h!]
\caption{Description of variables.}
\begin{tabular}{ p{1.5cm} p{13cm} }
\hline                  
        Variable & Description \\
  \hline   
        $c_j$ & Condition effect, for example the congruent condition ($j=1$) produces faster RTs so that $c_1<0$ and $c_2=1-c_1>0$. \\
        $c_{N,K,q^2}$ & Constant relating $t$ values to the estimated sensitivity in the indirect task, $d'_{\text{true}} = c_{N,K,q^2} \cdot t$. It depends on $N$, $M$ and $q$. \\
       
        $d' $& Observed, average sensitivity index, estimates the true sensitivity $d'_\text{true}$.  \\
        $d'_i$ & Observed, individual sensitivity indices, estimates the true, individual sensitivities $d'_{\text{true},i}$. \\

        $d'_\text{true}$ & True sensitivity, $d'_\text{true} = \frac{\Delta}{\sigma_\epsilon}$. \\
        $d'_\text{estimated}$ & Estimated sensitivity from the reported summary statistics in the direct ($d'_\text{estimated,direct}$) or indirect task ($d'_\text{estimated,indirect}$). \\
        
        $d'_{\text{true},i}$ & Individual sensitivity, $d'_{\text{true},i} = \frac{\Delta_i}{\sigma_\epsilon}$. \\

        $\Delta$ & The true difference between conditions, $\Delta =
        c_2 - c_1$. \\
        $\hat{\Delta}$ & The observed, mean difference between conditions. \\
        
        $\Delta_i$ & True, individual effects, $\Delta_i = c_2 + (p\times c)_{i2} - (c_1 + (p\times c)_{i1})$, for example the expected congruency effect between conditions of participant $i$.\\    
        $\hat{\Delta}_i$ & The observed difference between conditions of participant $i$. \\
        $\epsilon_{ijk}$ & trial-by-trial error, noise due to
        measurement error or random neuronal fluctuations. \\
        $f_{opt}(x)$ & Optimal classifier taking indirect measures $x$ (e.g., RTs) and predicting the condition (congruent/incongruent). \\
        $f_{t}(x)$ & Threshold classifier predicting one condition for indirect measures $x\leq t$ (e.g., RTs) and the other for $x>t$. \\
        $h$ & Linear approximation used to translate between sensitivities and accuracies.\\
        $i$ & Index for participant $i \in \{1,2,...,N\}$.\\
        
        $j$ & Index for condition $j \in \{1,2\}$, for example indicator for congruent ($j=1$) and incongruent ($j=2$) conditions. \\

        $K$ & Total number of trials per particpant, $K = 2M$. \\
        $k$ & Index for trial $k \in \{1,2,...,M\}$. Since there are two conditions, the number of observed trials per participant is $2M=K$. \\
        $M$ & Number of trials per participant $\times$ condition. The total
        number of trials per participant is $2M=K$. \\
        $\mu$ & Grand mean, for example the overall expected value of RTs. \\
        $N$ & Number of participants. \\
        
        $\Omega(x)$ & Marginal, cumulative density distribution (CDF) over indirect measures $x$. \\
        $p_i$ & Participant effect, for example participants with a faster RTs than average have a negative $p_i$ while slower participants have a positive $p_i$. \\
        $(p\times c)_{ij}$ & Interaction effect, for example some participants have different reaction time effects. \\  
        $\pi$ & True accuracy. \\
        $\hat{\pi}$ & Observed, mean accuracy. \\

        $\pi_i$ & True accuracy of
        participant $i$. It can be translated into a sensitivity by
        $d'_{\text{true},i} = 2\Phi^{-1}(\pi_i)$ where $\Phi$ is the        
        cumulative normal distribution. \\
        $\hat{\pi}_i$ & Observed, individual accuracy.\\   
        $q^2$ & Ratio between effect variance and trial-by-trial error variance, $q^2~=~\frac{\sigma^2_\text{effect}}{\sigma^2_\epsilon}$. This is the variance of true sensitivities across individuals, $q^2 = \text{Var}[d'_{\text{true},i}]$. A reasonable value in our setting is $q^2 = 0.0225$ implying $\text{SD}[d'_{\text{true},i}] = 0.15$. \\
  \hline  
\end{tabular}
\end{table}

\begin{table}[h!]
\addtocounter{table}{-1}
\caption{(continued).}
\begin{tabular}{ p{1.5cm} p{13cm} }
\hline                  
        Variable & Description \\
  \hline   

        $SE$ & Estimated standard error of the estimated sensitivity. \\
        $\sigma^2_{{\Delta}_i}$ & Variance of true individual effects, for example, to which degree participants vary in their congruency effect.\\
        $\sigma^2_{\hat{\Delta}_i}$ & True variance of observed individual effects, for example, variance of the observable congruency effects.\\
        $\hat\sigma^2_{\hat{\Delta}_i}$ & Estimated variance of observed individual effects. This is what scientists get when computing the variance on the observable congruency effects across participants.\\
        $\sigma^2_{p\times c}$ & Variance of the interaction effect, 
        $(p \times c)_{ij}$. \\
        $\sigma^2_\text{effect}$ & Variance of the effects $\Delta_i$, $\sigma^2_\text{effect} = 4\sigma^2_{p\times c}$. \\
        $\sigma^2_\epsilon$ & Variance of the trial-by-trial error,
        $\epsilon_{ijk}$. \\
        $t$ & $t$ value, in our context it comes from paired-$t$-tests
        between the two conditions of the indirect task. \\
        $Y_{ijk}$ & Response of participant $i$ in condition $j$ trial $k$ from the direct ($Y_{ijk}^{\text{dir}}$) or indirect task ($Y_{ijk}^{\text{indir}}$). The standard repeated measures ANOVA model is $Y_{ijk}~=~\mu~+~p_i~+~c_j~+~(p\times c)_{ij}~+~\epsilon_{ijk}$. \\
  \hline  
\end{tabular}
\end{table}

\newpage 
\begingroup

\section{References}
\let\clearpage\relax
\label{supp:references}
\bibliographystyle{apacite}

\makeatletter
\let\st@rtbibsection\@bibnewpage
\let\st@rtbibchapter\@bibnewpage
\makeatother


\endgroup

\end{document}